\documentclass[a4paper]{cas-sc}
\usepackage{etoolbox} 
\usepackage{dsfont} 
\usepackage{lastpage}
\usepackage[colorlinks]{hyperref}
\usepackage{moreverb}
\usepackage{xurl}
\usepackage[svgnames,dvipsnames]{xcolor}
\usepackage{afterpage}
\usepackage{amscd}
\usepackage{upgreek}
\usepackage{amsmath,amsfonts,amssymb}
\usepackage{verbatim}
\usepackage[utf8]{inputenc}
\usepackage{geometry} 
\DeclareMathOperator{\E}{\mathds{E}}
\usepackage{graphicx}
\usepackage[numbers]{natbib}
\usepackage{rotating}
\usepackage{mathtools}
\usepackage{hvfloat,lipsum}
\usepackage[ruled,vlined]{algorithm2e}
\usepackage{hhline}
\usepackage{threeparttable}
\usepackage{adjustbox}
\usepackage{blindtext}
\usepackage[]{caption}
\usepackage[]{appendix}
\usepackage{algpseudocode}
\usepackage{subcaption}

\def\tsc#1{\csdef{#1}{\textsc{\lowercase{#1}}\xspace}}
\tsc{WGM}
\tsc{QE}

\begin{document}
\let\WriteBookmarks\relax
\def\floatpagepagefraction{1}
\def\textpagefraction{.001}

\shorttitle{}

\title [mode = title]{Multi-agent deep reinforcement learning with centralized training and decentralized execution for transportation infrastructure management}

\affiliation[1]{organization={Department of Civil and Environmental Engineering},
            addressline={The Pennsylvania State University}, 
            city={University Park},
            postcode={16802}, 
            state={PA},
            country={USA}}

\author[1]{M. Saifullah}

\author[1]{K. G. Papakonstantinou}
\cormark[1]
\ead{kpapakon@psu.edu}

\author[1]{A. Bhattacharya}

\author[1]{S. M. Stoffels}

\author[2]{C. P. Andriotis}

\affiliation[2]{organization={Faculty of Architecture \& the Built Environment},
            addressline={Delft University of Technology}, 
            postcode={2628 BL}, 
            state={Delft},
            country={The Netherlands}}

\cortext[1]{Corresponding author}

\setcounter{page}{1}
\begin{abstract}
Life-cycle management of large-scale transportation systems is a computationally intensive task. It requires determining a sequence of inspection and maintenance decisions to minimize long-term risks and costs while dealing with multiple uncertainties and constraints that lie in high-dimensional spaces. Traditional approaches, such as static age- or condition-based maintenance and risk-based or periodic inspection plans, have been widely applied but often suffer from limitations related to optimality, scalability, and the ability to properly handle uncertainty in the loop of optimization. Moreover, many existing methods rely on unconstrained formulations that overlook critical hard and soft operational constraints. We address these issues in this work by casting the optimization problem within the framework of constrained Partially Observable Markov Decision Processes (POMDPs), which provide a robust mathematical foundation for stochastic sequential decision-making under observation uncertainties, in the presence of risk and resource limitations. To tackle the high dimensionality of state and action spaces, we propose DDMAC-CTDE, a Deep Decentralized Multi-Agent Actor-Critic (DDMAC) reinforcement learning architecture with Centralized Training and Decentralized Execution (CTDE). To demonstrate the utility of the proposed framework, we also develop a new comprehensive benchmark environment representing an existing transportation network in Virginia, U.S., with heterogeneous pavement and bridge assets undergoing nonstationary degradation. This environment incorporates multiple practical constraints related to budget limits, performance guidelines from State and Federal agencies, traffic delays, and risk considerations. On this benchmark, DDMAC-CTDE consistently outperforms standard transportation management baselines, producing more reliable, cost- and risk-efficient policies that respect operational constraints. Together, the proposed framework and benchmark provide (i) a scalable, constraint-aware methodology for sequential decision-making under uncertainty, and (ii) a realistic,  rigorous testbed for comprehensive evaluation of Deep Reinforcement Learning  (DRL) for transportation infrastructure management.

\end{abstract}


\maketitle

\section{Introduction}\label{intro}

\noindent
The effective management of transportation infrastructure remains a significant challenge, requiring the development of advanced inspection and maintenance policies that address stochastic degradation processes while adhering to stringent performance and resource constraints. This issue is particularly pressing in light of infrastructure conditions in many countries worldwide, as also revealed by the 2025 ASCE Infrastructure Report Card \cite{ASCE2025}. Several economic analyses also highlight a substantial funding deficit in preserving infrastructure to acceptable societal standards, both at the Federal and State levels \cite{ASCE2025, vdot2021a}. For example, and only illustratively, the Virginia Department of Transportation (VDOT) reported that 50\% of the state’s bridges have exceeded their intended service lives and the projected costs to replace these aging structures over the next 50 years are estimated to be five times greater than the anticipated available funding \cite{vdot2021a}. These facts further underscore the urgent need for innovative and resource-efficient strategies to ensure the sustainability, safety, and performance of transportation infrastructure systems.

Intelligent inspection and maintenance (I\&M) planning is key for the efficient allocation of economic and other resources in such systems. Developing optimal policies in complex and uncertain environments poses significant computational challenges, stemming from the heterogeneity of asset classes, the large population of components leading to vast state and action spaces, the imperfect and unreliable observations, the scarcity of available resources, and the multiple associated risks. Despite progress, there remains a shortage of scalable computational frameworks that can jointly handle features such as (i) online and offline data-driven learning, (ii) imperfect information support, (iii) stochastic action outcomes, and (iv) optimization of long-term goals under multiple constraints (e.g., safety targets or resource constraints). Computational challenges, such as the curse of dimensionality and history, further complicate this objective \cite{bellman1957a, pineau2003a}. Overcoming all these challenges necessitates the development of advanced computational frameworks capable of integrating these diverse elements while ensuring scalability and computational feasibility. At the same time, standardized, sufficiently realistic, detailed, and rigorous benchmarks that can reflect the dynamic, stochastic, and constrained nature of transportation infrastructure management are also currently lacking in the literature. This impedes fair comparison, reproducibility, and collective progress. Developing comprehensive planning environments, with realistic heterogeneity, observation noise, resource limits, and risk targets, would enable systematic evaluation and accelerate the design of appropriate, safe, and scalable I\&M policies.

Traditional computational methods face significant limitations in addressing these multifaceted challenges within I\&M optimization. Over the years, numerous advanced techniques have been proposed, including genetic algorithms, threshold-based heuristics, risk-informed approaches, renewal processes, and decision-tree frameworks \cite{bocchini2011a, sanchez2016maintenance, AJRUA60001104, nguyen2015multi, yang2019life}. Among these, risk-based approaches have garnered particular attention, yielding practical solutions \cite{bismut2021optimal, yang2021risk, han2022risk, han2023life}, but they often struggle with large-scale systems, computational inefficiencies, and suboptimal scalability. Additionally, many rely on static formulations that fail to capture the dynamic and stochastic characteristics of infrastructure deterioration and maintenance needs, limiting their applicability for real-world, non-stationary systems. These shortcomings are also evident in transportation asset management studies, such as \cite{saydam2015risk, CHU20121123, frangopol2007, azad2016disruption}. A broader review of multi-asset approaches, including several transportation-focused studies, is presented in \cite{petchrompo2019review}. Although such multi-asset frameworks provide a more holistic perspective on network-level optimization, as emphasized in \cite{petchrompo2019review} and the extensive references therein, current state-of-the-art methods remain unable to inclusively address all the previously mentioned challenges and the core concepts (i)–(iv) above. In addition, none of the currently available studies offer a truly comprehensive planning environment capable of managing a large number of components with dynamic, adaptive planning needs, diverse optimization objectives, and various uncertainties, all while respecting real-world operational performance and resource constraints.

Stochastic optimal control approaches based on Markov Decision Processes (MDPs) and Partially Observable Markov Decision Processes (POMDPs) can tackle many of the aforementioned challenges \cite{frangopol2007bridge, papakonstantinou2014corrod, papakonstantinou2018a}. POMDPs' dynamic programming principles alleviate the curse of history and enable adaptive reasoning in the presence of noisy real-time data \cite{kaelbling1998a}. Numerous research efforts, including \cite{papakonstantinou2014a, papakonstantinou2014partII, papakonstantinou2016point, memarzadeh2016aIntegrated, schoebi2016a, memarzadeh2015optimal}, have explored and proved the successful POMDP use and efficient performance in I\&M planning problems. In  \cite{MORATO2022102140}, POMDPs are also combined with dynamic Bayesian networks, to balance the costs of maintenance actions and observations against the risk of structural failures, whereas in \citep{arcieri2023bridging} a robust POMDP approach is suggested, considering model uncertainty. POMDPs have also been used for pavement management, as in \citep{madanat1993a, madanat1994a, guillaumot2003adaptive}, while in  \cite{ZHANG2023104054} a stochastic condition-based I\&M optimization is suggested, and in \cite{SHON2021103328} a bottom-up dynamic programming solution is presented for the optimization problem. There are many ways to efficiently solve POMDPs to global optimality, e.g. \cite{pineau2003a,papakonstantinou2018a, spaan2005a, shani2013a,lovejoy1991survey}. However, unfortunately, most traditional algorithms still fail to efficiently scale to large systems with multiple components. 

To address relevant scalability issues, neural networks parameterizations can be utilized, through the field of Deep Reinforcement Learning (DRL) \citep{mnih2015human, schulman2017proximal,andriotis2019managing}. Several works have thus been recently published, examining DRL performance and capabilities for I\&M planning problems. For example, in \cite{lei2022deep}, DQN learning is used for life-cycle maintenance planning of deteriorating bridges under different budget constraints, while in  \cite{DU2022102221} DQN is used for sequential maintenance decisions for highway bridge portfolios. In \cite{HAMIDA2023109214} hierarchy in actions and a Dueling DQN is used for maintenance planning of bridges, whereas a Double DQN with prioritized replay buffer is used in \cite{MOHAMMADI2022108615} for rail renewal and maintenance planning, and a POMDP-DRL framework for inference and robust solutions is presented in \cite{arcieri2023pomdp}, together with an application to railway optimal maintenance. Several other of many notable relevant approaches can be seen in \citep{yang2022deep,lei2023sustainable, Lai2024SynergeticDRL, sresakoolchai2023railway}. Other diverse applications include aircraft maintenance \citep{LEE2023108908}, traffic and signal planning at intersections \cite{ZHANG2023104063, YAZDANI2023103991}, dynamic bus scheduling \cite{wu2024dynamic}, and many more.

Most DRL approaches for transportation I\&M planning are designed within single-agent frameworks, which face significant scalability challenges in multi-component systems due to the exponential growth of joint state and action spaces \cite{fardyuan}. Attending to this issue, one of the early Multi-Agent DRL (MADRL) techniques for I\&M, the Deep Centralized Multi-agent Actor-Critic (DCMAC) algorithm, has been proposed in \cite{andriotis2019managing, andriotis2019lifecycle}, which is a subset of the larger family of actor-critic DRL approaches \cite{wang2016a, degris2012off}. DCMAC addresses partial observability by making use of belief-state MDPs, and therefore operates directly on the posterior probabilities of system states given previous actions and observations. DCMAC has also been applied to address a bilevel optimization problem of scheduling roadway improvements while considering traffic impacts associated with maintenance in \citep{zhou2022reinforcement}. \citep{EM194378890002028} suggested a DCMAC variant by incorporating an entropy term, applying the approach to the life-cycle maintenance of bridge girder systems, while \citep{zhang2024multi} also utilized a centralized multi-agent framework using PPO \citep{schulman2017proximal} to optimize life-cycle maintenance decisions for bridge networks, considering the interdependencies between bridge components. 

To enable decentralized parameter training, the Deep Decentralized Multi-agent Actor-Critic (DDMAC) is proposed in \cite{andriotis2021deep}, where now a decentralized independent actor unit represents each component, and their outputs are utilized to build a centralized value function, which updates the actor and critic networks using policy gradients and temporal difference error estimates. DDMAC is also used to compute optimal system level I\&M policies, for systems with correlated fatigue deterioration processes among components \cite{MORATO2023109144}, and a new Maintainer-Inspector architecture for sequential actions is presented in \cite{andriotis2022optimizing, SaifIcasp14}, based on the DDMAC framework, found to be notably efficient for settings with large combinations of available I\&M actions. Beyond DDMAC, alternative multi-agent deep reinforcement learning algorithms have also led to advancements in maintenance policies and cost-effectiveness. In \cite{nguyen2022artificial, nguyen2022weighted, do2024multi} a MADRL method based on the value decomposition framework of \cite{rashid2020weighted} is presented, while \cite{lee2023risk} presents a parallelized multi-agent deep Q-network algorithm for risk-informed lifeline system maintenance. The Value-Decomposition Actor-Critic (VDAC) algorithm \cite{su2021value} is employed in \cite{SU2022116323}, for preventive maintenance policies of manufacturing systems, and \cite{wang2016multi} devises multi-agent maintenance strategies for resource-constrained flow line systems. Similarly, \cite{rodriguez2022multi} capitalizes on MADRL for predictive maintenance on parallel machines, showcasing substantial performance enhancements. Several MADRL approaches and benchmarks for asset management can also be seen in \cite{leroy2023imp}. 

In this work, we advance research on deep reinforcement learning (DRL) for transportation infrastructure applications by  developing the first multi-asset transportation infrastructure management environment accounting for realistic, high-fidelity system modeling and several, diverse, deterministic and probabilistic agency constraints. This environment is designed for optimizing I\&M under various operational metrics and uncertainties. Next, to enable and scale the DDMAC architecture for even larger systems, we introduce it here to the Centralized Training and Decentralized Execution (CTDE) paradigm, a notable framework in cooperative multi-agent reinforcement learning (MARL) built on the Decentralized-POMDP (Dec-POMDP) formalism \cite{oliehoek2016concise,bernstein2002complexity}. This DDMAC-CTDE architecture provides now only local information as input to decentralized actors, reflecting the condition states of individual components. This new design substantially enhances scalability capabilities while preserving centralized gradient information flow through a single critic, consistent with the original DDMAC framework. The improved algorithm is evaluated in a realistic, large-scale setting. We formulate the POMDP optimization problem within the benchmark environment representing an existing real-world transportation network at Hampton Roads, Virginia, U.S.A, with two classes (pavements and bridges) and multiple subclasses of assets  under numerous resource, performance, and risk constraints. A detailed modeling environment of both pavements and bridges is developed based on their corresponding condition state indices. Stationary and non-stationary Markovian transition probabilities are obtained for different classes and subclasses of assets. Four types of maintenance actions are considered, namely Do-Nothing, Minor Repair, Major Repair, and Reconstruction. Observation actions are, in turn, defined as no-inspection, low-fidelity, and high-fidelity inspection for both asset types. The maintenance and inspection action costs and effects are obtained from actual data and complemented by up-to-date literature, where needed. Extensive numerical results are presented demonstrating the effectiveness and applicability of the proposed DDMAC-CTDE formulation for this type of large-scale, extremely challenging system optimization problems, with derived solutions outperforming state-of-the-art heuristics and currently in-place policies by a significant margin, while satisfying the imposed performance and budget constraints. Overall, this work importantly delivers (i) a scalable computational framework for DRL-driven I\&M, and (ii) a realistically complex benchmark for fair, demanding, reproducible evaluation. These two contributions lay together a solid and necessary foundation for even further future advances in real-world DRL-driven management and engineering decision-making, broadly defined.

\section{Background}\label{background}
\subsection{Partially Observable Markov Decision Processes}\label{POMDP}
\noindent
MDP is represented by a tuple $<S, A,  \mathds{P}, \mathds{C}, \gamma>$ for environment $E$, where $S$ is the state space, $A$ is a control action space, $\mathds{P}$ is the environment transition model, $\mathds{C}$ denotes the cost (or negative reward) model, and $\gamma \in [0,1]$ is a scalar discount factor. At each decision step \( t \), given the current state \( s_t \in S \), the agent takes an action \( a_t \in A \), incurs a cost as a result of this state and action, \( c(s_t, a_t) \), and the system transitions to state \( s_{t+1} \in S \) based on the Markovian probability \( p(s_{t+1} \mid s_t, a_t) \). The agent chooses actions according to its policy $\pi$,  which can be deterministic,  $\pi(s_t)$: $S$ $\rightarrow A$ or stochastic, i.e., a distribution over actions conditioned on $s_t$, i.e., $\pi(a_t|s_t)$: $S\rightarrow P(A)$. Policy $\pi(s_t)$ is thus a scalar value for a deterministic policy, while for a stochastic policy, $\pi$ is a probability mass and density function for discrete and continuous action spaces, respectively. The corresponding cost function ${C_t} = \sum\nolimits_{i = t}^T {{\gamma ^{i - t}}c\left( {{s_i},{a_i}} \right){\mkern 1mu} }$ is the total cost incurred from time step $t$ to the end of the planning horizon $T$ under the policy $\pi$.
Further, we can define the action-value function, $Q^\pi$, as the expected cost over all potential future states and actions, conditioned on the current state-action pair, $s_t$, $a_t$ as:
\begin{equation}
\begin{split}
Q^\pi(s_t,a_t) &  = \mathds{E}_{s_{i>t} \sim E,a_{i>t}\sim \pi }\Big[ {C_t}|s_t,a_t \Big]\\
 &  = c(s_t,a_t)+\gamma\mathds{E}_{s_{t+1}\sim E,a_{t+1}\sim \pi }\Big[ Q^\pi(s_{t+1},a_{t+1}) \Big]
\end{split}
\label{eq:1}
\end{equation}
Similarly, a value function of a policy $\pi$ is the cost expectation from state $s_t$. It can also be defined as the expectation of the action-value function over all possible actions at the current step:
\begin{equation}
\begin{split}
V^\pi \left( s_t \right) & = \mathds{E}_{a_t\sim \pi } \Big[ Q^\pi(s_{t},a_{t})\Big]\\
\end{split}
\label{eq:2}
\end{equation}
The goal of the agent is to select the optimal sequence of actions that minimize the value function, constituting the optimal policy $\pi^*$. For the optimal policy, with a given model of transitions $p\left( {{s_{t + 1}}|{s_t},{a_t}} \right)$ and the aid of Eq. \eqref{eq:1}, the optimal action-value ${Q^{\pi^*}}\left( {{s_t},{a_t}} \right)$ follows the Bellman equation \cite{bellman1957a}, and can be obtained as:

\begin{equation}
Q^{\pi^*}\left( {{s_t}, {a_t}} \right)  =   c(s_t,a_t)+\gamma\mathds{E}_{s_{t+1}\sim E}\Big[\mathop{\min} \limits_{{a} \in A} Q^\pi(s_{t+1},a_{t+1}) \Big]
\label{eq:3}
\end{equation}
Similarly, the optimal value function ${V^{\pi^*}}\left( {{s_t}} \right)$ can be estimated as:
\begin{equation}
\begin{split}
V^{\pi^*} \left( s_t \right) & = \mathop{\min} \limits_{{a} \in A}  \left\{ {Q\left( {{s_t},{a_t}} \right)} \right\} \\
 &  = \mathop{\min} \limits_{{a} \in A}  \left\{c(s_t,a_t)+\gamma\mathds{E}_{s_{t+1}\sim E}\Big[\mathop{\min} \limits_{{a} \in A} Q^\pi(s_{t+1},a_{t+1}) \Big]  \right\}\\
 &  = \mathop{\min} \limits_{{a} \in A}  \left\{c(s_t,a_t)+\gamma\mathds{E}_{s_{t+1}\sim E}\Big[V^{*}(s_{t+1}) \Big]  \right\}\\
 &  = \mathop{\min} \limits_{{a} \in A}  \left\{c(s_t,a_t)+\gamma\sum\limits_{s_{t+1} \in S}p(s_{t+1}|s_t, a_t)V^{*}(s_{t+1})  \right\}
\end{split}
\label{eq:4}
\end{equation}
The objective function of MDP can be thus described by Eq. \eqref{eq:3} or \eqref{eq:4}, which can be solved using value iteration, policy iteration, or linear programming formulations, as presented in  \cite{papakonstantinou2014a}.

A natural extension to an MDP is the partially observable MDP (POMDP), which is a generalization to the case when the environment is not perfectly observable. This is relevant to a wide variety of engineering applications where the available information is incapable of revealing the true state of the system.  Analogous to MDPs, in the POMDP framework the decision-maker/agent starts at a hidden state, $s_t$ at a time step, $t$, takes an action $a_t$, incurs a cost, $c$, transitions to the next hidden state, $s_{t+1}$, and receives an observation, $o_{t+1} \in \Omega$, based on the observation probability model, $p(o_{t+1}|s_{t+1}, a_t)$ for a given state and action. More formally, a POMDP is a tuple $\left\langle {S,\,A,{\mathds{P}},\Omega ,{\mathds{O}},{\mathds{C}},\gamma } \right\rangle$ where $S$, $A$ and $\Omega$ are sets of states, actions, and possible observations, respectively, and ${\mathds{P}}$, ${\mathds{O}}$, and ${\mathds{C}}$ are the transition, observation, and cost models, respectively.

Due to the uncertainty associated with observations, the agent does not know the actual state, hence it can only form a belief ${\bf{b}}_t$ about its state, where ${\bf{b}}_t$ is a probability distribution over the set $S$ of all possible discrete states. A Bayesian update can be used to calculate the new belief ${\bf{b}}_{t+1}$ for a given action and observation, \cite{papakonstantinou2014a}: 

\begin{equation}
\begin{split}
b\left( {{s_{t + 1}}} \right) & = p\left( {{s_{t + 1}}|{o_{t + 1}},{a_t},{{\bf{b}}_t}} \right)\\
& = \frac{{p\left( {{o_{t + 1}}|{s_{t + 1}},{a_t}} \right)}}{{p\left( {{o_{t + 1}}|{{\bf{b}}_t},{a_t}} \right)}}\sum\limits_{{s_t} \in S} {p\left( {{s_{t + 1}}|{s_t},{a_t}} \right)} \,b\left( {{s_t}} \right)
\end{split}
\label{belief_update}
\end{equation}
where probabilities $b\left( {{s_t}} \right)$, for all ${s_t} \in S$, form the belief vector ${{\bf{b}}_t}$ of length $|S|$, and the denominator of Eq. \eqref{belief_update}, $p\left( {{o_{t + 1}}|{{\mathbf{b}}_t},{a_t}} \right),$  is the standard normalizing constant. As such, beliefs can be seen as alternative states of this environment, and POMDPs can be accordingly regarded as belief-MDPs. Further, the optimal value function $V^{\pi^*}$, following the optimal $\pi^*$, takes an expression similar to Eq. \eqref{eq:3}:	
\begin{equation}
    V^{\pi^*}\left( {{{\bf{b}}_t}} \right) = \mathop {\min }\limits_{{a_t} \in A} \left\{ {\sum\limits_{{s_t} \in S} {b\left( {{s_t}} \right)c\left( {{s_t},{a_t}} \right)}  + \gamma \sum\limits_{{o_{t + 1}} \in \Omega } {p\left( {{o_{t + 1}}|{{\bf{b}}_t},{a_t}} \right)V^{\pi^*}\left( {{{\bf{b}}_{t + 1}}} \right)} } \right\}
    \label{V_belief}
\end{equation}

Despite the similarity with MDPs, POMDPs are more difficult to solve. The new belief-state space is continuous rather than discrete, producing a ($\lvert S \rvert - 1$) – dimensional simplex. The optimum value function has been proved however to be piece-wise linear and convex, and as so it can be represented by a finite number of affine hyperplanes \citep{sondik1971a}. Thus, the decision problem is reduced to choosing a finite number of vectors, commonly known as the $\alpha$-vectors. To solve this problem, point-based algorithms seek to iteratively determine a set of $\alpha$-vectors  that can adequately support the belief space\cite{spaan2005a,kurniawati2008sarsop}. However, such point-based solvers can become impractical for large POMDP problems, particularly for infrastructure system with many components \citep{papakonstantinou2018a}. As already mentioned, the curse of dimensionality associated with POMDPs can be alleviated using a deep reinforcement learning framework.  

\subsection{Deep Reinforcement Learning}
\noindent
Reinforcement learning (RL) is a computational framework for evaluating and automating goal-directed learning and decision-making, and is especially well-suited for solving MDP/POMDP problems as it is mathematically structured around them \cite{sutton2018reinforcement}. RL is theoretically able to alleviate the curse of dimensionality related to the state space, either under model-free approaches that do not utilize prior environment information on transition dynamics, or model-based approaches based on an underlying transition model of the environment or hybrid approaches thereof. 

Deep Reinforcement Learning (DRL) utilizes deep neural network representations to approximate the value functions or policies, showing capabilities of discovering meaningful parameterizations in immense state spaces, outperforming human experts \cite{sutton2018reinforcement,sutton2000policy}. DRL algorithms are generally categorized into value-based and policy-based learning methods. Value-based methods learn the state or state-action value function and determine the optimal action for each state \cite{mnih2015human,sutton2018reinforcement,van2016deep}. In contrast, policy-based methods directly learn a policy, \( \pi: S \to A \) or \( \pi: S \to P(A) \), using policy gradient techniques \cite{williams1992simple}. Actor-critic methods combine both approaches by simultaneously learning a policy (actor) and a value function (critic), enhancing stability and convergence \citep{konda2000actor}. This hybrid strategy is widely adopted in DRL algorithms, including Proximal Policy Optimization (PPO) \citep{schulman2017proximal}, Asynchronous Advantage Actor-Critic (A3C) \citep{mnih2016asynchronous}, and Deep Deterministic Policy Gradient (DDPG) \citep{lillicrap2016continuous}. 

Multi-Agent Deep Reinforcement Learning (MADRL) for optimal decision-making in managing infrastructure systems has also been introduced in the literature  \cite{andriotis2019managing, andriotis2021deep}. However, MADRL methods have primarily been evaluated so far in environments with tens of agents or components. In this work, we aim to advance the application of MADRL by implementing it in a more complex, real-world scenario  with nearly one hundred of heterogeneous components and several constraints. In the following section, we are detailing the MADRL formulation employed in our study.

\section{DDMAC-CTDE}\label{DCMAC_DDMAC}
\noindent
Designing optimal I\&M policies for large-scale, long-horizon, multi-component systems is inherently challenging due to the exponential growth of joint state and action spaces with the number of components. This combinatorial complexity renders traditional single-agent planning and learning algorithms computationally infeasible for such high-dimensional problems. Cooperative MADRL alleviates this complexity by addressing decision-making as a shared task among multiple autonomous agents. The agents optimize a shared
long-term objective conditionally independently in the same environment, thereby, making the joint policy scale linearly with the number of components. This paradigm is particularly well-suited for cross-asset life-cycle management of infrastructure systems, where coordinated planning of I\&M actions across diverse components, such as bridges, pavements, and other assets is essential. By fostering cooperation among agents, this approach can potentially minimize overall risks and achieve various cost-related objectives, ensuring sustained performance and reliability of critical infrastructure.

In \cite{andriotis2019managing}, the Deep Centralized Multi-agent Actor-Critic (DCMAC) approach was introduced to address multi-component decision-making challenges within the framework of off-policy actor-critic DRL. This approach leverages a replay buffer, a memory that stores past experiences collected by agents as they interact with their environment \cite{degris2012off, sutton1988learning}. These stored experiences are utilized for stable batch training of the neural networks through random sampling at each training step \cite{mnih2015human}. DCMAC uses policy gradient methods \cite{sutton2000policy} to update the policy parameters and has shown significant life-cycle cost improvements compared to heuristically optimized policies \cite{andriotis2019lifecycle}. DCMAC is designed as a centralized agent with respect to access to the full state space across all agents, predicting decentralized actions for every agent in the system. While information centralization in principle benefits coordinated behavior, it can, in certain cases, hinder training efficiency. Parametrization over the full state space introduces significant computational complexity, particularly as the number of agents increases. 

To alleviate some of the computational challenges, the DDMAC architecture, proposed in \cite{andriotis2021deep}, represents each agent with a separate actor/policy network. This design allows for a sparser policy parameterization by eliminating inter-agent interactions in the hidden layers, thereby improving training efficiency in certain cases. In \cite{andriotis2021deep}, DDMAC is further integrated into an algorithmic framework that supports planning under operational constraints, such as budget limitations. While DDMAC can offer improvements over DCMAC, it assumes that each agent has access to the entire system's state distribution. As a result, each actor receives the full state space as input, which may add computational burden that is disproportionate to the expected performance gains. This is because each actor must optimize over the full state space, which again becomes increasingly computationally expensive as the number of agents grows. Furthermore, it also limits applicability in true decentralized execution settings, where agents need to make decisions based solely on local observations rather than the global state.

To be able to address the large systems examined in this work, we herein introduce a new version of DDMAC, equipped with the Centralized Training and Decentralized Execution (CTDE) paradigm of cooperative reinforcement learning \cite{sunehag2017value, oliehoek2008optimal, foerster2018counterfactual, rashid2018qmix, gronauer2022multi}. This modified architecture, referred to as DDMAC-CTDE, builds upon the Decentralized-POMDP (Dec-POMDP) framework \cite{oliehoek2008optimal}, modeling agents under fully decentralized policies, each conditioned on its corresponding component’s access to  local information/ belief state. This ensures that agents operate independently during execution, eliminating the need for centralized control and enabling seamless scalability to larger systems with a greater number of agents. 

The training process in the proposed framework employs a centralized critic, which has access to the full system belief state, guiding each agent’s policy during training. This centralized critic evaluates the global state and joint behavior of all agents, ensuring that policy updates are informed through a global environment signal emitted through joint interaction among agents. However, the centralized critic is used exclusively during the training phase and is discarded during the execution or deployment phase, allowing for fully decentralized decisions in inference time. The centralized critic also helps mitigate the non-stationarity issue that arises in multi-agent settings, where simultaneous updates to agents' policies can make the environment's dynamics appear unstable, by leveraging global state information to provide a stable training signal, thereby reducing inconsistencies and accelerating convergence \cite{ foerster2018counterfactual, lowe2017multi}. 

This combination of Centralized Training and Decentralized Execution strikes a delicate balance between leveraging centralized information for effective training and maintaining decentralized control for scalability. The CTDE framework is particularly powerful in cooperative multi-agent reinforcement learning, where agents share similar tasks and require coordination. By enabling agents to optimize actions within their local observation space while benefiting from centralized guidance during training, CTDE ensures scalability and robustness in environments with a large number of agents and complex interdependencies \cite{gronauer2022multi}. Further details on the DDMAC-CTDE architecture, including its implementation and performance, are provided in the following sections.

\subsection{Formulation}\label{formulation}
\noindent
A fully cooperative multi-agent task can be described as a Decentralized POMDP (Dec-POMDP) introduced in \citep{oliehoek2016concise}. A Dec-POMDP consists of an environment with multiple agents, in which the agents do not have complete observations of the environment’s state and the actions taken by the other agents. More formally, Dec-POMDP is defined by a tuple $ \left\langle {\mathcal{N}_a, \mathcal{N}_c, \{S_i\},\{A_i\},\{\Omega_i \},{\mathds{P}} ,{\mathds{O}},{\mathds{C}},\gamma } \right\rangle$, where $\mathcal{N}_a$ is a set of agents from 1 to $N_a$, $\mathcal{N}_c$ is a set of components from 1 to $N_c$, $S_k$ is a set of states for the $k$-th component and $S = {S_1} \times  \cdots  \times {S_{N_c}}$ is a global state of the system, $A_i$ is a set of actions for agent $i$ and $A = {A_1} \times  \cdots  \times {A_{N_a}}$ is the joint action space, $\Omega_i$ is a set of observations for the $i$-th agent and $\Omega  = {\Omega _1} \times  \cdots  \times {\Omega_{N_a}}$ is the joint observation space, and ${\mathds{P}}$, ${\mathds{O}}$, and ${\mathds{C}}$ are same as defined earlier in Section \ref{POMDP}.

Without any loss of generality and for presentation simplicity, we henceforth assume that each agent controls exactly one corresponding system component, i.e., $N_a = N_c = N$. This assumption may easily be modified if an agent controls multiple components \cite{andriotis2019managing}. Then, the policy $\pi_i:\mathbf{b}^{(i) }\rightarrow P(A_i)$ is a policy followed by the $i$-th agent and it is a function of the local belief state vector $\mathbf{b}^{(i)}$, determined by the observation and action history of each component and calculated via Bayesian update, as described in Eq. \eqref{belief_update}. For all agents, the joint policy $\pi = {\pi_1}\cdot{\pi_2}\cdots{\pi_{N-1}}\cdot{\pi_{N}}$ may be defined as $\pi:{\mathbf{\hat b}} \rightarrow P(A)$, where ${\mathbf{\hat b}} = \left[ {{\bf{b}}^{(i)}} \right]_{i = 1}^N$ is a belief matrix of the system state. The joint objective of the agents is to identify an optimal policy $\pi^*$ that minimizes the discounted cumulative global cost $C_0$ over the planning horizon $T$. For concise notation, we are also using \(\left(\cdot\right)'\) for \(\left(\cdot\right)\) at $t$+1 time step.

Similar to DDMAC, a decentralized actor-critic architecture is selected within the premises of off-policy actor critic DRL, where each agent is assigned an actor network for policy output independent to other agents. During training, the centralized critic is using the system state distribution as input and the value function $V^{\pi}({\mathbf{\hat b}})$ as the output. The expression in Eq. (\ref{V_belief}) gives the optimal value function $V^{\pi^*}({\mathbf{\hat b}})$.
The actor network is updated using the policy gradient theorem \cite{sutton2000policy} extended to multi-agent settings \citep{andriotis2021deep}. The off-policy gradient ${{\bf{g}}_{{{\boldsymbol{\uptheta }}_{\pi}^{(j)} }}}$ for the $j^{th}$ actor can be estimated using samples generated by a behavior policy $\mu \neq \pi$ as:
 
\begin{equation}
\begin{split}
    {\bf{g}}_{{\boldsymbol{\uptheta }}_{\pi}^{(j)}} & = {\mathds{E}_{{{\mathbf{b}}}\sim{\boldsymbol{\uprho }},{{\bf{a}}}\sim{\boldsymbol{\upmu }}}}\left[ {{w}\left( {\sum\limits_{i = 1}^N {{\nabla _{{{\boldsymbol{\uptheta }}_\pi^{(j)} }}}\log {\pi _i}\left( {a^{(i)}|{\mathbf{b}^{(i)}},{{\boldsymbol{\uptheta }}_\pi^{(i)} }} \right)} } \right){A^\pi }\left( {{{\bf{\hat b}}},{{\bf{a}}}} \right)} \right]\\
   & = {\mathds{E}_{{{\mathbf{b}}}\sim{\boldsymbol{\uprho }},{{\bf{a}}}\sim{\boldsymbol{\upmu }}}}\left[ {{w}\left( {{{\nabla _{{{\boldsymbol{\uptheta }}_\pi^{(j)} }}}\log {\pi _j}\left( {a^{(j)}|{\mathbf{b}^{(j)}},{{\boldsymbol{\uptheta }}_\pi^{(j)} }} \right)} } \right){A^\pi }\left( {{{\bf{\hat b}}},{{\bf{a}}}} \right)} \right]
 \end{split} 
 \label{gradient2}
\end{equation}
where, ${{\bf{a}}} = \left\{ {a^{(i)}} \right\}_{i = 1}^N$ are the actions for $N$-agents, $\boldsymbol{\uptheta}_\pi^{(j)}$ is a weight parameter for the $j^{th}$ actor network, $w = min\{\delta,\prod_{i=1}^{N}\pi_i(a^{(i)}|\mathbf{b}^{(i)}) \slash\mu_i(a^{(i)}|\mathbf{b}^{(i)})\}$ is a truncated importance sampling weight, $\boldsymbol \upmu$ is a $N$-dimensional vector of agents’ behavior policies, $\boldsymbol \uprho$ is the $N$-dimensional limiting state distribution under these policies,  $\delta=2$ is a cut-off value if the ratio becomes too large due to very small $\mu_i$, and ${{A^\pi }\left( {{{\bf{\hat b}}},{{\bf{a}}}} \right)}$ is an advantage function defined as ${A^\pi }\left( {{{\bf{\hat b}}},{{\bf{a}}}} \right) = {Q^\pi }\left( {{{\bf{\hat b}}},{{\bf{a}}}} \right) - {V^\pi }\left( {{{\bf{\hat b}}}} \right)$, hence being a zero mean measure, i.e., ${\mathds{E}_{{{\bf{a}}}\sim \boldsymbol \pi }}\left[ {{A^\pi }\left( {{{\bf{\hat b}}},{{\bf{a}}}} \right)} \right] = 0$, estimating how advantageous the given action is.  Advantage function can further be approximated as in \cite{mnih2016asynchronous, andriotis2019managing}, similar to the TD-learning:
\begin{equation}
    {A^\pi }\left( {{{\bf{\hat b}}},{{\bf{a}}}|{{\boldsymbol{\uptheta }}_V}} \right) \simeq c\left( {{{\bf{\hat b}}},{{\bf{a}}}} \right) + \gamma {V^\pi }\left( {{{\bf{\hat b}}'}|{{\boldsymbol{\uptheta }}_V}} \right) - {V^\pi }\left( {{{\bf{\hat b}}}|{{\boldsymbol{\uptheta }}_V}} \right)
    \label{advantage2}
\end{equation}
where $\boldsymbol{\uptheta}_V$ are weight parameters of the value/critic network, and ${\bf{\hat b}}'$ is the system belief at the next time step. The mean squared error is considered as the loss function, $L_V$, for the critic network, given as: 
\begin{equation}
        {L_V({{{\boldsymbol{\uptheta }}_V }})} = {\mathds{E}_{{{\bf{\hat b}}}\sim{\boldsymbol{\uprho }},{{\bf{a}}}\sim{\boldsymbol{\upmu }}}}\left[ {{w}{{\Big( {c\left( {{{\bf{\hat b}}},{{\bf{a}}}} \right) + \gamma {V^\pi }\left( {{{\bf{\hat b}}'}|{{\boldsymbol{\uptheta }}_V}} \right) - {V^\pi }\left( {{{\bf{\hat b}}}|{{\boldsymbol{\uptheta }}_V}} \right)} \Big)}^2}} \right]
    \label{DCMAC:MSE}
\end{equation}
and its gradient for updating $\boldsymbol{\uptheta}_V$ can be estimated as:
\begin{equation}
    {{\bf{g}}_{{{\boldsymbol{\uptheta }}_V }}} = {\mathds{E}_{{{\bf{\hat b}}}\sim{\boldsymbol{\uprho }},{{\bf{a}}}\sim{\boldsymbol{\upmu }}}}\left[ {{w}{\nabla _{{{\boldsymbol{\uptheta }}_V}}}{V^\pi }\left( {{{\bf{\hat b}}}|{{\boldsymbol{\uptheta }}_V}} \right){A^\pi }\left( {{{\bf{\hat b}}},{{\bf{a}}}|{{\boldsymbol{\uptheta }}_V}} \right)} \right]
    \label{gradient2_V}
\end{equation}
Both actor and critic networks are trained using batch sampling from a buffer memory ${\cal M}$, containing experience tuples $\left( {{\bf{\hat b}},{\bf{a}},[{{\boldsymbol{\pi}} _i}]_{i = 1}^{{N}},{c},{\bf{\hat b}}'} \right)$. 

\section{Constrained Optimization}
\noindent
The life-cycle management problem is often accompanied by several constraints, e.g., long or short-term budgets, risk tolerance, and other imposed restrictions by various agencies. Depending on their nature, constrains are handled in this work through state augmentation or Lagrangian multipliers.

\subsection{Objective function with constraints}
\noindent
The objective of the decision-maker (agent) may be defined as determining an optimum policy $\pi = \pi^*$ that minimizes the entire cumulative operational costs and risks given several constraints \cite{andriotis2021deep}:
\begin{equation}
\begin{split}
\pi ^* & = \mathop {\arg \min }\limits_{\pi  \in {\Pi} } {\mathds{E}_{{s_{0:T}},{o_{0:T}},{a_{0:T}}}}\left[ {\left. {\sum\limits_{t = 0}^T {{\gamma ^t}{c}\,} } \right|{a_t} \sim \pi \left( {{o_{0:t}},{a_{0:t - 1}}} \right),{s_0}\sim{{\bf{b}}_0}} \right]\\
{\rm{ s.t.}} \; \;{Z_{h,k}} &  =  \sum\limits_{t = 0}^T {{\gamma_h ^t}{z_{h,k}}\left( {{s_t},{a_t}} \right)\,}  - {\alpha _{h,k}} \le 0,\,\,k = 1,..,K\\
\; \; \;{Z_{s,m}} &  =  {\mathds{E}_{{s_{0:T}},{o_{0:T}},{a_{0:T}}}}\left[ {\sum\limits_{t = 0}^T {{\gamma_s ^t}{z_{s,m}}\left( {{s_t},{a_t},{s_{t + 1}}} \right)\,} } \right] - {\alpha _{s,m}} \le 0,\,\,m = 1,..,M
\end{split}
\label{eq:objective2}
\end{equation}
where $c = c(s_t, a_t, s_{t+1})$ is the cost incurred at time $t$ by taking action $a_t$, transitioning from state $s_t$ to state $s_{t+1}$ and taking an observation $o_t$; $\gamma$, $\gamma_h$, and $\gamma_s \in [0,1]$ are the discount factors for cost, hard constraints, and soft constraints, respectively; $\mathbf{b}_0$ is an initial belief; $T$ is the length of the planning horizon; $Z_{h,k}$ and $Z_{s,m}$ are the {\it{hard}} and {\it{soft}} constraints, respectively; and $z_{h,k}$ and $z_{s,m}$ are their respective auxiliary costs; $\alpha_{h,k}$ and $\alpha_{s,m}$ are real-valued scalars, with the former determining the allowable values of $K$ hard constraint thresholds that need to be satisfied at all time steps, and the latter representing the values for $M$ soft constraint thresholds which need to be satisfied in expectation. More details on soft and hard constraints will be provided subsequently. The policy function belongs to a space $\pi \in \Pi_c$, which contains all possible policies that are admissible under the existing constraints of the problem. $\Pi_c$ is a subset of $\Pi$, the policy space of the unconstrained problem.

The constraint structure in Eq. \eqref{eq:objective2} is adaptable to a wide variety of constraint types important to infrastructure management.
Hard constraints, for example, may be used to simulate a broad range of fixed resource allocation and control action availability problems, such as those with budget restrictions. In turn, soft constraints in the form of Eq. \eqref{eq:objective2} may be used to represent constraints that need to be satisfied in expectation, such as risk-based or performance-based constraints (e.g., restricting poor condition infrastructure levels, imposing goals for good condition levels, and many others). Hard constraints can be handled through state augmentation, however, this is impractical for soft constraints since one must then track the full distribution of the cumulative discounted value of $z_{s,m}$. As a result, probabilistic constraints are handled using Lagrangian relaxation \cite{bertsekas1997nonlinear}. Based on the above, the optimization problem can be restated as in \cite{andriotis2021deep}:
\begin{align}
    \begin{split}
       V(\mathbf{b}_0,\mathbf{y}_0) & = \begin{multlined}[t][0.6\textwidth]
            \mathop {\max }\limits_{{\lambda _1},\cdots,{\lambda _M} \ge 0} \mathop {\min }\limits_{\pi  \in \Pi} {\mathds{E}_{s_{0:T},{{\bf{y}}_{0:T}},{o_{0:T}},{a_{0:T}}}} \left[ {\sum\limits_{t = 0}^T  \Bigg( {{\gamma ^t}\bar c\left({s_t,{a_t},{{\bf{y}}_t}} \right)} \Bigg.} \right.\\
           \left. { + \Bigg. {\gamma_s ^t}\sum\limits_{m = 1}^M {{\lambda _m}{\left(z_{s,m}-\alpha _{s,m} \right)}} \Bigg) \left| {{a_t}\sim \pi \left( {{o_{0:t}},{a_{0:t - 1}},{{\bf{y}}_t}} \right),{s_0}\sim {{\bf{b}}_0},{{\bf{y}}_0}} \right.} \right]
         \end{multlined}\\
         & = \begin{multlined}[t][0.1\textwidth]
         \mathop {\max }\limits_{{\lambda _1},...,{\lambda _M} \ge 0} \mathop {\min }\limits_{\pi  \in \Pi } V_\lambda ^\pi ({{\bf{b}}_0},{{\bf{y}}_0}),\end{multlined}\\  
   \end{split}
   \label{eq:value_func}
\end{align}
\begin{equation*}
    \text{s.t.}\,\,\,{{\bf{y}}_t} = \left\{ {{y_{kt}}} \right\}_{k = 1}^K,\,
         {y_{kt}} = \sum\limits_{\tau  = 0}^{t - 1} {{\gamma_h ^\tau }{z_{h,k}}\left( {{s_\tau },{a_\tau }} \right)},\,{y_{k0}} = 0, {y_{kt}}\, \in \left[ {0,{\alpha _{h,k}}} \right]\,,\,k = 1,...,K
\end{equation*}
where, variables $y_{kt}$ track the discounted cumulative value of the function related to hard constraints, $Z_{h,k}$, up to time step $t-1$, and $\bar c\left( {s_t,{a_t}, s_{t+1},{{\bf{y}}_t}} \right)$ is the total cost accumulated at a time step $t$ taking ${\bf{y}}_{t}$ into consideration. Variables $y_{kt}$ are upper bounded by $\alpha_{h,k}$. Lagrange multipliers, $\lambda_m$, constitute the dual variables of the max-min dual problem, they are positive scalars and are associated with the soft constraints.  \vspace{10pt}\\
{\it{Hard constraints:}}\vspace{5pt}\\
Depending on the operational and resource allocation plan, available funds for inspection and maintenance must meet particular short- or long-term objectives associated with a certain budget cycle length, $T_B$. The total cost of inspection actions ($c_I$) and maintenance actions ($c_M$) at time instant $\tau$ is given as follows:
\begin{align}
    z_h^{}\left( {{s_\tau },{a_\tau }} \right) &= \left( {{c_M} + \gamma {c_I}} \right){{\bf{1}}_{\tau  \in {\Lambda _t}}}
    \label{accumulted_cost_of_inm}
\end{align}
where, ${\Lambda _t} = \left( {\left\lfloor {t/{T_B}} \right\rfloor {T_B},\left( {\left\lfloor {t/{T_B}} \right\rfloor  + 1} \right){T_B}} \right]$ is a budget interval of length $T_B$, $\left\lfloor x \right\rfloor$ is integer part of x. For a given budget cap $\alpha_h$, the maintenance and inspection costs at each time are adjusted as:
\begin{equation}
\begin{array}{l}
{{\bar c}_M} = {{\bf{1}}_{y_{}^{} + {z_h} \le {\alpha _h}}}{c_M}\\
{{\bar c}_I} = {{\bf{1}}_{y_{}^{} + {z_h} \le {\alpha _h}}}{c_I}
\end{array}
\label{budget_affected_cost}
\end{equation}
Eqs. \eqref{accumulted_cost_of_inm} and \eqref{budget_affected_cost} essentially affect inspection and maintenance costs, such that the accumulated cost $z_h^{}\left( {{s_\tau },{a_\tau }} \right)$ in a given budget interval $T_B$ does not exceed the budget limit $\alpha_h$ for the interval. Introducing a new state variable $y = y_t$ allows for the inclusion of budget constraints, which are assumed to be fully observable. This enables the agent to justify control actions based on the available budget, $\alpha_h - y_t$, at each time step. \vspace{6pt}\\
{\it{Soft constraints:}}\vspace{5pt}\\
Soft constraints are the constraints that need to be satisfied in expectation over the life-cycle of the considered system. The generic formulation provided in \cite{andriotis2021deep} is as follows:
\begin{equation}
\begin{split}
Z_s & = \mathds{E}_{{s_{0:T}},{o_{0:T}},{a_{0:T}}}\,\left[ \sum\limits_{t = 0}^T \gamma_s^tJ_t^{\pi} \right] - {\alpha _s}\\
  & = {J^\pi} - {\alpha _s}
\end{split}
\end{equation}
where, $Z_s$ is a constraint function, $J^\pi$ is a cumulative discounted cost (or any generic parameter that needs to be bounded) accrued by following policy $\pi$. $J_t^\pi$ is a cost (or the parameter value) at time step $t$, $\alpha_s$ is the cap that needs to be satisfied over the life-cycle.

Many reliability-related decision problems are focused in limiting the likelihood of failure (i.e., the probability of reaching a failure state from a non-failure state) throughout the system's operational life. For transportation network management, it is common to limit the poor condition states of pavement and bridge components based on state and federal criteria and thresholds. This study will demonstrate the incorporation of soft constraints in such cases. The details of the performance criteria will be discussed in Section \ref{sec:cons}.

\subsection{Costs related to infrastructure management}
\noindent
Costs at different time steps for a selected action can be decomposed into inspection cost, $c_I (a_{I,t})$, maintenance cost, $c_M (a_{M,t})$, and user delay cost, $c_D(a_{M,t})$, incurred in this case due to the traffic delays from maintenance actions, and risk cost $c_{ {\mathfrak{R}^\pi }}(s_t,a_t,s_{t+1})$. In addition, other costs could be considered, such as vehicle operating costs possibly incurred due to the system's current state, for example, extra fuel and tire costs due to the bad ride quality, environmental costs due to fuel emissions which will also depend on the system's state, and various other costs. Therefore, the total cost at each decision step can be generally expressed as:

\begin{equation}
   c(s_t,a_t,s_{t+1}) = {c_M}({a_{M,t}}) + {c_D}({a_{M,t}}) + c_{ {\mathfrak{R}^\pi }}(s_t,a_t,s_{t+1}) + \gamma {c_I}({a_{I,t}})
   \label{eq:cost_func1}
\end{equation}
where $a_{I,t} \in A_I$ is the selected inspection and $a_{M,t} \in A_M$  is the selected maintenance action at time step $t$. Under this distinctive consideration of actions, the total action can be defined as $a_t \in A = A_I \times A_M$. For finite horizon problems, often a terminal cost ($c_{s_T}$) is added to guide actions at the end of the horizon, i.e., at time step $t=T$. After incorporating the budget constraint at time $t$ and $c_{s_T}$ at time step $T$, the cost can be modified as:
\begin{equation}
     {\bar c}(s_t, a_t, s_{t+1},{\bf{y}}_t) = {\bar c_M}\left({a_{M,t}} \right) + \gamma {\bar c_I}\left({a_{I,t}} \right)+{\bar c_D}(a_{M,t})+{\bar c_{ {\mathfrak{R}^\pi }}}(s_t,a_t,s_{t+1})+1_{t = T}{\bar c_{s_T}}(s_t)
     \label{eq:cost_func}
\end{equation}
The transition and observation probabilities are also affected by the budget constraint, because after utilizing the budget in a given budget cycle, the agents can only choose actions with no costs, as no maintenance and inspection actions can be performed. Further, the costs in Eqs. \eqref{eq:cost_func1} and \eqref{eq:cost_func} can readily be obtained  by calculating cost in expectation with respect to the belief state {\bf{b}} as in \citep{andriotis2021deep}:

\begin{equation}
\begin{split}
{{\bar c}_{b,t}} & = {{\bar c}_{b,t}(\textbf{b}_t,\textbf{a}_t, {\bf{y}}_t) }\\
           & = \E_{s_t,s_{t+1}}\left[{\bar c}(s_t, a_t, s_{t+1},y_t)\right]\\
           & = \sum\limits_{s_t \in S}b(s_t)\sum\limits_{s_{t+1} \in S}P\left(s_{t+1}|s_t,a \right){\bar c}(s_t, a_t, s_{t+1},y_t)
\end{split}
     \label{eq:cost_beleif}
\end{equation}

\noindent where ${{\bar c}_{b,t}}$ is the estimated total cost at time $t$, obtained using belief state $\textbf{b}_t$.

\subsection{DDMAC-CTDE formulation with constraints}
\noindent
The architecture given in Section \ref{DCMAC_DDMAC} can readily be extended to a constrained case. Eqs. \eqref{gradient2}-\eqref{gradient2_V} are modified to incorporate Eqs. \eqref{eq:value_func}-\eqref{eq:cost_beleif}. 
Similarly, here we also have $N$ components and $N$ number of agents, where one agent has control over exactly one component. For such settings, the system policy function can be written as: 
 
\begin{figure}[t]
\centering
\includegraphics[width=.68\textwidth]{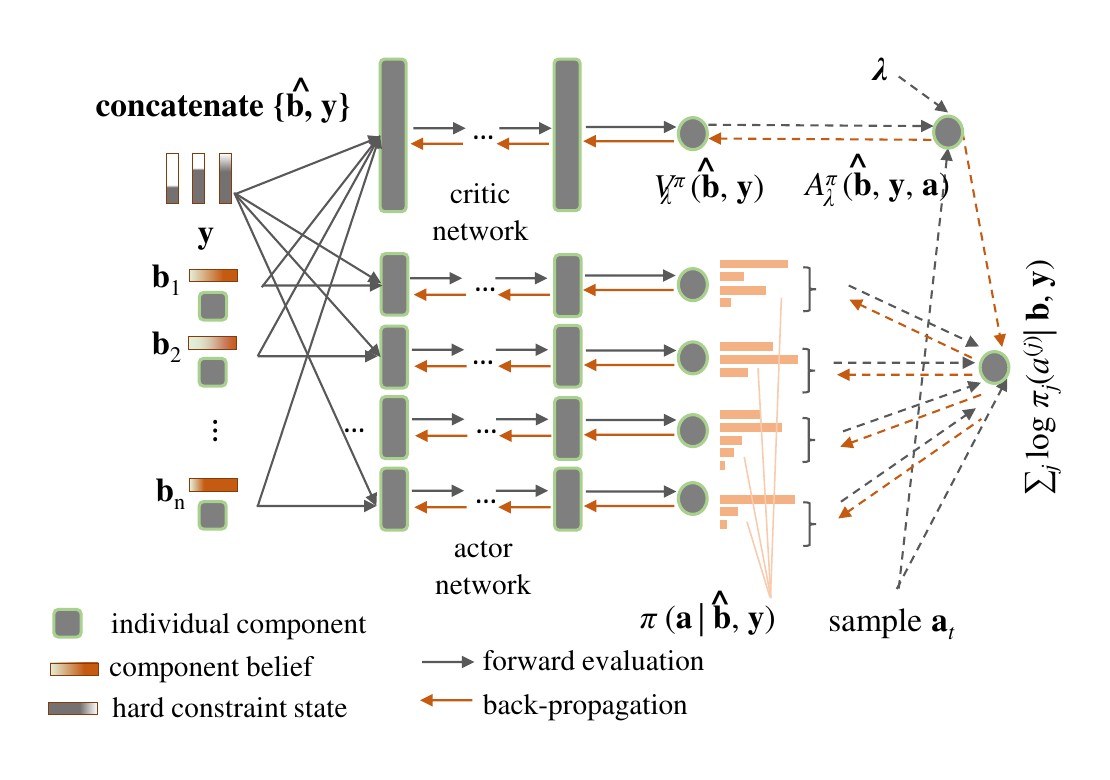}
 \caption{Constrained Deep Decentralized Multi-agent Actor Critic (DDMAC) with Centralized Training Decentralized Execution (CTDE) architecture.}
\label{consDDMAC}
\end{figure}
 
\begin{equation}
    \pi \left( {{\bf{a}}|{\bf{\hat b}},{\bf{y}}} \right) = \prod\limits_{i = 1}^{N} {{\pi _i}\left( {{a^{(i)}}|{\bf{b}}^{(i)},{\bf{y}}} \right)} 
    \label{eq:policy_func}
\end{equation}
where ${\bf{y}}= \left[ {y_{k}} \right]_{k = 1}^{K}$ is an augmented state vector of {\it hard constraints}. It is important to note that each agent selects its policy based on its belief and the global state of {\it hard constraints}. The decentralized policy given in Eq.~\eqref{eq:policy_func} and the centralized Lagrangian value function can be parameterized as:
\begin{equation}
    \begin{array}{l}
{\pi _i}\left( {a_{}^{(i)}|{\bf{b}}^{(i)},{\bf{y}}} \right) \simeq {\pi _i}\left( {a_{}^{(i)}|{\bf{b}}^{(i)},{\bf{y}},{\boldsymbol{\uptheta }}_\pi ^{(i)}} \right)\\
V_\lambda ^\pi \left( {{\bf{\hat b}},{\bf{y}}} \right) \simeq V_\lambda ^\pi \left( {{\bf{\hat b}},{\bf{y}}|{{\boldsymbol{\uptheta }}_V}} \right)
\end{array}
\end{equation}
where, ${\boldsymbol{\uptheta }}_\pi ^{(i)}$, ${\boldsymbol{\uptheta }}_V$ are weight parameters for policy and value network, respectively. Training is performed off-policy with experience replay ${\cal M}$. The replay buffer contains agents' collected experiences as $\left( {{\bf{\hat b}},{\bf{y}},{\bf{a}},[{\pi _i}]_{i = 1}^{{N}},{{\bar c}_b},[{z_{s,m}}]_{m = 1}^M,{\bf{\hat b}}',{\bf{y}}'} \right)$ tuples, ${{\bar c}_b}$ is a cost collected based on the beliefs instead of states. The modified off-policy gradients are now given:
\begin{equation}
    {\nabla _{{\boldsymbol{\uptheta }}_\pi ^{(j)}}}V_\lambda ^\pi  = {\mathds{E}_{\cal M}}\left[ {w\,\,\left( {\sum\limits_{i = 1}^{{N}} {{\nabla _{{\boldsymbol{\uptheta }}_\pi ^{(j)}}}\log {\pi _i}\left( {{a^{(i)}}|{\bf{b}}^{(i)},{\bf{y}},{\boldsymbol{\uptheta }}_\pi ^{(i)}} \right)} } \right)A_\lambda ^\pi \left( {{\bf{\hat b}},{\bf{y}},{\bf{a}}} \right)} \right]
\end{equation}

\begin{equation}
    {\nabla _{{\boldsymbol{\uptheta }}_V}}V_\lambda ^\pi  = {\mathds{E}_{\cal M}}\left[ {w\,{\nabla _{{{\boldsymbol{\uptheta }}_V}}}V_\lambda ^\pi \left( {{\bf{\hat b}},{\bf{y}}|{{\boldsymbol{\uptheta }}_V}} \right)A_\lambda ^\pi \left( {{\bf{\hat b}},{\bf{y}},{\bf{a}}} \right)} \right]
\end{equation}
\begin{equation}
    {\nabla _{{\lambda _m}}}V_\lambda ^\pi  \simeq \sum\limits_{t = 0}^T {{\gamma_s ^t}{z_{s,m}}}  - {\alpha _{s,m}}
\end{equation}
where, $\lambda_m$ is a Lagrange multiplier and $A_\lambda ^\pi$ is an advantage function, computed here as:
\begin{equation}
    A_\lambda ^\pi \left( {{\bf{\hat b}},{\bf{y}},{\bf{a}}|{{\boldsymbol{\uptheta }}_V}} \right) \simeq {\bar c_b} + \sum\limits_{m = 1}^M {{\lambda _m}{z_{s,m}}} + \gamma V_\lambda ^\pi \left( {{\bf{\hat b}}',{\bf{y}}'|{{\boldsymbol{\uptheta }}_V}} \right) - V_\lambda ^\pi \left( {{\bf{\hat b}},{\bf{y}}|{{\boldsymbol{\uptheta }}_V}} \right)
\end{equation}
The full DDMAC-CTDE with constraints schematic is presented in Fig. \ref{consDDMAC}, and the implementation steps are described in Algorithm \ref{algo:DDMAC_cons}. In the subsequent sections the specific modeling of pavements, bridges and the network system are described.

\setcounter{algocf}{0}
\begin{algorithm}[t]
\footnotesize
\SetAlgoLined
Initialize replay buffer (empty) \;
Initialize actor, critic network weights $\left[ {{\boldsymbol{\uptheta }}_\pi ^{(j)}} \right]_{j = 1}^{{N}},{\boldsymbol{\uptheta }}_V^{}$ randomly, and Lagrange multipliers $\left[ {{\lambda _m}} \right]_{m = 1}^M$ with zeros\;
\For{episode = 1,$N_{ep}$}{
\For{t = 1,T}{
Select action ${\bf{a}}_t$ at random according to exploration noise\;
Otherwise select action ${{\bf{a}}_t}\sim{{\boldsymbol{\upmu }}_t} = \left[ {{\pi _j}\left( { \cdot |{{{\bf{b}}}_t}^{(j)},{{\bf{y}}_t},{\boldsymbol{\uptheta }}_\pi ^{(j)}} \right)} \right]_j^{{N}}$ \;
Estimate costs ${\bar c}_{b,t}$ as:\\
\uIf {$t\neq T$}{${\bar c_{b,t}} = {\bar c_b}$\;} 
\uElse{${\bar c_{b,t}} = {\bar c_b}+{\bar c}_{s_T}$\;} 
calculate ${z_{s,mt}} = {z_{s,m}}$ given ${{\bf{\hat b}}_t}$ and ${{\bf{a}}_t}$\;
Observe $o_{t + 1}^{(l)}\sim p\left( {o_{t + 1}^{(l)}|{\bf{b}}_t^{(l)},{{\bf{a}}_t}} \right)$ for $l = 1,2,\cdots, N$\;
Compute beliefs ${\bf{b}}_{t + 1}^{(l)}$ for $l = 1,2,\cdots, N$\;
Store experience $\left( {{{{\bf{\hat b}}}_t},{{\bf{y}}_t},{{\bf{a}}_t},{{\boldsymbol{\upmu }}_t},{{\bar c}_{b,t}},[{z_{s,mt}}]_{m = 1}^M,{{{\bf{\hat b}}}_{t + 1}},{{\bf{y}}_{t + 1}}} \right)$ to replay buffer\;}
Sample batch of $\left( {{{{\bf{\hat b}}}_i},{{\bf{y}}_i},{{\bf{a}}_i},{{\boldsymbol{\upmu }}_i},{{\bar c}_{b,i}},[{z_{s,mi}}]_{m = 1}^M,{{{\bf{\hat b}}}_{i + 1}},{{\bf{y}}_{i + 1}}} \right)$ from the buffer\;
If ${{\bf{\hat b}}_{i}}$ is terminal state ${A_i} = {\bar c_{b,i}}+\sum\limits_{m = 1}^M {{\lambda _m}{z_{s,mi}}} - V_\lambda ^\pi\left( {{{{\bf{\hat b}}}_i}|{{\boldsymbol{\uptheta }}_V}} \right)$\;
Otherwise: $A_{\lambda ,i}^\pi  = {\bar c_{b,i}} + \sum\limits_{m = 1}^M {{\lambda _m}{z_{s,mi}}}  + \gamma V_\lambda ^\pi \left( {{{{\bf{\hat b}}}_{i + 1}},{{\bf{y}}_{i + 1}}|{{\boldsymbol{\uptheta }}_V}} \right) - V_\lambda ^\pi \left( {{{{\bf{\hat b}}}_i},{{\bf{y}}_i}|{{\boldsymbol{\uptheta }}_V}} \right)$\;
Update actor parameters ${\boldsymbol{\uptheta }}_\pi ^{(j)}$ according to gradient: 
${\nabla _{{\boldsymbol{\uptheta }}_\pi ^{(j)}}}V_\lambda ^\pi  \simeq \sum\limits_i {{w_i}\left( {\sum\limits_{u = 1}^{{N}} {{\nabla _{{\boldsymbol{\uptheta }}_\pi ^{(j)}}}\log {\pi _u}\left( {a_i^{(u)}|{{{\bf{b}}}_i}^{(u)},{{\bf{y}}_i},{\boldsymbol{\uptheta }}_\pi ^{(u)}} \right)} } \right)A_{\lambda ,i}^\pi }$\;
Update critic parameters ${{\boldsymbol{\uptheta }}_V}$  according to gradient: 
${\nabla _{{\boldsymbol{\uptheta }}_V^{}}}V_\lambda ^\pi  \simeq \sum\limits_i {{w_i}{\nabla _{{{\boldsymbol{\uptheta }}_V}}}V_\lambda ^\pi \left( {{{{\bf{\hat b}}}_i},{{\bf{y}}_i}|{{\boldsymbol{\uptheta }}_V}} \right)} A_{\lambda ,i}^\pi$\;
Update Lagrange multipliers $\lambda_m, m=1,\cdots,M,$
according to gradient:
${\nabla _{{\lambda _m}}}V_\lambda ^\pi  \simeq \sum\nolimits_{t = 0}^T {{\gamma_s ^t}} {z_{s,mt}} - {\alpha _{s,m}}$\;
}
\caption{DDMAC-CTDE with constraints}
\label{algo:DDMAC_cons}
\end{algorithm}

\section{Pavement Modeling}
\noindent
There are various indicators to characterize the condition state of pavements, such as the Pavement Condition Index (PCI), Critical Condition Index (CCI), International Roughness Index (IRI), and Load Related Distress Index (LDR), among many others \cite{hasnat2023comparative}. CCI and IRI are used in this work, consistent with the Virginia state pavement management system.

\subsection{Critical Condition Index (CCI)}\label{subsubsec:CCI}
\noindent
The CCI model in this study is a modified version of the modeling described in VDOT report \cite{katicha2016a}, where a regression model is presented that estimates mean CCI, using collected data from the years 2007-12,  and 2014-15, based on the Virginia Department of Transportation (VDOT) Pavement Management System (PMS). The age of the pavement is calculated as the difference between the year of condition reporting and the last year of relevant recorded maintenance. The total dataset in \cite{katicha2016a} consists of 3,473 observations from the years 2007 to 2012, and 1,560 observations from the years 2014 and 2015. The mean regression model in \cite{katicha2016a} directly incorporates the effect of age of the pavement section and indirectly considers the impact of the traffic load and other pavement structural parameters. Hence, based on this modeling, the effective age of the pavement section also determines its deterioration rate in time. Fig. \ref{CCI} shows the mean CCI as a function of pavement age for different traffic levels (A to E), where level A to E indicates heavy to light traffic conditions, assuming other structural parameters are constant. 
In this context, while maintaining general applicability and without any loss of generality, we make the assumption that the depicted curves for various pavements serve as indicative examples for certain pavement design criteria and traffic levels. It is important to note that for alternative design criteria and traffic levels, the resulting curves may exhibit variations. Here, we have thus illustratively assigned deterioration curve A to the representative interstate pavement type, curve C to the primary pavement type, and curve E to the secondary pavement type. Then, a nonstationary gamma process is fitted in the following section to estimate the relevant nonstationary transition probabilities.

\subsubsection{Fitting a nonstationary gamma process}\label{subsubsec:gamma_process}
\label{sec:rate}
\begin{figure}[t]
\centering
\includegraphics[width=.45\textwidth]{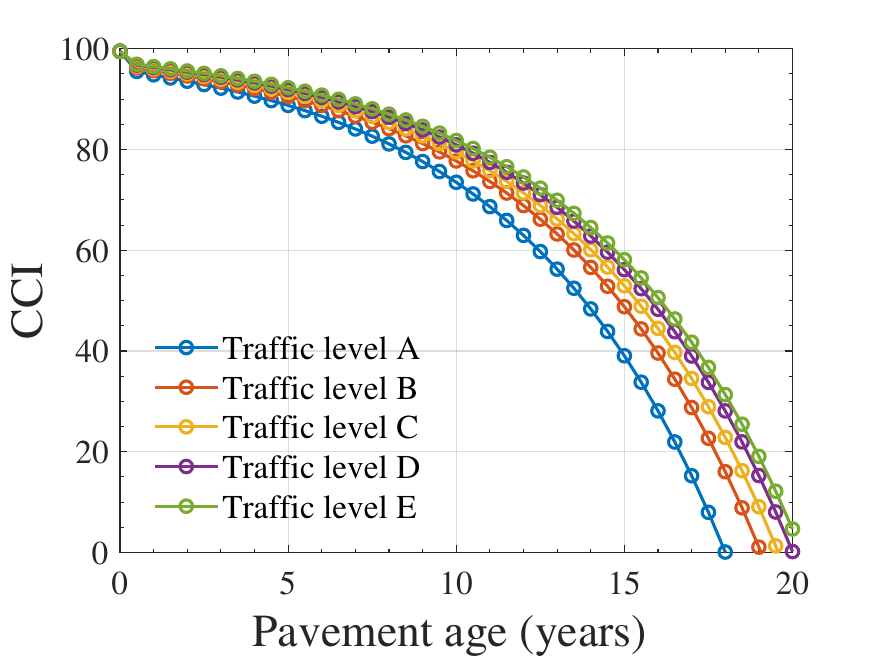}
\caption[Modeled mean CCI for different levels of traffic]{Modeled mean CCI for different levels of traffic.}
\label{CCI}
\end{figure}

\noindent
A gamma process is utilized in this section, with its mean in time equal to the modified mean CCI predictions in Fig. \ref{CCI} and a relevant model variance $\sigma _s^2\left( t \right),$  primarily determined based on \cite{katicha2016a} and the observed data trends there. The marginal probability of CCI at every step $t$ can be obtained by estimating the damage index $DI,$ where $DI = 100 - {\rm{CCI}},$ using the gamma distribution with probability density function:
\begin{equation}
    Ga\left( {DI|f\left( t \right),g\left( t \right)} \right) = \frac{{g{{\left( t \right)}^{f\left( t \right)}}}}{{\Gamma \left( {f\left( t \right)} \right)}}D{I^{f\left( t \right) - 1}}{e^{ - g\left( t \right)DI}}
\end{equation}

The gamma process is parametrized here by a non-negative time varying scale parameter function $g(t)$ and a non-negative time-varying shape parameter function $f(t)$, and $\Gamma (u) = \int\limits_0^\infty  {{v^{u - 1}}{e^{ - v}}dv} $. The relevant parameters are estimated in time based on the mean $DI$ prediction model and the model variance $\sigma _s^2\left( t \right)$.The relationships between the gamma process parameter functions and ${\mu _{DI}}$, ${\rm{ }}{\sigma _s}$  are given as:
\begin{equation}
    {\mu _{DI}}\left( t \right) = \frac{{f\left( t \right)}}{{g\left( t \right)}},{\rm{ }}{\sigma _s}\left( t \right) = \frac{{\sqrt {f\left( t \right)} }}{{g\left( t \right)}}
\end{equation}

\begin{figure}[]

\centering
   \begin{subfigure}[b]{0.4\textwidth}
        \centering
    \includegraphics[width=\textwidth]{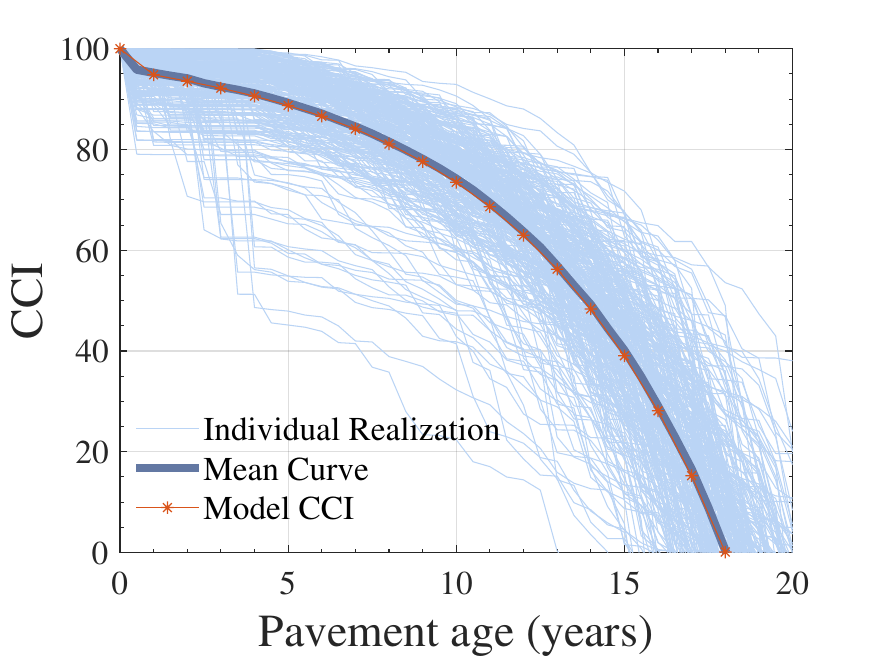}
        \caption{}
    \end{subfigure}%
    ~ 
    \begin{subfigure}[b]{0.4\textwidth}
        \centering
  \includegraphics[width=\textwidth]{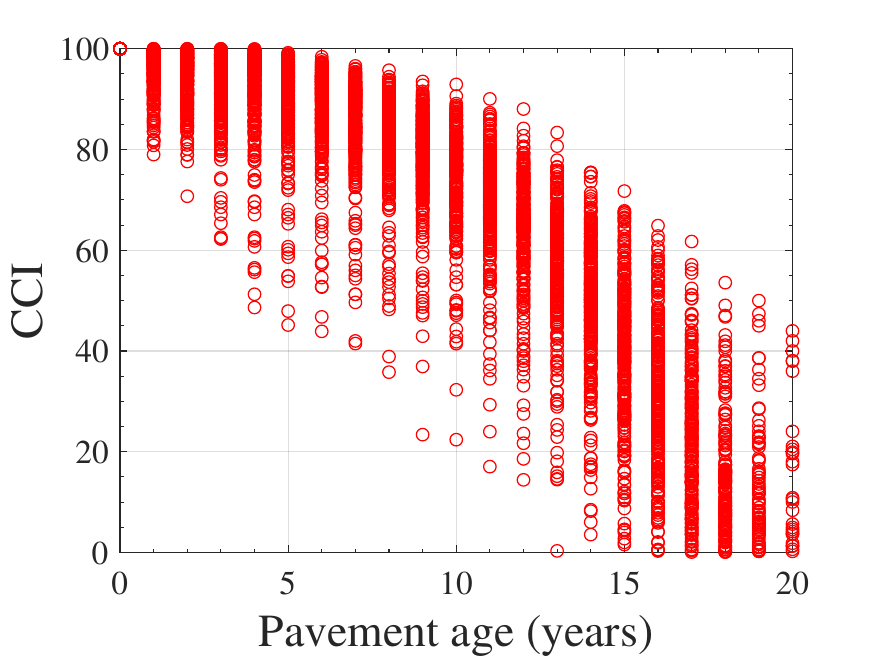}
        \caption{}
    \end{subfigure}

\caption[(a) Fitted gamma model, (b) Scatter plot for CCI for traffic level A]{(a) Fitted gamma model, (b) Scatter plot for CCI corresponding to traffic level A.}
\label{fig:3.3}
\end{figure}

Due to their monotonicity, gamma processes are readily used as a suitable modeling choice in stochastic deterioration engineering applications and can describe continuous Markovian transitions. For time instant ${t_1} < {t_2}$, the increment of $DI$ follows a gamma distribution:
\begin{equation}
    DI\left( {{t_2}} \right) - DI\left( {{t_1}} \right)\sim Ga\left( {.|f\left( {{t_2}} \right) - f\left( {{t_1}} \right),{\rm{ }}g\left( {{t_2}} \right)} \right)
\end{equation}
where $g(t)$ is assumed to be constant in the interval $ \left(t_1, t_2 \right]$. In Fig. \ref{fig:3.3}, relevant simulation results are indicatively shown for traffic level A with 300 different realizations. All corresponding $f(t)$ and $g(t)$ values are shown in Table \ref{tab:gammacoef} in the Appendix, for all different traffic levels.

\subsubsection{Determining transition probabilities}\label{subsec:TP}
\noindent
To calculate the transition probabilities, one must first specify the discrete condition states, as given in Table \ref{tab:3.1}. The discretized condition states in the table are mostly based on the VDOT's prescribed maintenance recommendations in \cite{v2016a}, which are as follows for interstate highways: For CCI values greater than 89, the treatment category is always Do Nothing (DN); for values greater than 84, the treatment category is always DN or Preventive Maintenance (PM); for values between 60 and 84, the treatment category is PM; for values less than 60, the treatment category is Corrective Maintenance (CM), Restorative Maintenance (RM), or Rehabilitation/Reconstruction (RC); for values less than 49, the treatment category is always RM or RC; and for values less than 37, the treatment category is always RC. The detailed definitions of VDOT maintenance actions are provided in Table \ref{tab:actions} in the Appendix.

\begin{figure}[]

\centering
   \begin{subfigure}[b]{0.4\textwidth}
        \centering
    \includegraphics[width=\textwidth]{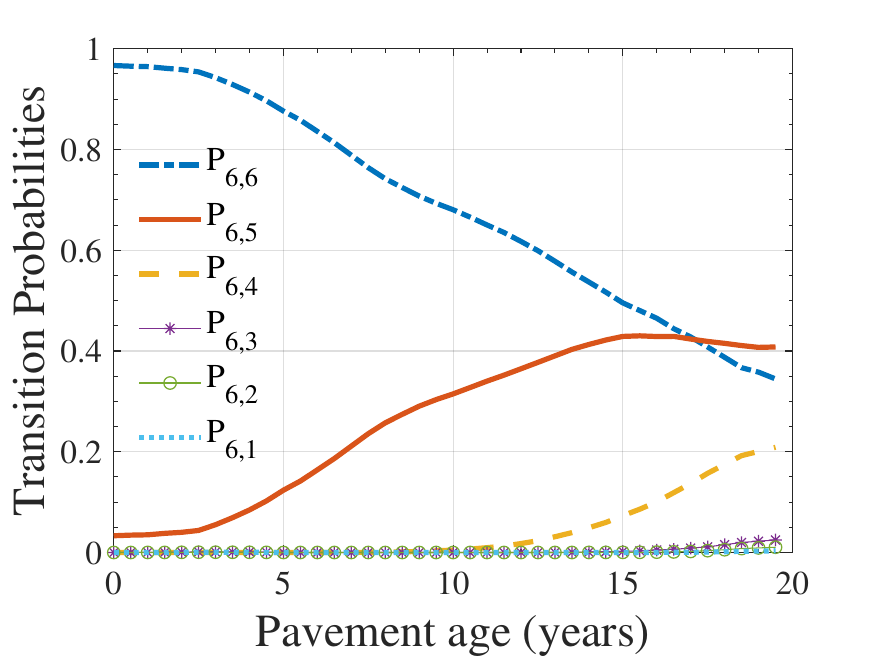}
        \caption{}
    \end{subfigure}%
    ~ 
    \begin{subfigure}[b]{0.4\textwidth}
        \centering
   \includegraphics[width=\textwidth]{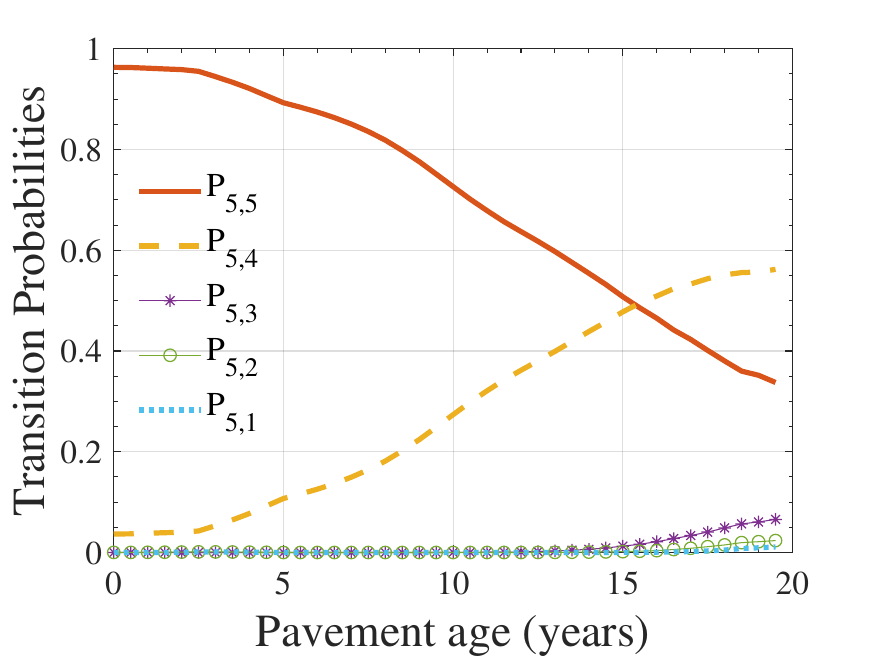}
        \caption{}
    \end{subfigure}

\caption[Transition probabilities for Traffic level A, with (a) starting state = 6, (b) starting state = 5, smoothed over time with a 5-point window]{Transition probabilities for Traffic level A, with (a) starting state = 6, (b) starting state = 5, smoothed over time.}

\label{fig:3.4}
\end{figure}

Monte Carlo sampling is then performed by generating CCI and pavement age pairs, propagated in time following the described gamma process. A million sequences are generated in total to obtain the transition probabilities for a given traffic level. Fig. \ref{fig:3.4} indicatively shows computed transition probabilities for traffic level A. Various advanced sampling techniques, like Hamiltonian Monte Carlo \cite{arcieri2023bridging, papakonstantinou2023hamiltonian}, can be naturally used for more challenging distributions.

\begin{table}[b]
\centering
\begin{threeparttable} 
\caption[State discretization based on CCI values]{\em{State discretization based on CCI values.}}
\label{tab:3.1}
\begin{tabular}{ccc}
\hline \noalign {\smallskip}

CCI state (s) & CCI values & Pavement Condition\\
\hline \noalign {\smallskip}
$s = 6$ & 100-90 & Excellent\\
$s = 5$ & 89-80 & Very Good \\
$s = 4$ & 79-61 & Good\\
$s = 3$ & 60-50 & Fair\\
$s = 2$ & 49-37 & Poor\\
$s = 1$ & $<$37 &Very Poor\\
\hline \noalign {\smallskip}

\end{tabular} 
\end{threeparttable} 
\end{table}

\subsubsection{Observation probabilities for CCI inspection action}\label{sec:obs_prob_CCI}
\noindent
Observation uncertainty can be appropriately modeled by the likelihood functions $p({o_t}|{s_t})$ which quantify the probability of observing an observation $o_t$ at time $t$ given a state $s_t$ at that instant. To calculate the observation probability $p({o_t}|{s_t})$ a normal distribution is considered with mean equal to the actual CCI value and an error variance $\sigma _{error}^2 = 72$, based on \cite{katicha2016a}, and is considered as a low fidelity inspection due to the significant observation uncertainty. The error variance is linked to the variability in the measurement of CCI and the error in the reported pavement condition by an inspector. These conditional observation probabilities, $p({o_t}|{s_t})$, are generated by calculating the area under the curve of the normal distribution.

It has been observed that the observation probabilities are largely independent of different traffic levels, so the average values have been reported in Table \ref{tab:3.2}, which can be used for all traffic levels values. High-fidelity inspections can also be considered, which rely on higher precision, expensive devices, and multi-modality. These techniques may also require significant post-processing which increases the overall inspection cost. Hence, a much smaller error variance, $\sigma _{error}^2 = 18$, is used to compute the observation probabilities, as compared to the previous case. Using these inspection techniques, we derive the observation probabilities as shown in Table \ref{tab:3.3}. The no-inspection case is also mathematically equivalent to when one has equal probability of observing each state i.e., $p({o_t}|{s_t}) = 1/6$ $\forall {\rm{ }}{o_t},{\rm{ }}{s_t} \in \{ 1,2 \cdots ,6\} .$ In Table \ref{tab:3.2}, we notice observation probabilities less than 50\% for observing the actual states, which are attributed here to the non-uniform CCI partitioning, as mentioned in Section \ref{subsec:TP}.

As discussed earlier, the cost of inspections depends on the techniques used for observing the pavement condition along with the post-processing time/complexity. The inspection techniques have been categorized into three precision categories of no-inspection, low-fidelity inspection, and high-fidelity inspection, as previously mentioned, and the associated costs are informed by the Michigan DOT survey \cite{m2014a} for monitoring highway assets. These costs are summarized in Table \ref{tab:3.4} in the Appendix.

\subsubsection{Maintenance, rehabilitation, and reconstruction actions in relation to CCI}\label{mainte}
\noindent
For ease of notation and to simplify expressions, the word maintenance is used here forth as a general term encompassing all actions in relation to preventive treatments, repairs, restorations, and reconstructions. There are various guidelines for pavement maintenance for different agencies/states. Along similar lines to VDOT \cite{v2016a}, four different maintenance actions are utilized here, i.e.,  ({\it Do Nothing, Minor Repair, Major Repair, and Reconstruction}). {\it Do Nothing} is the same as DN by VDOT, and {\it Minor Repair} is more aligned with VDOT's Preventive Maintenance (PM) category. {\it Major Repair} is, accordingly, similar to Corrective Maintenance (CM) and Restorative Maintenance (RM), and lastly {\it Reconstruction} is the same as VDOT's Reconstruction action. Further details related to VDOT's maintenance actions are provided in the Appendix Table \ref{tab:actions}.

The maintenance action transition probabilities for CCI are shown in Tables~\ref{tab:3.5}-\ref{tab:3.7} similar to the line of work presented in \cite{madanat1994a}. It is assumed that the {\it{Minor Repair}} (crack filling, moderate patching, etc.) cannot change the rate of deterioration, or else the effective age of the component, but only its condition state. The change in condition states is considered based on Table \ref{tab:3.5}. 
For a {\it{Major Repair}} maintenance action, deterioration rate is however reduced to that of an asset younger by 5 years or to that of a newly built pavement component, i.e., ${({T_{age}})_{updated}} = \max (0,{T_{age}} - 5)$ years. {\it{Major Repair}}s also alter the CCI conditions, with the transition probabilities shown in Table \ref{tab:3.6}. 
The {\it{Reconstruction}} maintenance action brings the CCI condition back to the intact state (i.e., $s_t = 6$) with certainty, as shown in  Table \ref{tab:3.7}, and alters the deterioration rate of the component to the initial intact value, or equivalently assigns it an effective age of 0. Transitions in these tables do not include the transition due to the environment and its deterioration effects. Decoupled environmental transitions can essentially allow us to include action duration and probability effects of successfully completed actions, among others. In Table~\ref{tab:3.5}, we can observe that after applying maintenance actions the system can still transition to a worse state with a non-zero probability. This happens because we are accounting for the inherent uncertainty present in the outcome of the applied {\it{Minor Repair}} action, which also includes the possibility of failed actions.

\begin{table}[]
\centering
\begin{threeparttable} 

\caption[]{{\em{Maintenance action costs for asphalt pavements, reported in}} USD/m\textsuperscript{2}.}
\label{tab:3.8}
\begin{tabular}{@{}ccccc@{}}
\toprule
Actions        & Description                                                                                                                                          & \begin{tabular}[c]{@{}c@{}}Cost\\  (USD/m\textsuperscript{2})   \\ Interstate\end{tabular} & \begin{tabular}[c]{@{}c@{}}Cost \\ (USD/m\textsuperscript{2})\\ Primary\end{tabular} & \begin{tabular}[c]{@{}c@{}}Cost\\  (USD/m\textsuperscript{2}) \\ Secondary\end{tabular} \\ \midrule
Do Nothing     & NA                                                                                                                                                   & 0.00                                                                        & 0.00                                                                  & 0.00                                                                     \\ \midrule
Minor Repair   & \begin{tabular}[c]{@{}c@{}}Moderate patching (\textless{}10\%), \\surface treatment, \\ partial depth patching, \\thin Asphalt Concrete \\(AC) overlay\end{tabular}     & 20                                                                          & 16                                                                    & 10                                                                       \\ \midrule
Major Repair   & \begin{tabular}[c]{@{}c@{}}Heavy patching (\textless{}20\%\\  of the pavement area),\\ full depth patching,   \\ structural overlay\end{tabular} & 75                                                                          & 68                                                                    & 52                                                                       \\ \midrule
Reconstruction & \begin{tabular}[c]{@{}c@{}}Replacement of the entire\\  pavement section\end{tabular}                                                                     & 350                                                                         & 330                                                                   & 250                                                                      \\ \bottomrule

\end{tabular} 
\end{threeparttable} 
\end{table}

\subsubsection{Maintenance action costs and durations}
\noindent
The cost of different maintenance actions is considered from \cite{v2016a, a2020a, f2020a, penndot2017a} and provided for interstate, primary, and secondary roads in Table~\ref{tab:3.8}. It is reported in USD/m\textsuperscript{2} and the total cost can be calculated using the components’ length and width (which is considered as ~3.7 m or 12 ft per lane). Here, it is important to mention that the action descriptions provided in Table~\ref{tab:3.8} for {\it{Minor Repair and Major Repair}} are indicative actions of the corresponding category and these actions are a subset of actions used in \cite{v2016a}, as presented in the Appendix in Table \ref{tab:actions}.

The duration of different maintenance actions is inferred from \cite{a2018a, penndot2019a}. It is further assumed that no action takes more than 2 years, of which only {\it{Reconstruction}} actions can take more than one year to complete for a component. This can be justified by assuming multiple maintenance activities at a time for longer and multi-lane components. These durations are provided in Table \ref{tab:3.9}. If {\it{Reconstruction}} action taking more than a year is performed by the agent, the component will not be available for the next time-step as part of the active component set. Thus, no action is recommended for this component at the next time-step, and this can be tracked in our DRL framework using an additional state.


\subsection{International Roughness Index (IRI)}
\noindent
IRI (m/km) is another metric that quantifies the functional pavement condition in terms of the roughness experienced by vehicle passengers. IRI is calculated from measured longitudinal road profiles using the simulated vertical displacement response of a quarter-car \cite{sayer1986a}. IRI (m/km) in this work is discretized into 5 states, as in \cite{faddoul2013a, f1999a} presented in Table~\ref{tab:3.10}. The transition probabilities for the {\it{Do Nothing}} action are now given in Table~\ref{tab:3.11}.
\subsubsection{Maintenance actions in relation to IRI}
\noindent
The same four maintenance actions are considered for IRI: {\it{Do Nothing, Minor Repair, Major Repair, and Reconstruction}}. The IRI transitions for {\it{Minor Repair to Reconstruction}} actions are given in  Tables~\ref{tab:3.12}-\ref{tab:3.14} respectively, similar to the CCI case. Further, the cost of maintenance actions in relation to IRI is the same as the one reported in Table~\ref{tab:3.8}. Maintenance actions taken at any given time will improve both CCI and IRI indices simultaneously.

\begin{table}[]
\centering
\begin{threeparttable} 

\caption[State classification based on IRI (m/km) values]{\em{State classification based on IRI (m/km) values.}}
\label{tab:3.10}
\begin{tabular}{@{}ccc@{}}
\toprule
Condition State (s) & IRI (m/km)      & Pavement Condition \\ \midrule
$s = 5$             & \textless{}0.95 & Very Good          \\
$s = 4$             & 0.95-1.56       & Good               \\
$s = 3$             & 1.57-2.19       & Fair               \\
$s = 2$             & 2.20-3.14       & Mediocre           \\ 
$s = 1$             & \textgreater{}3.15 & Poor               \\ \bottomrule
\end{tabular} 
\end{threeparttable} 
\end{table}

\begin{table}[]
\centering
\begin{threeparttable} 

\caption[Do-Nothing transition probabilities for 5 IRI states]{\em{Do-Nothing transition probabilities for 5 IRI states.}}
\label{tab:3.11}
\begin{tabular}{@{}cccccc@{}}
\toprule
Condition State & $s_{t+1} = 5$ & $s_{t+1} = 4$ & $s_{t+1} = 3$ & $s_{t+1} = 2$ & $s_{t+1} = 1$ \\ \midrule
$s_t = 5$       & 0.840         & 0.121         & 0.039         &               &               \\
$s_t = 4$       &               & 0.788         & 0.142         & 0.070         &               \\
$s_t = 3$       &               &               & 0.708         & 0.192         & 0.01          \\
$s_t = 2$       &               &               &               & 0.578         & 0.422         \\
$s_t = 1$       &               &               &               &               & 1.000         \\ \bottomrule
\end{tabular} 
\end{threeparttable} 
\end{table}

\subsubsection{Observation probabilities for IRI inspection actions}
\noindent
Three assumed inspection actions of different fidelities are again considered in this case, and the measurement errors related to the respective inspection technologies are assumed to be normally distributed with zero mean and standard deviations (SD) of 0.08, 0.32, and $\infty$ m/km, respectively, as reported in \cite{faddoul2013a}. The observation probabilities are summarized in Table~\ref{tab:3.15}, where $o_t$ is the observed state and $s_t$ is the (hidden) actual state. The underlying assumption is that more costly inspections are high-fidelity inspections that provide more accurate information. For the last inspection technique, an infinite standard deviation is considered. Thus, this inspection is completely uninformative, equivalent to a no-inspection action.

Inspection costs are considered independent of the pavement state, as also considered in \cite{faddoul2013a}, and depend on the fidelity of the inspection technology that is utilized. In Table~\ref{tab:3.15} costs are equal to 0.10, 0.03, 0 USD per m\textsuperscript{2}, respectively. Table~\ref{tab:3.4} and \ref{tab:3.15} provides the costs of individual inspection actions for CCI and IRI, respectively, while in Table~\ref{tab:3.16} the joint cost of pavement inspection is considered when both IRI and CCI are observed, as adopted in this work when an inspection action is chosen.

\begin{table}[h!]
\centering

\begin{threeparttable} 

\caption[]{\em{Pavement inspection cost (i.e., combined cost of IRI and CCI).}}
\label{tab:3.16}
\begin{tabular}{@{}ccc@{}}
\toprule
Inspection  Technique & Description   & Cost (in USD/m\textsuperscript{2}) \\ \midrule
$i_2$                                                             & High fidelity & 0.20          \\
$i_1$                                                             & Low fidelity  & 0.10          \\
$i_0$                                                             & No inspection & 0.00          \\ \bottomrule
\end{tabular} 
\end{threeparttable} 
\end{table}

\section{Bridge Modeling}
\noindent
Decks are considered in this study to characterize the bridge condition and inform action selection, due to their expensive and frequent maintenance needs, in relation to other bridge subsystems. Decks often have the most intense rate of deterioration, among bridge components, with factors such as corrosion, cracking, freezing and thawing cycles, and direct exposure to traffic loads contributing toward their degradation, their condition becomes the dominant factor in determining the overall bridge condition rating.

\subsection{Bridge decks}
\begin{figure}[]
\centering
\includegraphics[width=.45\textwidth]{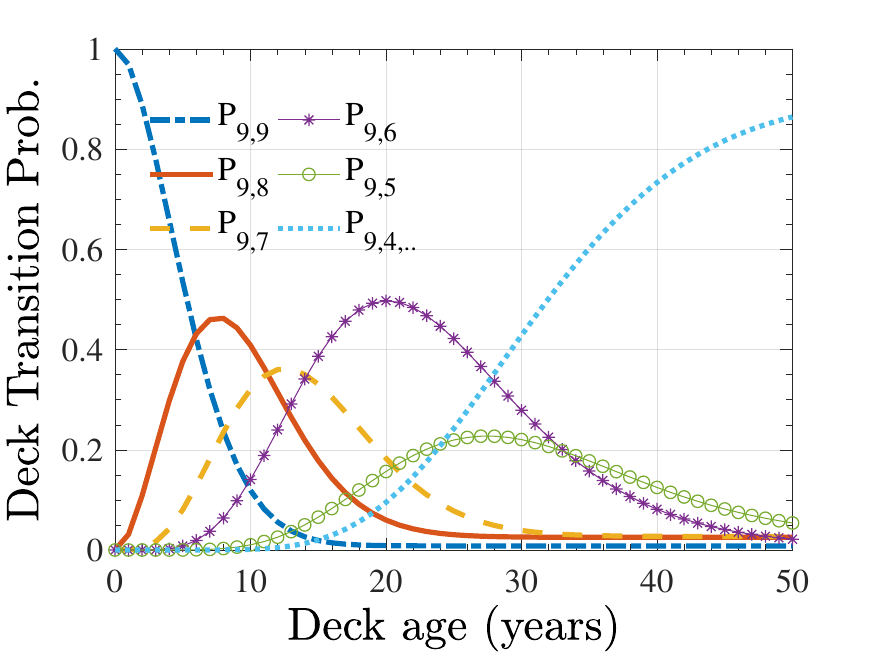}
\includegraphics[width=.45\textwidth]{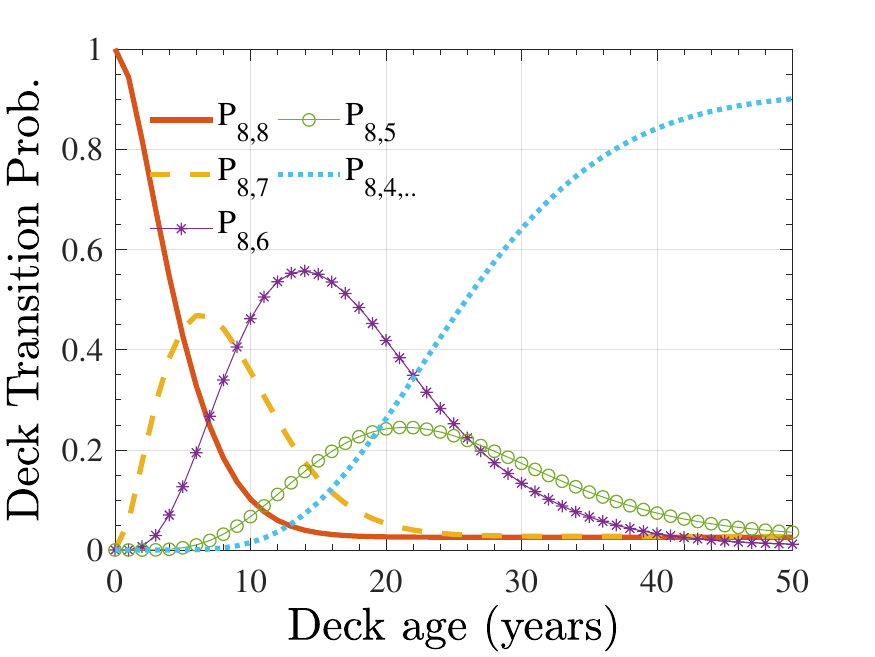}
  
\caption[Transition probabilities in time, moving from state 9 (left) and 8 (right) to lower states]{Transition probabilities in time, moving from state 9 (left) and 8 (right) to lower states.}

\label{fig:3.5}
\end{figure}

\noindent
For bridge deck condition, 9 total states are generally considered, with state 9 being the undamaged state, as per the convention adopted by PennDOT \cite{penndot2009a} and other DOTs. For this report, state 4,... characterizes all subsequent states, as also considered in \cite{manafpour2018a}. Thus, 6 bridge deck states along with a failure state are used. The transition probabilities related to these 6 condition states are based on approximately 30 years of in-service performance data for over 22,000 bridges in Pennsylvania, as analyzed in \cite{manafpour2018a} for different exogenous parameters, i.e., deck length, location, protective coating, span type, etc. Fig.~\ref{fig:3.5} shows an indicative case based on the following exogenous parameters: Deck length = 90 ft; Location (district) type = 4; Reinforcement bar protection = 1 (no protective coating); Type of span interaction for main unit = 1 (simple, non-composite); Primary load carrying members = 1 (reinforced concrete); Deck surface type = 1 (concrete); Total number of spans in main unit = 2 (multi-span); NHS = 1 (interstate route); Maintenance type = type 1 (no maintenance). Apart from these 6 nonstationary transitions, stationary failure probabilities are also considered, as shown in Table~\ref{tab:3.17}.  Note that these failure probabilities are indicative; the actual failure probabilities depending on multiple factors, such as bridge type, location, and hazard and other factors, can be computed using methods presented in \cite{papakonstantinou2023hamiltonian, papakonstantinou2023hamiltonian2}.

\begin{table}[]
\centering
\begin{threeparttable} 

\caption[Bridge failure probability given bridge condition state]{\em{Bridge failure probability given bridge condition state.}}
\label{tab:3.17}
\begin{tabular}{@{}cccccccc@{}}
\toprule
Condition State & $s_t = 9$ & $s_t = 8$ & $s_t = 7$ & $s_t = 6$ & $s_t = 5$ & $s_t = 4,...$ & $s_t = failed$ \\ \midrule
\begin{tabular}[c]{@{}c@{}}Failure prob.\\ ($s_{t+1} = failed$)\end{tabular} & 0.001     & 0.001     & 0.005     & 0.005     & 0.005     & 0.01          & 1.000             \\ \bottomrule
\end{tabular} 
\end{threeparttable} 
\end{table}

\subsubsection{Maintenance actions for bridge decks}
\noindent
Similar to pavements, four maintenance actions are considered for maintaining the bridge decks:  {\it{Do Nothing, Minor Repair, Major Repair, and Reconstruction}}. It is again assumed here that the {\it{Minor Repair}} action does not change the rate of deterioration of the deck, but it can change the condition state of the structure, as per the transition probabilities given in Table~\ref{tab:3.18}. A {\it{Major Repair}} action can, however, improve the deterioration rate to that of a deck younger by five years or a reset to a newly made deck, i.e., ${({T_{age}})_{updated}} = \max (0,{T_{age}} - 5)$ years, and the change in condition state is now considered as reported in Table~\ref{tab:3.19}. The transition probabilities for the {\it{Reconstruction}} action are shown in Table~\ref{tab:3.20} and reset the state and deterioration rate of the deck to a newly built one.  

\subsubsection{Maintenance action costs and durations for bridge decks}
\noindent
The cost of maintenance actions has small dependence upon the current state of the bridge deck. As a result, maintenance costs are considered independent of the state of the deck here. The cost for a reconstruction action is obtained from \cite{f2019a} and the costs for minor and major maintenance actions are considered as 15\% and 45\% of that, respectively, as inferred from \cite{wells1995a}. Table~\ref{tab:3.21} shows the costs of performing individual actions in USD/m\textsuperscript{2}. It can be observed that the unit costs of actions for bridges are higher than the unit costs of pavement actions.

The duration of different maintenance actions is mainly inferred from \cite{a2018a, goodspeed2017a, oakgrove2013a}.  For large bridges, it is assumed that maintenance activity is performed at multiple locations simultaneously to increase efficiency. The action durations are provided in  Table~\ref{tab:3.22} for 3 different types of bridges, types I-III, categorized based on their sizes. Type I are the largest size bridges with heavy traffic, thus, it takes the longest to repair, type II are intermediate size bridges with medium traffic, thus it takes less time than type I bridges to repair, and type III are small size bridges with the lowest traffic, needing the least amount of time to repair. More details about different types of bridges are provided in Section~\ref{sec:topo}. 

\begin{table}
\centering
\begin{threeparttable} 
\caption[]{ {\em{ Cost of maintenance actions in}} USD/m\textsuperscript{2}.}
\label{tab:3.21}
\begin{tabular}{ccc}
\toprule
Actions        & Description                                                                                                                                                 & Cost (USD/m\textsuperscript{2}) \\ \midrule 
Do Nothing     & NA                                                                                                                                                          & 0.00             \\
Minor Repair   & \begin{tabular}[c]{@{}c@{}}Moderate cracks \\ filling and patching area \textless{}10\% of the deck\\ area, minor replacement of reinforcement\end{tabular} & 400.00           \\
Major Repair   & \begin{tabular}[c]{@{}c@{}}Fixing Spalls/delamination \\ with deck   area \textless{}25\%, \\ major replacement of reinforcement\end{tabular}               & 1,200.00         \\
Reconstruction & Replacement of the entire deck                                    & 2,650.00\\ \bottomrule        
\end{tabular} 
\end{threeparttable} 
\end{table}

\subsubsection{Inspection actions and observation probabilities}
\noindent
There is a variety of destructive and non-destructive inspection techniques that are used for bridge decks, like visual inspection, acoustic sensing, infrared/thermal imaging, ground penetrating radar, coring and chipping, and half-cell potential test, among many others. Based on their accuracy, inspection techniques can be characterized as $i_0, {i_1},$ and $i_2$ corresponding to uninformative, low-fidelity, and high-fidelity inspections, respectively, and the relevant inspection costs are related to their accuracy. As assumed for pavements, we here again relate the inspection costs to the reconstruction cost of the bridge and we consider 0.1\% and 0.5\% of the rebuild cost for low- and high-fidelity inspection techniques, respectively. The resulting costs are shown in Table~\ref{tab:3.23}.

\begin{table}[]
\centering
\begin{threeparttable} 
\caption[]{ \em{Inspection action costs for 3 fidelity levels.}}
\label{tab:3.23}
\begin{tabular}{@{}ccc@{}}
\toprule
 Inspection Technique & Description & Cost (USD/m\textsuperscript{2})                \\ \midrule
$i_2$                                       & High fidelity      & 1.20 \\
$i_1$                                       & Low fidelity       & 0.50                         \\
$i_0$                                       & No Inspection      & 0.00  \\ \bottomrule                      
\end{tabular} 
\end{threeparttable} 
\end{table}

The observation probabilities for the bridge decks are considered similarly to \cite{madanat1993a, madanat1994a, faddoul2013a} and for the no-inspection case  $p({o_t}|{s_t}) = 1/6$    $\forall {\rm{ }}{o_t},{\rm{ }}{s_t} \in \{ 1,2 \cdots ,6\} $ is used, as the failure state is assumed to be self announcing. Table \ref{tab:3.24} and Table \ref{tab:3.25} describe the observation probabilities for low- and high-fidelity inspection techniques, respectively.

\section{Transportation Network}
\subsection{Topology and functionality}\label{sec:topo}
\begin{figure}
\centering
\includegraphics[width=.45\textwidth]{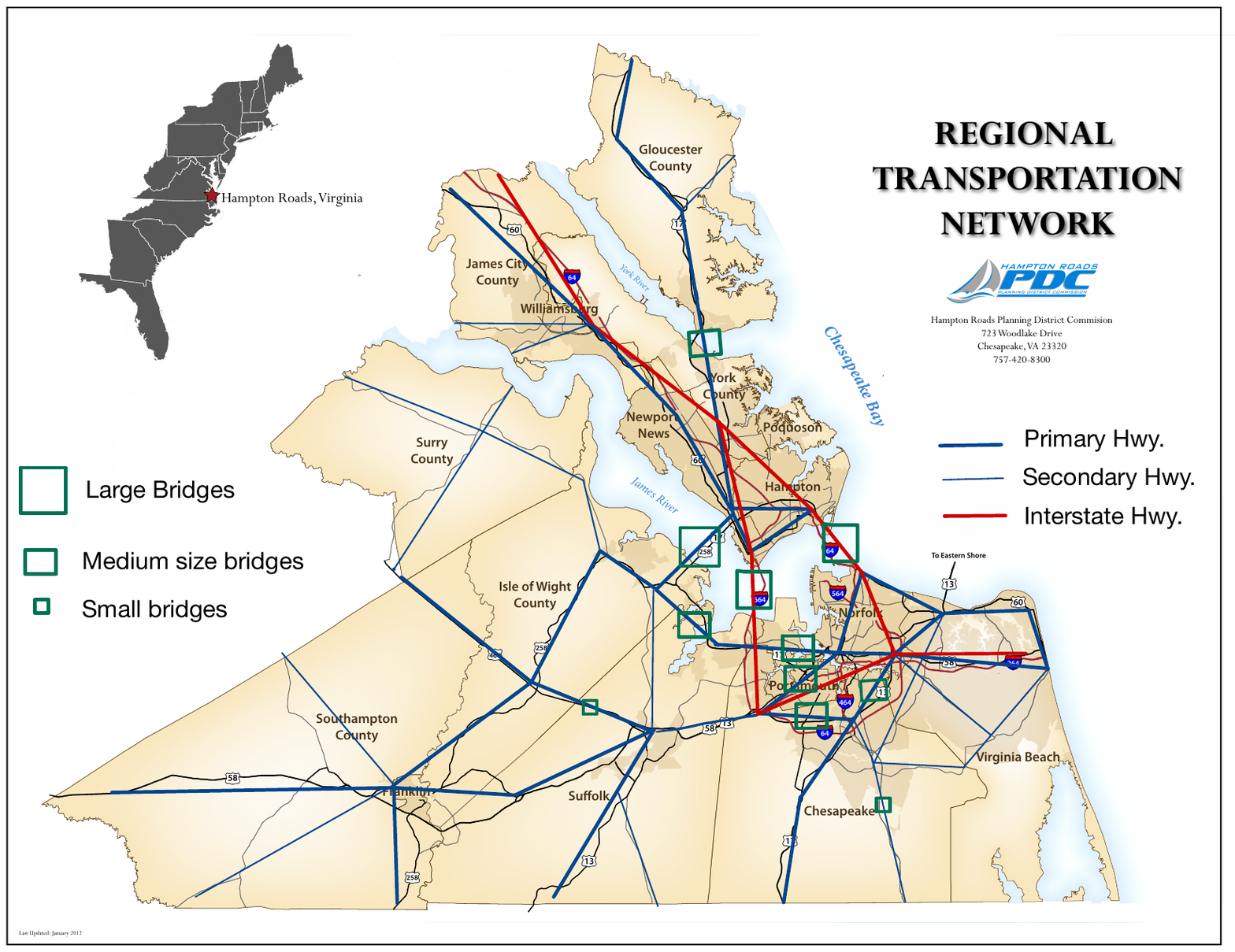}
\includegraphics[width=.42\textwidth]{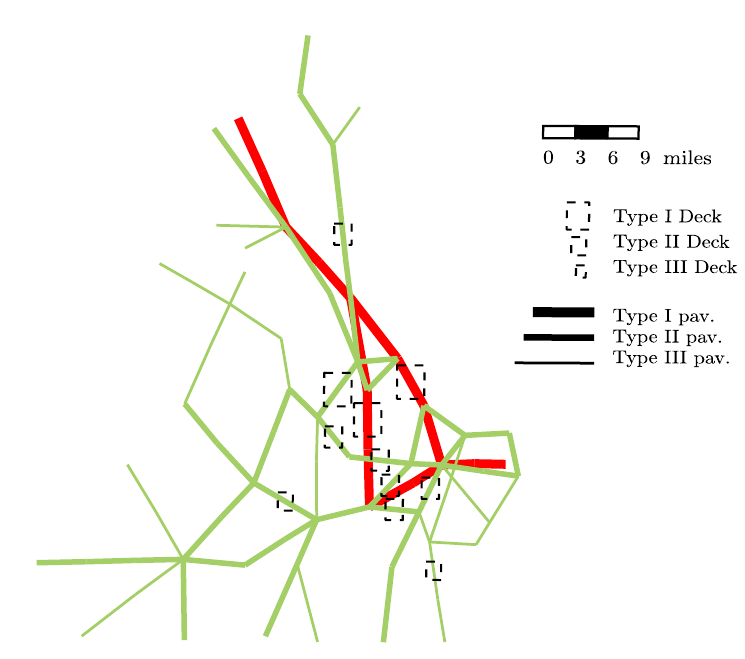}
  
\caption[]{(Left) Modified Hampton Roads network map \cite{hamptonmaps}. (Right) Network topology modeling of Hampton Roads: Green links are primary and secondary pavements, and red links are interstate highways.}
\label{fig:3.6}
\end{figure}

\noindent
As a testbed, we consider a modified version of the Hampton Roads transportation network, in Virginia, USA. The original topology, road characteristics, and daily traffic data of the network are used, together with 11 main network bridges for this cross-asset environment, that are related to important network routes. Each bridge is bidirectional, with the same number of lanes as the original network, as illustrated in Fig.~\ref{fig:3.6}. Different decks, types I-III, are categorized based on their relative sizes and traffic volume in descending order. Type I bridges are large bridges, typically having a length of more than 2 miles, with the highest volume of traffic; type II bridges range between 0.4-2 miles with intermediate traffic volume; and type III bridges usually have lengths less than 0.4 miles and have the lowest traffic volume. An exception to this classification is the Great Bridge bypass, which should be of type II based on its size, but as it is on a secondary highway with no nearby interstates or high-traffic volume highways, it is categorized as type III. More details on bridge characteristics are shown in Table \ref{tab:Deck_curr}. Traffic volumes, which can be inferred from \cite{trafficVDOT2019}, are also related to user delay costs associated with maintenance actions, as also explained later in more detail. 

Similarly, the network has various pavement components categorized as type I-III. Type I pavements are associated with interstate highways with bidirectional traffic having four lanes in each direction, thus, constituting the highest vehicular mileage per mile. Type II pavements correspond to the primary state highways with bidirectional medium-level traffic, having two lanes in each direction. Lastly, type III pavements characterize the secondary state highways with low-level bidirectional traffic, with one lane in each direction. Fig.~\ref{fig:3.6} (left) shows the Hampton roads map, and Fig.~\ref{fig:3.6} (right) represents the schematic of the network with green links pertaining to primary and secondary highways and red links representing interstate highways. Further, the deterioration rates of pavements are being selected from Fig.~\ref{CCI} based on the traffic levels in this work, without loss of generality, as mentioned in Section \ref{subsubsec:CCI}.
\subsection{Network level risk}
\noindent
In the context of reliability-based decision problems, the risk is usually defined as an expected cumulative discounted failure state cost over the life-cycle, as in \cite{andriotis2021deep}. Using the same definition, we will associate risk with bridge failures only for this network, and the associated cost can be calculated as:
\begin{equation}
    c_{\mathfrak{R}^\pi } = {\mathds{E}_{{o_{1:T}}}}\left[ {\sum\limits_{t = 0}^T {{\gamma ^t}\left({c_{F_1}}{P_{{F_{t + 1}}|{a_{0:t}}{o_{0:t}}}}+{c_{F_2}\left({P_{{F_{t + 1}}|{a_{0:t}}{o_{0:t}}}}-{P_{{F_{t}}|{a_{0:t}}{o_{0:t}}}}\right)}\right)} } \right]
\end{equation}
where $c_{\mathfrak{R}^\pi }$ is the risk cost which has two components: (1) accruable cost ($c_{F_1}$) and (2) instantaneous cost ($c_{F_2}$). Instantaneous failure risk cost is the cost incurred upon the system entering a damage state and is not collected for the duration of the system's sojourn time in this damage state. It is herein taken as ten times the rebuilding cost of the bridge. The accruable cost is taken twice as the rebuilding cost of the bridge and is imposed at every time step during the relevant sojourn time. Two types of risks are considered: (i) risk of individual bridge failures (for all network bridges) and (ii) system-level risk. The risk associated with bridge failures can straightforwardly be calculated based on their failure probability and the two costs mentioned above. This risk can be directly added to the cost function or objective function.

\begin{figure}[t]
\centering
    \begin{subfigure}[b]{0.32\textwidth}
        \centering
   \includegraphics[width=\textwidth]{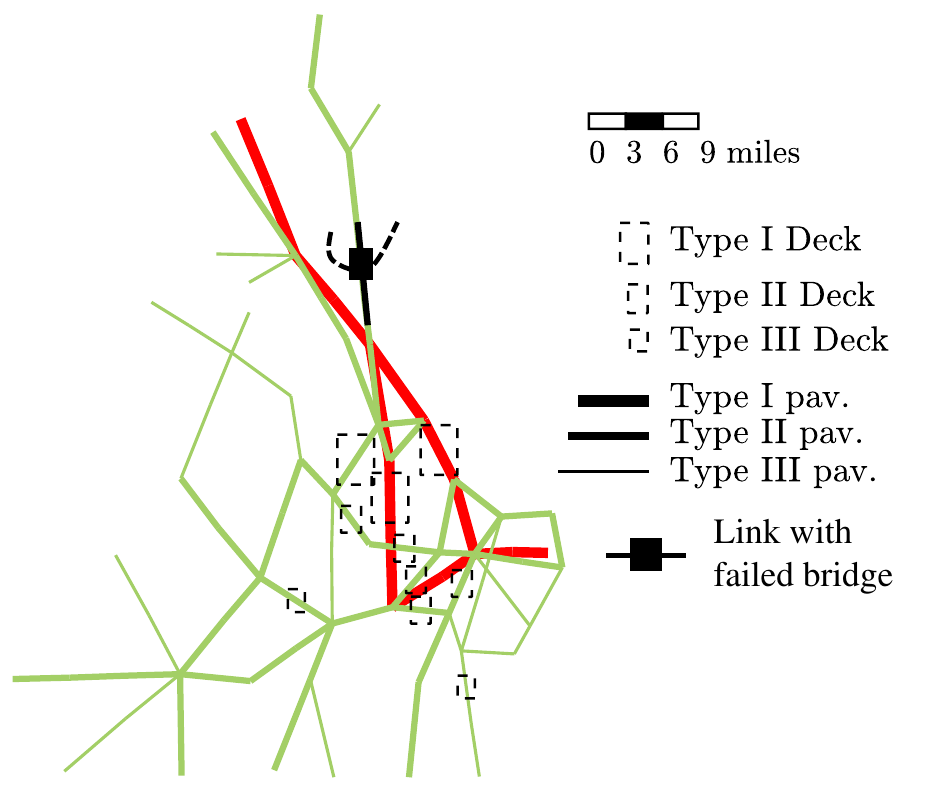}
        \caption{}
    \end{subfigure}%
    ~ 
    \begin{subfigure}[b]{0.32\textwidth}
        \centering

   \includegraphics[width=\textwidth]{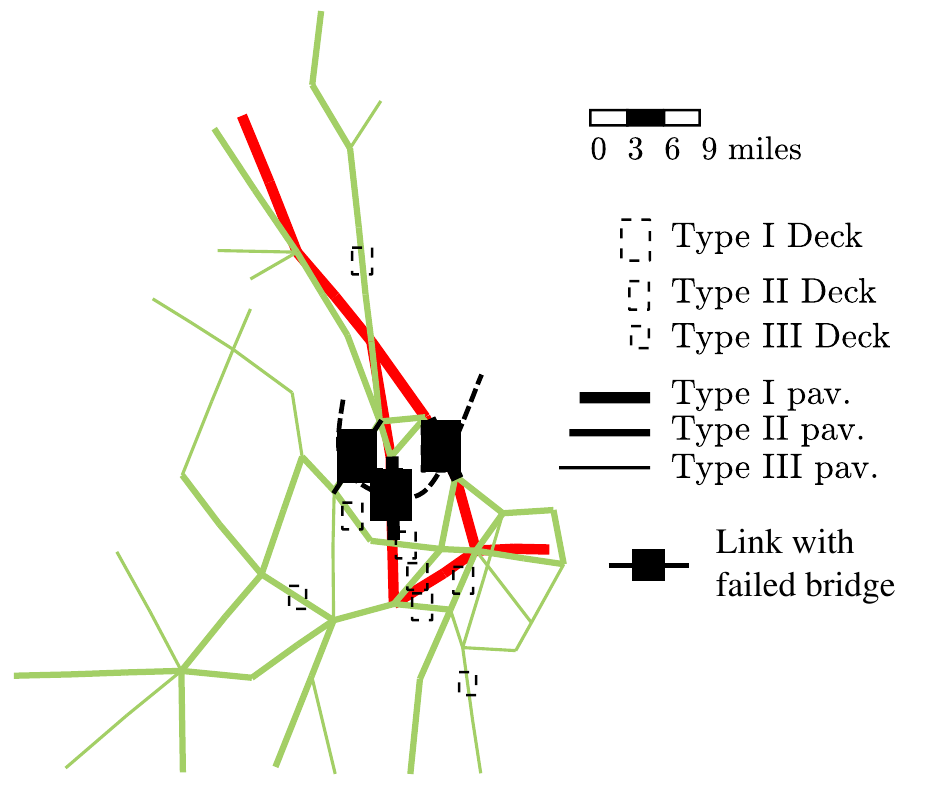}
        \caption{}
    \end{subfigure}
    ~
    \begin{subfigure}[b]{0.32\textwidth}
        \centering
   \includegraphics[width=\textwidth]{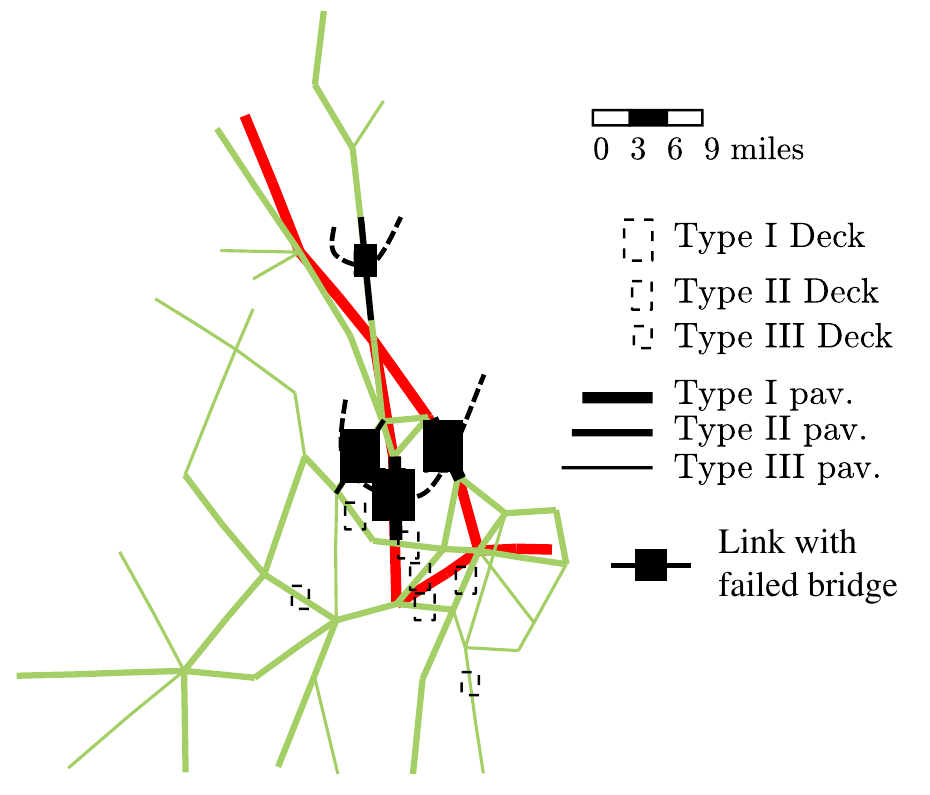}
        \caption{}
    \end{subfigure}
   
\caption[Different configurations of system failures (a) Mode 1, (b) Mode 2, (c) Mode 3]{Different configurations of system failures (a) Mode 1, (b) Mode 2, (c) Mode 3}
\label{fig:3.7}
\end{figure}

For the system risk, first, we define the system level risk and its failure modes 1-3. The loss of connectivity due to bridge failures of different parts of the network is considered a system failure. Mode 1 failure is considered when the network loses the connection between Gloucester County to Hampton connected via a single bridge. Mode 2 failure is defined as the failure of all three bridges connecting Hampton to the southern part of the network. Finally, mode 3 failure is defined when mode 1 and mode 2 failures occur simultaneously. These failure modes are shown in Fig. \ref{fig:3.7}. Failed links and bridges are indicated in black, and the dotted line indicates the network partitioning due to disconnection.

It is critical to note that maintenance actions do not result in link closures; a link closure only occurs when a bridge between two nodes fails. Based on the failure probability of bridges, the failure probability of the three modes can be obtained in closed form using a combination of closed links. The two costs associated with each mode can be calculated as the sum of the failure costs of the involved bridges. The estimation of failure cost using instantaneous and accrued costs can be seen in \citep{andriotis2021deep}.

\begin{table}[]
\centering
\begin{threeparttable} 
\caption[]{ \em{Average Daily Vehicular Miles Traveled (DVMT) of the modeled Hampton Roads\\ network \cite{trafficVDOT2019}.}}
\label{tab:DVMT}
\begin{tabular}{cccc}
\hline
Pavements            & Total miles & \begin{tabular}[c]{@{}c@{}}DVMT milage in 2019 \\ (million miles)\end{tabular} & \begin{tabular}[c]{@{}c@{}}Truck traffic\\ (\%)\end{tabular} \\ \hline
Interstate (type I)  & 68.9        & 6.040                                                                          & 5.52                                                         \\
Primary (type II)    & 267.67      & 7.560                                                                          & 10.96                                                        \\
Secondary (type III) & 145         & 0.486                                                                          & 1.85                                                         \\ \hline
\end{tabular}
\end{threeparttable}
\end{table}

\subsection{Delay costs}\label{sec: delay_costs}
\noindent
For transportation networks, maintenance and rehabilitation activities have associated user delay costs due to work zone-related restrictions (like lane closures, speed reductions, and others), which result in increased travel time. FHWA defines the user delay cost as the additional costs borne by commuters and the community at large due to work zone activity. User delay cost is an important ingredient of the life-cycle cost analysis of pavements and bridges, which also incorporates agency costs such as inspection and maintenance costs and other user-related costs. This study estimates user delay cost by first calculating the additional user time to commute due to the work zones. The model developed in \citep{vadakpat2000road} is used for estimating the additional travel time during the maintenance period as:
\begin{equation}
T_d = exp{\left(1.67+0.0129(Trucks) + 0.000003(X-X')^2+0.00395(X-X')+0.602(Lane)\right)} 
\label{eq:delay}
\end{equation}
\begin{equation*}
 X'  = (X+48,600)/5
\label{eq:delay2}
\end{equation*}
where $T_d$ is the additional travel time (in vehicle-hours) incurred by all vehicles as a result of the work zone, $Trucks$ is a truck percentage in the vehicle stream, $X$ is the volume for which the delay is computed (as vehicles per hour), and $Lane$ is an indicator of which lane is blocked (1 if the right lane is closed, 0 if the left lane is closed).

To calculate the delay cost for the network using this model requires traffic distribution and USD costs/hr value for different vehicles. The traffic distribution is calculated based on the average Daily Vehicular Mile Traveled (DVMT) for the Hampton roads as provided in \cite{trafficVDOT2019} for the year 2019. The data for 2020 onwards is not used due to the pandemic-based traffic disruption. Table \ref{tab:DVMT} summarizes the traffic distribution for interstate, primary, and secondary pavements, however, explicit bridge data are not present in \citep{trafficVDOT2019}. We thus extrapolated traffic based on the total DVMT and the number of bridge miles present in the network. The truck percentages are assigned to type I-III bridges, same as for interstate-secondary ones. Further, to use Eq. \eqref{eq:delay}, the daily vehicular traffic volume is converted to hourly volume per lane-mile for the individual components. A pair of lanes are considered at a time to calculate the additional delay time, i.e., (a) $T_{d, R}$ when the right lane is closed, (b) $T_{d, L}$ when the left lane is closed. The following equation translates the obtained delay time to USD cost as:
\begin{equation}
 c_D = \left(T_{d,L}+T_{d,R} \right)T_m\left(c_{car}(1-Trucks)+c_{truck}Trucks\right)n_pL
\label{eq:delay_cost}
\end{equation}
where $c_D$ is the delay cost in USD, $T_m$ is the maintenance duration in hours, $L$ is the component length in miles, $n_p$ is the number of lane pairs, for example, a 4-lane component would have two-lane pairs, $c_{car} = \$ 21.89$ is an average hourly wage for passenger cars, and $c_{truck} = \$29.65$ is an average hourly wage for trucks. These hourly wage rates depend upon the consumer price index and other macroeconomic indicators for particular year; thus, the values above are based on the 2019 data reported by FHWA \cite{fhwaUserCost}.

\subsection{Network constraints}
\label{sec:cons}
\noindent
In the context of transportation asset management, performance measurement is the process of estimating progress toward preset goals or objectives using evidence such as aggregate measurements of conditions, smoothness of surfaces, and safety, as well as public perception (e.g., public surveys). To offer the best possible transportation service within the constraints of available resources, federal and state agencies have established goals for transportation asset performance with a clear definition of performance measures. For instance, under the FHWA objectives for bridges, 23 CFR (Code of Federal Regulations) 490.411(a) \cite{FHWAbridge} states that no more than 10\% of the National Highway System's (NHS) bridges by total deck area should be in poor condition (i.e., structurally deficient with a rating of $\leq4$). Similarly, 23 CFR part 490.315(a) subpart (c) \cite{FHWApavement} mandates that no more than 5\% of the state's NHS interstate lane miles be in poor condition. Additionally, the FHWA requires biannual inspections of bridges to have up-to-date bridge condition states, however, in this analysis, we are not adding this restriction, to solve the more challenging problem of simultaneously obtaining optimal inspection and maintenance policies.
  
Similarly, state DOTs impose performance targets on pavements in addition to the FHWA ones, such as the VDOT goal of no more than 18\% of interstate and primary highways and 35\% of secondary roadways being designated as deficient (i.e., CCI$<$60). In terms of serviceability, no more than 15\% of interstate and primary highways should be classified as poor in ride quality (i.e., IRI$>$2.2 m/km). VDOT wants also to reach a CCI of no less than 35 for the interstate system, \cite{VDOT2018pav, vdot2019a}. Apart from performance targets, there are also resource limits, such as the available budget, which cannot be exceeded. Hampton Roads districts have been granted a five-year budget of $\sim$ 1.3 billion for 2021-2026 \cite{hampton2020}.
Summarizing, we are considering six performance constraints and one budget constraint, as follows:
\begin{enumerate}

\item For NHS bridges, poor condition decks by area should be less than 10\% of total deck area. 
\item For NHS pavements, no more than 5\% of the interstate lane-miles should be classified deficient (i.e., CCI<60 and IRI >2.20 m/km).
\item No more than 18\% of interstate and primary roadway pavement conditions be classified as deficient (i.e., CCI<60).
\item No more than 15\% of interstate and primary roadways be classified as deficient in terms of ride quality (i.e., IRI>2.2 m/km).
\item No more than 35\% of secondary roadways pavement condition be classified as deficient (i.e., CCI<60).
\item No more than 2\% of interstate lane miles have CCI less than 37.
\item Hampton roads available budget for FY2021-2026 is ~\$ 1.3 billion, renewed every five years, discounted for every five year-cycle $\gamma$.
\end{enumerate}

All six performance-related targets imposed by FHWA and VDOT are integrated as soft constraints in our DRL framework, for the studied Hampton Roads network, i.e., these performance goals will be met in an average sense, during the life of the considered network system, as mentioned in previous sections. Since it is infeasible to satisfy stochastically the VDOT performance goal of no pavement with CCI less than 35, this is changed from 0 to 2\% in this work. Lastly, the available budget is considered a hard constraint and implemented as an augmented, fully observable state of the network. Additionally, the budget limit for each cycle is dynamically adjusted to account for the discounting of future values.

\subsection{Condition-based maintenance and VDOT policy baselines}
\label{CBM-VDOT}
\noindent
A Condition-Based Maintenance (CBM) policy is designed and optimized for performance comparisons in order to serve as a baseline against the obtained DRL solution. The most cost-effective policy is sought, to find the relevant thresholds that activate inspection and maintenance actions based on the condition of each component type, i.e., bridge, interstate, primary, and secondary pavements. The derived optimized policy involves the full suite of actions at every even time step, however, at every odd year, actions do-nothing and high-fidelity-inspection are deemed optimal. The CBM rules are given in Algorithm~\ref{algo:2} and \ref{algo:3} for intact and non-intact starting condition states, respectively, as explained subsequently. These condition-based rules are localized with respect to the individual components, and cannot achieve a global optimum for the entire network. However, these rules are chosen such that they can satisfy all the system constraints imposed in the previous section as tightly as possible.

\begin{figure}[]
\centering
\includegraphics[width=0.8\textwidth]{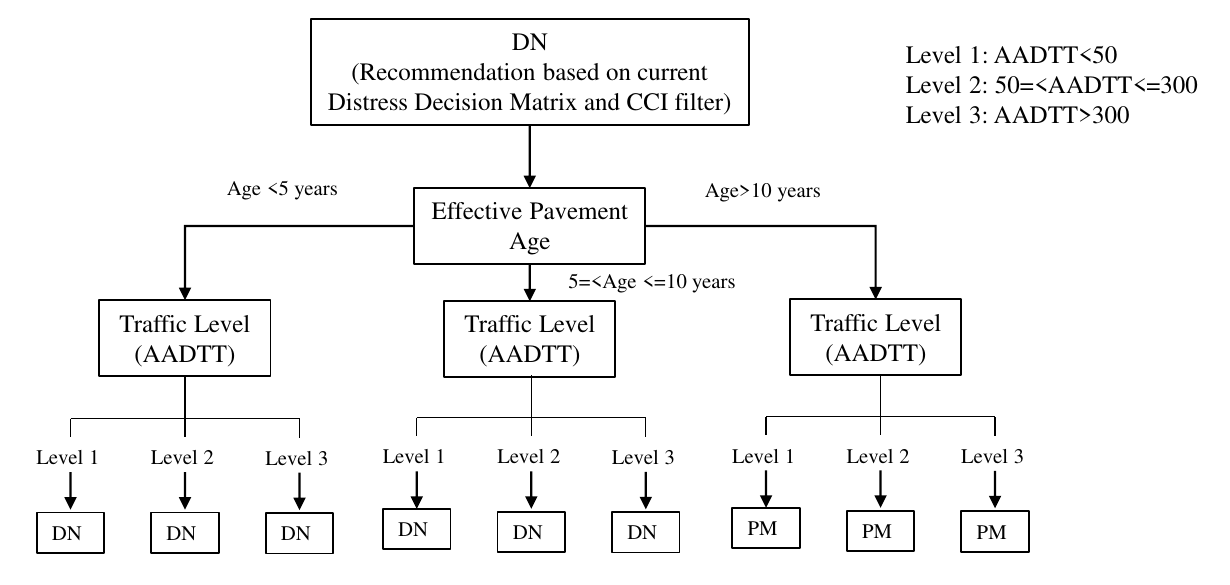}
\caption[]{Augmented decision tree for primary highways with Do Nothing (DN) as a stage one recommended action, along with trigger values based on effective age and Annual Average Daily Truck Traffic (AADTT).}
\label{fig:VDOT_pol}
\end{figure}

Another baseline largely based on the VDOT policy \cite{v2016a} is also added for comparison. VDOT policy assesses 100\% of the interstate and primary systems and 20\% of the secondary systems each year. The policy is divided into two stages. In the first stage, after the assessment of a component, depending upon the current distress decision matrix and CCI filter, an action is recommended \cite{v2016a}. In this study, due to the unavailability of a distress decision matrix, this first stage recommendation is solely based on the CCI filter, as reported in Section \ref{subsec:TP}. In the second stage, the recommended maintenance action may be updated based on the effective pavement age, Average Annual Daily Truck Traffic (AADTT), and Falling Weight Deflectometer (FWD). Different highway types have different decision trees with different sets of criteria and trigger values. For example, interstate highways utilize all three criteria (age, AADTT, and FWD, however due to the lack of data on FWD we are only using age and AADTT in this work), primary highways consider age and AADTT in their decision trees, and secondary highways decision trees are only based on AADT values. A second stage decision tree is shown in Fig.~\ref{fig:VDOT_pol} as an example case, with DN as a first-stage recommended action for primary highways \cite{v2016a}. Fig.~(\ref{fig:VDOT_pol}) shows the decision tree with starting node DN splitting to three different age groups, and the final decision taken based on the different AADTT levels. It may be worth reminding that our action categories {\it Do-Nothing, Minor, Major, Reconstruction} correspond to the VDOT defined actions DN, PM, CM/RM, and RC, respectively, as mentioned in Section \ref{mainte}.  Due to the introduced risk metrics for bridges in this work, we have again designed and optimized an annual CBM policy for bridge decks, to be used now in combination with the VDOT based rules for pavements. The derived most cost-effective policy rules in this case are given in Table \ref{tab:VDOT_brdg} in the Appendix, for both intact and non-intact starting condition states. Due to budget depletion in this VDOT-based policy case, for the non-intact starting conditions, a prioritization  scheme is also applied in this case, at applicable time steps, as also described in the Appendix Table \ref{tab:VDOT_brdg}.

\begin{figure}[t]
\centering
    \begin{subfigure}[b]{0.45\textwidth}
        \centering
   \includegraphics[width=\textwidth]{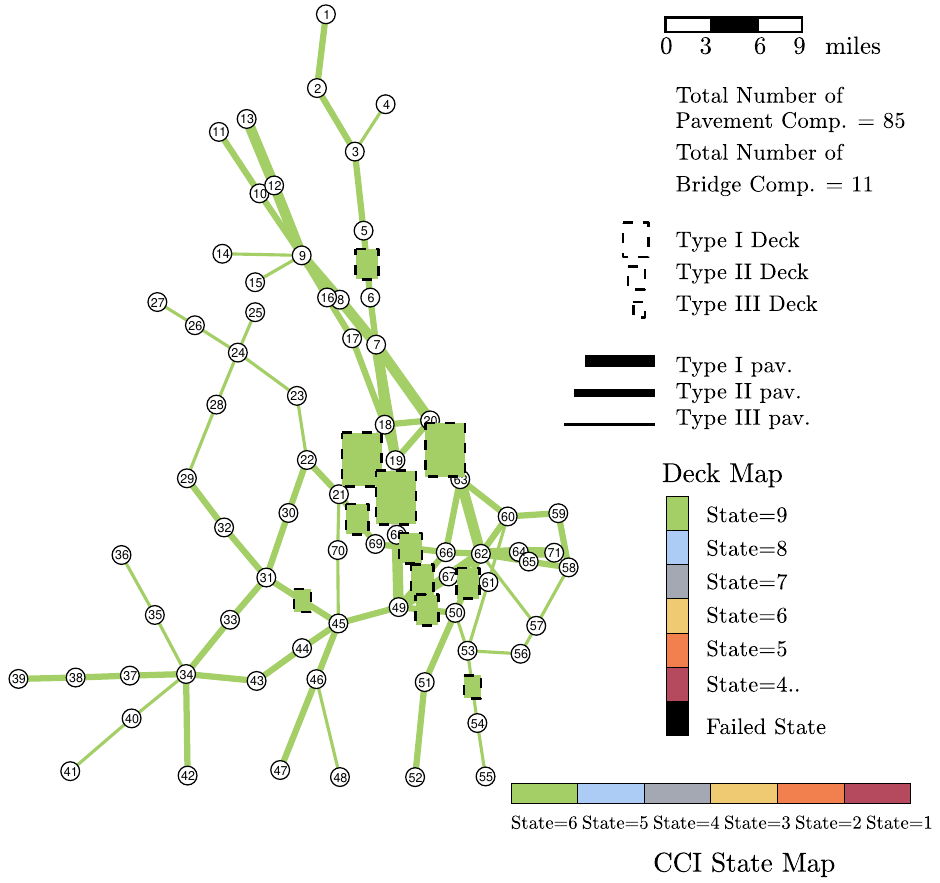}
        \caption{}
    \end{subfigure}%
    ~ 
    \begin{subfigure}[b]{0.45\textwidth}
        \centering
   \includegraphics[width=\textwidth]{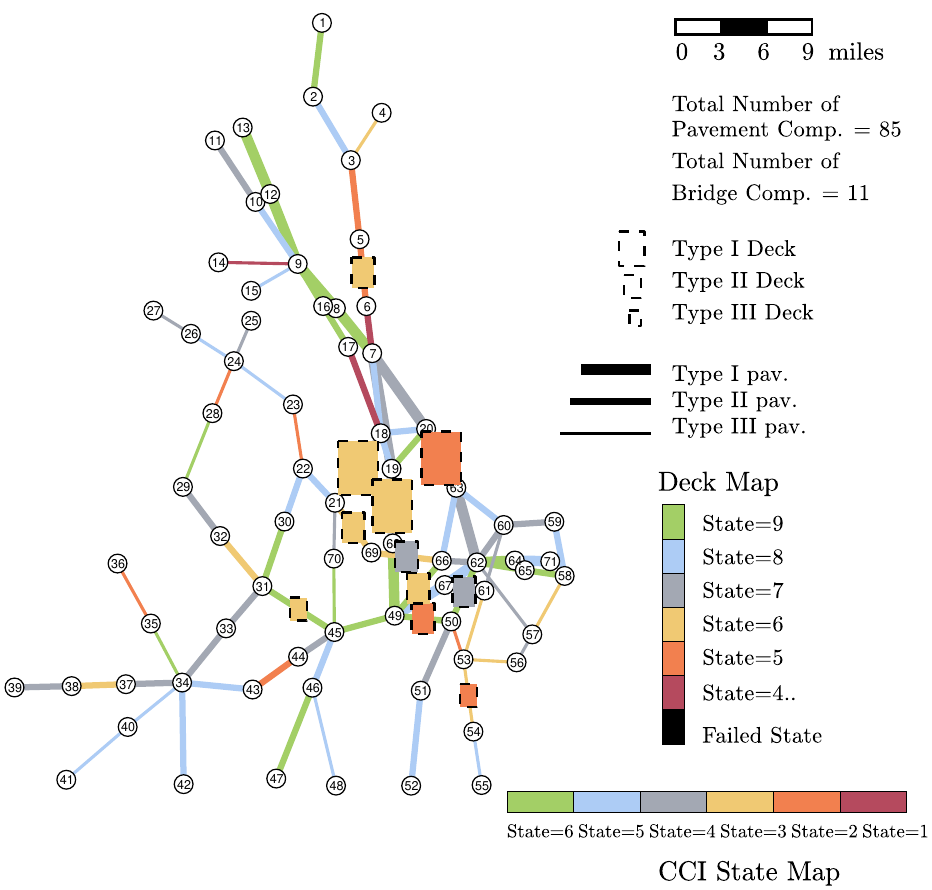}
        \caption{}
    \end{subfigure}
   
\caption[]{Discretized network with 96 components consisting of 11 bridges and 85 highway sections starting with initial conditions corresponding to (a) intact state and (b) non-intact actual state as of 2021.}
\label{fig:strat_networks}
\end{figure}

\begin{figure}[t]
\centering
\begin{subfigure}{.4\textwidth}
    \centering
    \includegraphics[width=.95\linewidth]{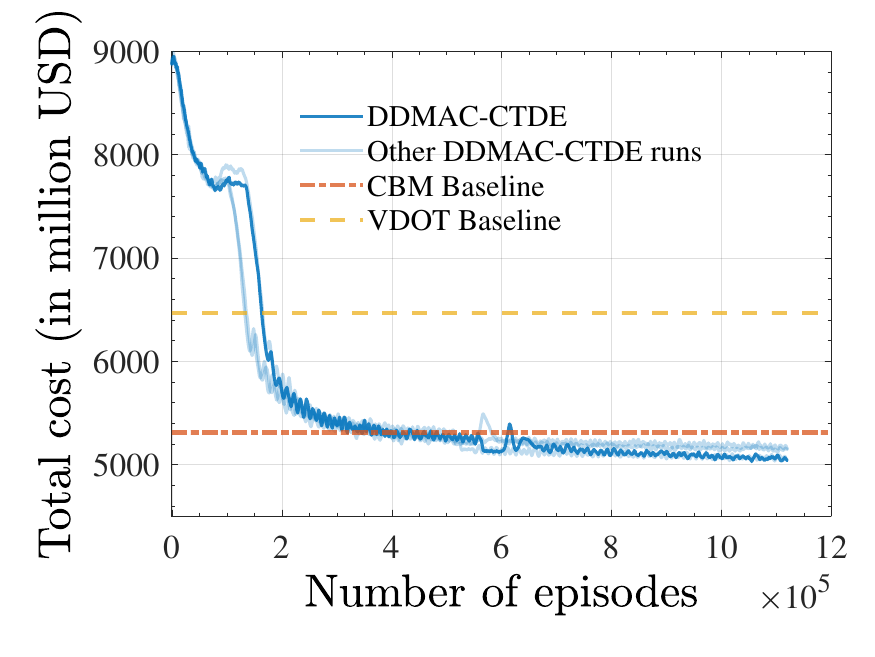}  
    \caption{}
    \label{train_cost_int}
\end{subfigure}
\begin{subfigure}{.4\textwidth}
    \centering
    \includegraphics[width=.95\linewidth]{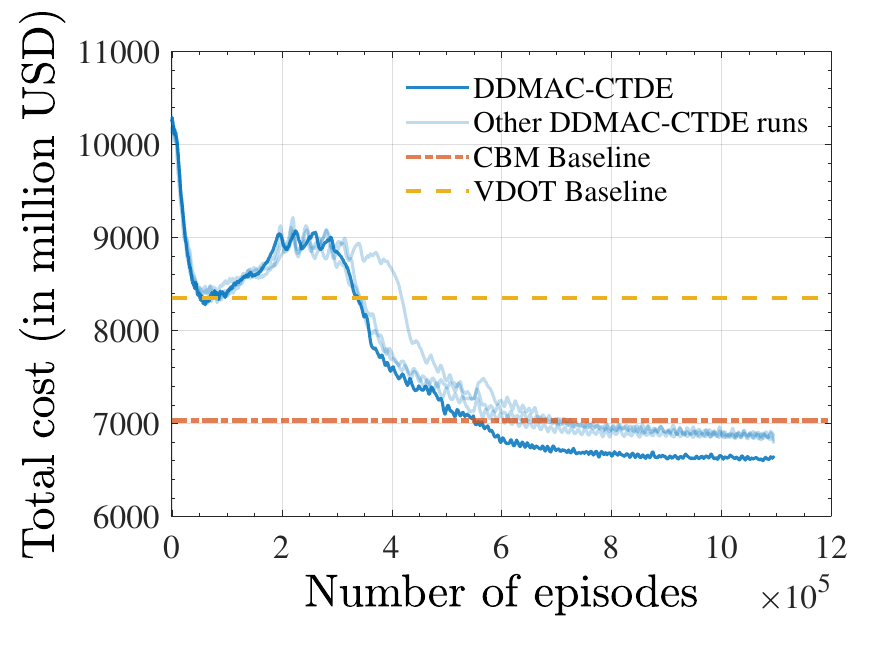}  
    \caption{}
    \label{train_cost_curr}
\end{subfigure}
\begin{subfigure}{.4\textwidth}
    \centering
    \includegraphics[width=.95\linewidth]{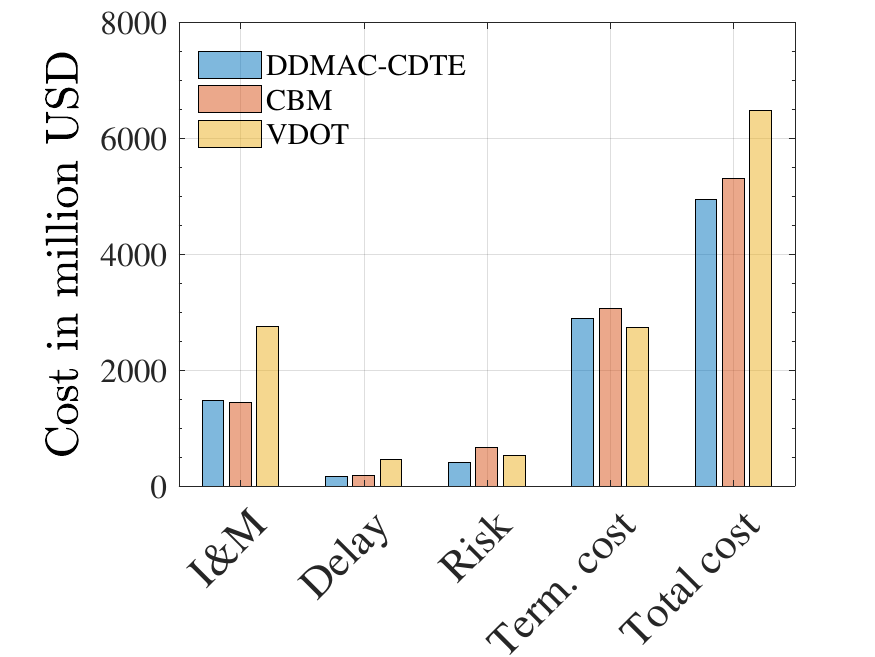}  
    \caption{}
    \label{bar_plot_int}
\end{subfigure}
\begin{subfigure}{.4\textwidth}
    \centering
    \includegraphics[width=.95\linewidth]{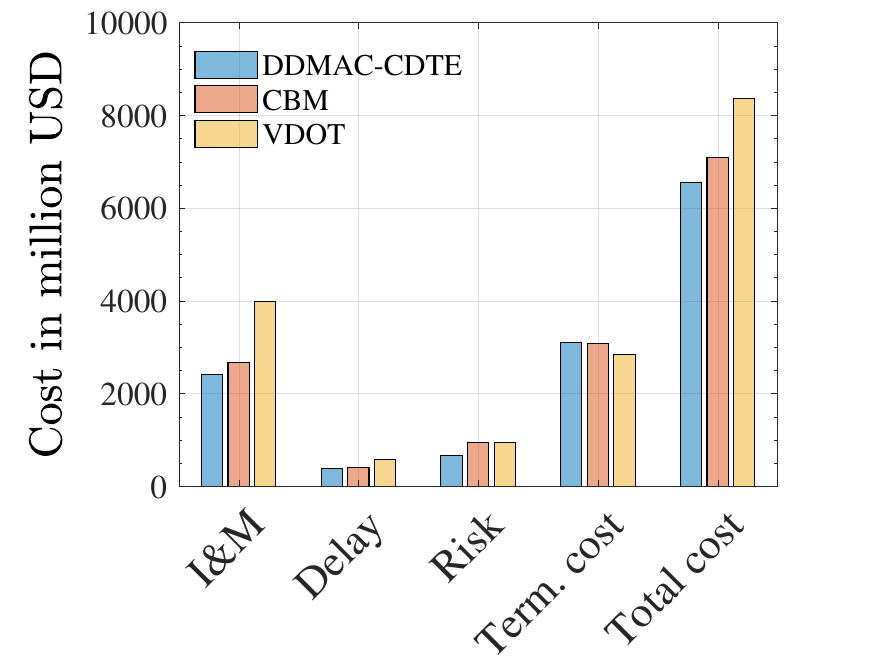}  
    \caption{}
    \label{bar_plot_curr}
\end{subfigure}
\caption{Total life cycle costs (excluding terminal cost) comparison of DDMAC-CTDE solution with CBM and VDOT policy baselines (top) for the network with  (a) intact starting state; (b) non-intact starting state. Comparison of the total cost and its constituents with CBM and VDOT policy baselines (bottom) for the network with (c) intact starting state; (d) non-intact starting state.}
\label{fig:3.8}
\end{figure}

\section{Results}
\noindent
This study considers a network from Hampton Roads, Virginia, with 96 components having a total number of $\sim 2\times 10^{248}$ possible system states at any given time instant. 10 actions per component are considered with action labels 0-9, where \{0,1,2\} represents {\em{Do Nothing, Minor}}, and {\em{Major Repair}} with no-inspection, respectively, \{3,4,5\} action codes are the same maintenance actions with low-fidelity inspection, and action codes \{6,7,8\} represent the three respective maintenance actions combined with high-fidelity inspection, and, finally, action code \{9\} indicates {\em {Reconstruction}}. The total number of available actions are $10^{96}$ for the entire network at each step. Out of 96 components, the network contains 85 pavement sections, and 11 important bridges with different sizes shown in Table \ref{tab:Deck_curr}. The pavements are discretized with lengths ranging from 2.7-8.9 miles and a constant lane-width of 12 feet, with 12 interstate highways, 47 primary highways, and 26 secondary highway components. At $t = 0$, the network starts from (a) intact state and (b) non-intact condition state (as of 2021; details can be found in Appendix \ref{sec:current_cond}), with an episode length of 20 years, and discount factors $\gamma = \gamma_h = 0.97$ and $\gamma_s = 1$. The two corresponding network cases are shown in Fig. \ref{fig:strat_networks}. These networks are trained using the budget and the six performance constraints mentioned in Section \ref{sec:cons}. A terminal cost, dependent on the component’s state, is added to the total cost to incentivize the agent to take maintenance actions even at the end of the policy horizon, adhering to engineering safety/serviceability requirements and assisting in avoiding  constraint violations at the end of the considered horizon. The method of calculating this terminal cost is included in Appendix \ref{sec:term_cost}. The training of the network is performed based on the details provided in the following subsection.

\subsection{Experimental details}
\noindent
For the purposes of this numerical investigation here, the actor and critic networks consist of fully connected layers. Input to pavement actors. Input to the critic network are the beliefs vectors of each component concatenated linearly, an fully observable extra state vector for pavement components, scalar budget, and current time step. Each actor network (one per component) has two hidden layers with 200 neurons each, utilizing ReLU activation, and an output layer with 10 neurons using a {\em{SoftMax}} activation function. The centralized critic network features two hidden layers with 500 neurons each, also employing ReLU activation, and an output layer with 1 neuron using a $Linear$ activation function. The training involves $11 \times 10^5$ episodes, each lasting 20 time steps, with a buffer memory size of 15,000 episodes. Exploration noise varies from 1.00 to 0.05 over 25,000 episodes, stabilizing at 0.02 after 100,000 episodes. Mini-batches of size 32 time steps are used, and network parameters are updated after each episode. The Adam optimizer is applied to both networks, with actor learning rate 0.0001, critic learning rate 0.001, and reducing the learning rate by half every 100,000 episodes until it stabilizes at 500,000 episodes. The loss function for the actor networks is the categorical cross-entropy, while the critic network uses a mean squared error function. The training is performed on a CPU with 96 cores and 364 GB of RAM. 

\subsection{DDMAC-CTDE and baseline solutions}
\noindent
Training for the DDMAC-CTDE network for $11\times 10^5$ episodes took 8-9 days with our available computational resources, while for the computed optimized baselines took 3-4 days for different instances. The original DDMAC approach could not be evaluated in this benchmark setting, as it could not converge reliably due to the dimensionality of the problem.  Figs.~(\ref{train_cost_int} \& \ref{train_cost_curr}) show a few DDMAC-CTDE training curves (solid lines) for this network problem, starting with intact and non-intact states, respectively, along with the CBM policy (orange dashed dot line) discussed in Section \ref{CBM} in the Appendix, and the VDOT policy (yellow dashed line) as explained before and largely specified in \cite{v2016a}. As seen in these figures, the DDMAC-CTDE solution surpasses the baselines in both cases with different starting states. 

Figs.~(\ref{bar_plot_int} \& \ref{bar_plot_curr}) give the histogram comparison of converged costs. The different costs such as inspection and maintenance, user delay, life-cycle risk, and terminal cost are different constituents of the total cost. These plots show that the DDMAC-CTDE has the lowest total cost metrics. The summarized results and comparisons are further shown in Table \ref{tab:Results}. The reported results are average values from $10^4$ simulated policies using the trained networks and optimized baselines. The solid dark blue instance in Figs.~(\ref{train_cost_int} and \ref{train_cost_curr}) represents the used trained network for the mentioned simulations.  Table~\ref{tab:Results} presents the objectives and six performance constraints in comparison with CBM and VDOT policies. DDMAC-CTDE demonstrates a substantial advantage, with an estimated total life-cycle cost of 4.94 billion USD for the intact starting state, as opposed to 5.31 billion USD for the   CBM policy (7.5\% more expensive) and 6.47 billion USD for the VDOT policy (31\% more expensive). In the case of the non-intact starting state, DDMAC-CTDE incurs a cost of 6.55 billion USD, while CBM and VDOT policies result in 7.3\% and 27\% higher life-cycle costs, respectively, compared to DDMAC-CTDE.
Excluding the terminal cost, the proposed method achieves even more compelling cost savings. Specifically, for the intact case, CBM and VDOT policies are approximately 10\% and 83\% more expensive, respectively, compared to DDMAC-CTDE. Similarly, for the non-intact case, CBM and VDOT policies are approximately 15\% and 59\% more expensive, respectively, than the DDMAC-CTDE policy. Additionally, DDMAC-CTDE results in terminal costs comparable to CBM- and VDOT-policies, without adhering to a conservative maintenance approach. Furthermore, it successfully meets the posed targets and constraints, overall.

\begin{sidewaystable}[hp]
\centering
\begin{threeparttable} 
\caption{\em{Comparison of objectives and imposed constraints using different policies for intact and non-intact starting condition states, where I-, P-, and S-Hwy are interstate, primary and secondary highways.}}
\label{tab:Results}
\begin{tabular}{ccccccc}
\hline
\multirow{3}{*}{\centering \begin{tabular}[c]{@{}c@{}}Objectives\\ \& constraints\end{tabular}}                   & \multicolumn{3}{c}{Intact starting state}                                                                                                                                & \multicolumn{3}{c}{Non-intact starting state}                                                                                                                           \\ \cline{2-7} 
                                                                                                       & \begin{tabular}[c]{@{}c@{}}DDMAC\\ -CTDE\end{tabular} & \begin{tabular}[c]{@{}c@{}}CBM\\ policy\end{tabular}    & \begin{tabular}[c]{@{}c@{}}VDOT\\ policy\end{tabular}  & \begin{tabular}[c]{@{}c@{}}DDMAC\\ -CTDE\end{tabular} & \begin{tabular}[c]{@{}c@{}}CBM\\ policy\end{tabular}   & \begin{tabular}[c]{@{}c@{}}VDOT\\ policy\end{tabular}  \\ \hline
\begin{tabular}[c]{@{}c@{}}Total cost incurred\\ (billion USD)\end{tabular}                            & \begin{tabular}[c]{@{}c@{}}4.94 \\ ($\pm$ $1.5x10^{-3}$)                                              \end{tabular}             & \begin{tabular}[c]{@{}c@{}}5.31(+{\textbf{7.5\%}})\\ ($\pm$ $3.4x10^{-3}$)                                             \end{tabular}  &\begin{tabular}[c]{@{}c@{}}6.47(+{\textbf{31\%}}) \\ ($\pm$ $4.7x10^{-3}$)\end{tabular}                                                    & \begin{tabular}[c]{@{}c@{}}6.55 \\ ($\pm$ $2.9x10^{-3}$)\end{tabular}                                                 & \begin{tabular}[c]{@{}c@{}}7.03(+{\textbf{7.3\%}}) \\ ($\pm$ $4.4x10^{-3}$)\end{tabular}                                                  & \begin{tabular}[c]{@{}c@{}}8.35(+{\textbf{27\%}}) \\ ($\pm$ $7.6x10^{-3}$)\end{tabular}                                                     \\
\hline
\begin{tabular}[c]{@{}c@{}}Total cost incurred w/o\\ terminal cost (billion USD)\end{tabular}          & \begin{tabular}[c]{@{}c@{}}2.04 \\ ($\pm$ $1.2x10^{-3}$)\end{tabular}                                                                                                    & \begin{tabular}[c]{@{}c@{}}2.25(+{\textbf{10\%}}) \\ ($\pm$ $3.4x10^{-3}$)\end{tabular} & \begin{tabular}[c]{@{}c@{}}3.74(+{\textbf{83\%}})\\ ($\pm$ $4.5x10^{-3}$)\end{tabular} & \begin{tabular}[c]{@{}c@{}}3.45 \\ ($\pm$ $2.7x10^{-3}$)\end{tabular}  & \begin{tabular}[c]{@{}c@{}}3.97(+{\textbf{15\%}}) \\ ($\pm$ $4.3x10^{-3}$)\end{tabular} & \begin{tabular}[c]{@{}c@{}}5.50(+{\textbf{59\%}}) \\ ($\pm$ $7.4x10^{-3}$)\end{tabular} \\
\hline
\begin{tabular}[c]{@{}c@{}}CCI\textless{}60 and IRI\textgreater{}2.2m/km\\ for I-Hwy (\%, cap 5\%)\end{tabular} & 2.94($\pm$ 0.03)                                                  & 2.82($\pm$ 0.04)    & 0.00($\pm$ 0.00)                                                 &  1.34($\pm$ 0.02)                                                 & 1.61($\pm$ 0.02)                                                   & 3.39($\pm$ 0.07)                                                    \\
\hline
CCI\textless{}35 of I-Hwy (\%, cap 2\%)                                                                         & 2.68($\pm$ 0.02)                                                  & 0.70($\pm$ 0.01)  & 0.02($\pm$ 0.00)                                                    & 1.50($\pm$ 0.02)                                                  & 0.68($\pm$ 0.01)                                                  & 3.80($\pm$ 0.07)                                                    \\
\hline
\begin{tabular}[c]{@{}c@{}}CCI\textless{}60 for I and\\ P-Hwy (\%, cap 18\%)\end{tabular}                        & 5.27($\pm$ 0.02)                                                  & 4.41($\pm$ 0.02)  & 0.05($\pm$ 0.00)                                                    & 17.65($\pm$ 0.04)                                                & 12.67($\pm$ 0.02)                                                                                                  & 5.80($\pm$ 0.04)                                                  \\
\hline
\begin{tabular}[c]{@{}c@{}}IRI\textgreater{}2.2 m/km for I\\ and P-Hwy (\%, cap 15\%)\end{tabular}               & 14.76($\pm$ 0.04)                                                 & 12.88($\pm$ 0.05) & 11.23($\pm$ 0.04)                                                  & 15.40($\pm$ 0.04)                                                    & 11.49($\pm$ 0.04)                                                                                                 & 15.65($\pm$ 0.07)                                                  \\
\hline
\begin{tabular}[c]{@{}c@{}}CCI\textless{}60 for S-Hwy\\(\%, cap 35\%)\end{tabular}                            & 10.13($\pm$ 0.04)                                                & 7.79($\pm$ 0.03)  & 0.72($\pm$ 0.01)                                                   & 33.00($\pm$ 0.08)                                                   & 28.18($\pm$ 0.04)                                                 & 37.86($\pm$ 0.13)  
\\
\hline
\begin{tabular}[c]{@{}c@{}}Decks with condition\\ rating $\leq$4 (\%, cap 10\%)\end{tabular}                          & 0.95($\pm$ 0.02)                                                   & 4.76($\pm$ 0.05) & 2.22($\pm$ 0.04)                                                   & 8.33($\pm$ 0.05)                                                    & 8.79($\pm$ 0.06)                                                  & 15.85($\pm$ 0.15)                                                  \\ \hline
\end{tabular}
\begin{tablenotes}[flushleft]\footnotesize 
\item {\emph{Note:}} {The estimated 95\% confidence bounds, assuming Gaussian distributions, are provided in parentheses, calculated based on $10^4$ simulations. The bold values indicate the difference in costs for CBM and VDOT with respect to the DDMAC-CTDE solution.}
\end{tablenotes} 
\end{threeparttable}
\end{sidewaystable}

Table~\ref{tab:Results} also compares the average performance over $10^4$ simulations in terms of the six different constraints detailed in Section \ref{sec:cons}. The performance constraints are in the rows of the table, and I, P, and S Hwy are the abbreviations of interstate, primary, and secondary highways, respectively. As noticed in the table, the DDMAC-CTDE solutions satisfy almost all the constraints for both starting states, except for the soft constraints CCI$<$35 of I-Hwy (2.68\% surpassing the 2\% target), for intact state case, and IRI $>$ 2.2 m/km for I and P-Hwy (15.4\% exceeding the 15\% target) for non-intact state case. In comparison, the CBM based policy satisfies all the constraints for both intact and non-intact cases, while the VDOT-recommended policy satisfies all the constraints for the intact initial state scenario. For the non-intact case, the VDOT policy closely meets the constraints or exceeds them mildly in certain cases related to pavements, and in particular for the targets for CCI $<$ 35 for I-Hwy (3.80\% exceeding the 2\% target), IRI $>$ 2.2 m/km for I and P-Hwy (15.65\% exceeding the 15\% target) and S-Hwy CCI$<$60 (37.86\% surpassing the 35\% target). For the bridge decks target of condition ratings $\leq$4, the implemented VDOT policy for the non-intact starting state significantly surpasses the 10\% target by 5.85\%, mainly as a result of depletion of the budget, which was only observed in this case among all utilized policies and starting conditions.

To better understand how DDMAC-CTDE policies change over time, detailed policy realizations for some representative components are given in Figs.~\ref{fig:pol_real_int} and \ref{fig:pol_real_current} for intact and non-intact starting states. The figures illustrate actions generated by one of the instances of the optimum policy and the evolution of belief states as a contour. Figs.~\ref{fig:pol_real_int} and \ref{fig:pol_real_current} also depict the discounted budget utilization over time, as well as the discounted 5-year budget for each cycle. The budget is a hard constraint that the agents are not allowed to exceed, and this requirement is satisfied by the resulting solutions. Further, it can be noticed that the agents do not utilize the full budget in both cases, since they can optimally satisfy the performance constraints with minimum life-cycle cost. In comparison, the VDOT policy for the non-intact starting state exhausted the entire budget and did not satisfy various performance constraints. Additionally, the temporal evolution of six performance constraints is presented in Figs.~\ref{fig:pol_real_int} and \ref{fig:pol_real_current}, in relation to pavement and bridge condition states. It is seen that the six performance thresholds are very well respected for the most part, except at the end of the life cycle. However, for these indicative realizations, the performance constraints are never violated on average over the life-cycle. The time variation of total risk cost associated with individual bridges and the three modes of system risk is shown as well. Finally, a pie chart illustrating the cost distribution for that realization among various types of pavements and bridges is also presented.

\afterpage{%
\begin{sidewaysfigure}
\captionsetup{labelfont=bf}
    \centering
    \includegraphics[width=1.0\textwidth]{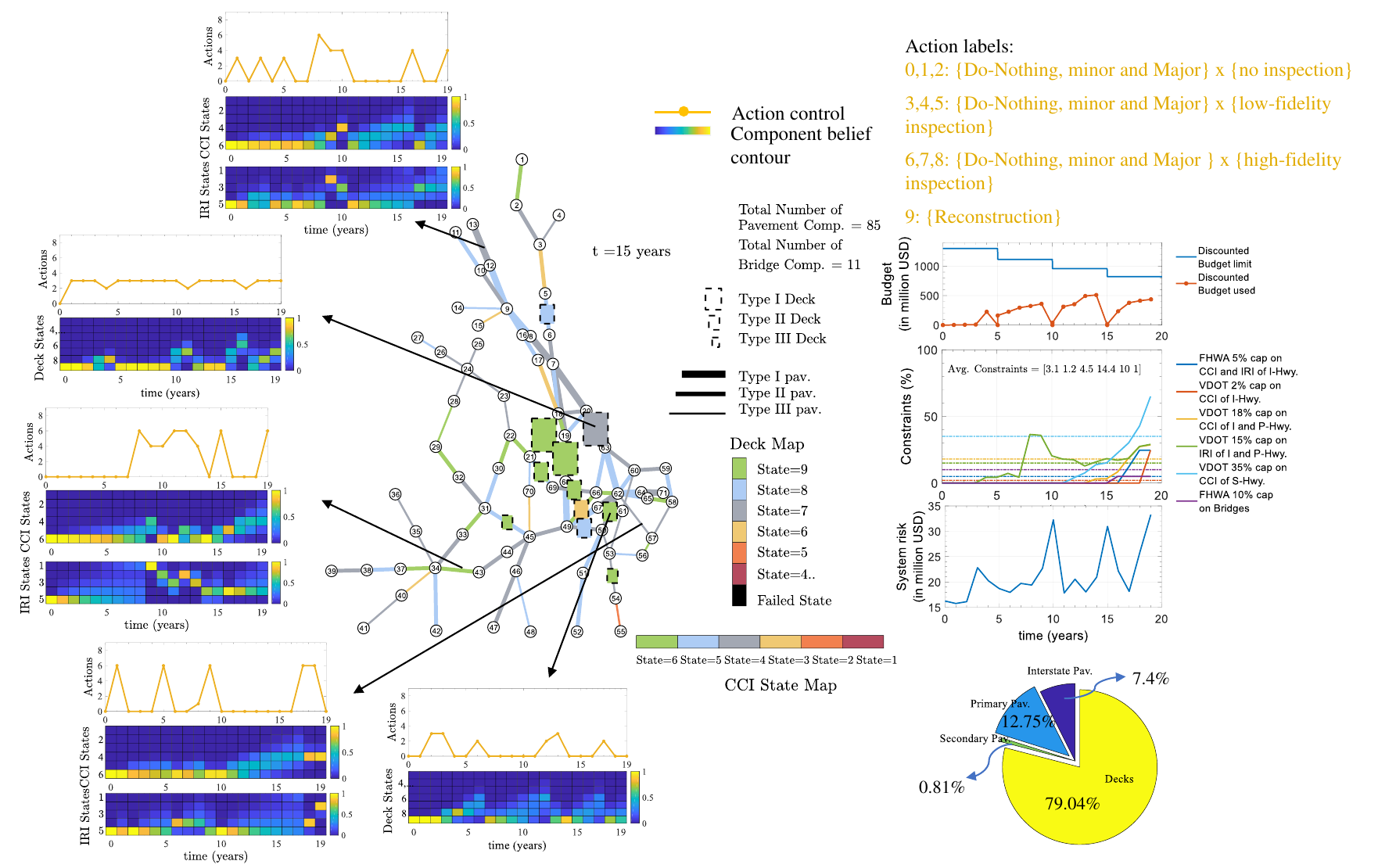}%
    \caption{Life-cycle realization of the learned DDMAC-CTDE policy for Hampton Roads network, starting from intact state.}
    \label{fig:pol_real_int}
\end{sidewaysfigure}
\clearpage
}

\afterpage{%
\begin{sidewaysfigure}
\captionsetup{labelfont=bf}
    \centering
    \includegraphics[width=1.0\textwidth]{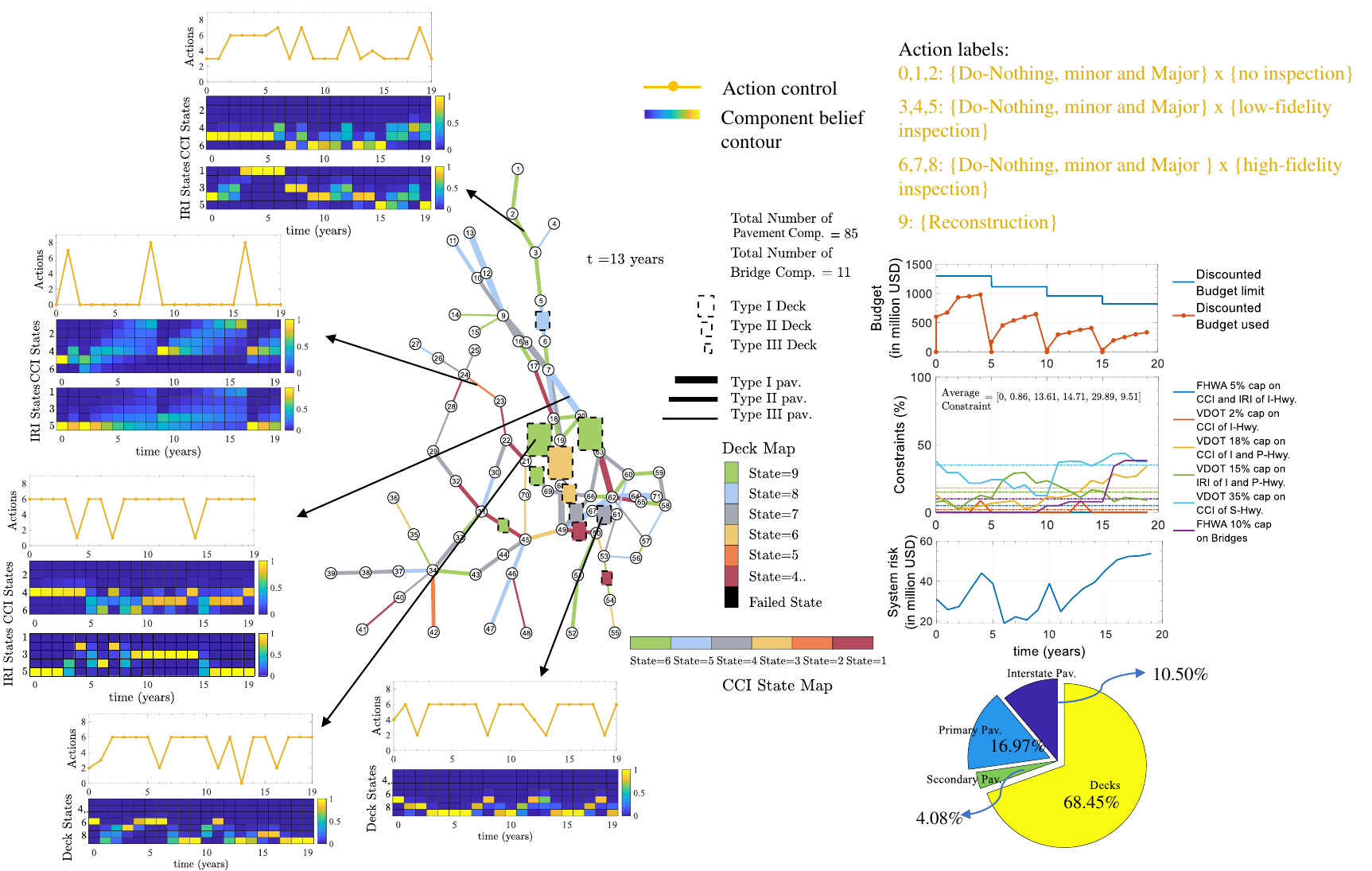}%
    \caption{Life-cycle realization of the learned DDMAC-CTDE policy for Hampton Roads network, starting from current state as of 2021.}
    \label{fig:pol_real_current}
\end{sidewaysfigure}
\clearpage
}

Plots with control actions represent the actions taken over time. The maintenance actions, taken at each time step, update the current belief of the system, as manifested in the next time step. The evolution of contour plots in the case of pavements shows current beliefs for both CCI and IRI states, and the current belief states at each step for two bridge decks are also shown. For example, in Fig.~\ref{fig:pol_real_int} the agent takes action 1 at $t = 8$ years for a secondary pavement, and the belief gets updated according to the {\em{Minor Repair}} action and the environment effect transitions. Accordingly in Fig.~\ref{fig:pol_real_current}, the agent is shown to take action 8 at $t = 8$ years for a secondary pavement, and then the updated beliefs of CCI and IRI are shown at $t = 9$ years, incorporating inspection outcome, maintenance actions, and environment transition effects. 

As seen in Figs.~\ref{fig:pol_real_int} and \ref{fig:pol_real_current}, control actions are compatible with belief states. For example, in Fig.~\ref{fig:pol_real_int} the agents initially choose {\em{Do-Nothing}} actions since the belief states for both pavements and bridges initiate in the intact condition. As the conditions gradually deteriorate, more interventions are considered. However, this is not the case with the non-intact starting state example, where interventions need to start from the beginning of planning horizon. Further, the inspection and maintenance actions at the end of the horizon are justified in both cases due to the state-associated terminal cost and the present risk for bridges.

It is also observed through the policy simulations that the agents maintain the type I bridges more systematically and consistently choose inspection actions for them to have better informed state estimates. This is because type I bridges have their individual failure risk, as well as mode 1 and mode 3 system failure risks associated with them. Similarly, importance is given to more essential pavement components. For example, in Fig.~\ref{fig:pol_real_current}, the maintenance and inspection actions are more frequent for interstate than primary components, as well as for primary pavements with respect to secondary ones.  

From the pie charts, shown in Figs.~\ref{fig:pol_real_int} and \ref{fig:pol_real_current}, it is observed that cost distributions are heavily skewed (as much as 79\% for intact start and 68.45\% for non-intact start) toward the bridge components, due to their high maintenance costs, associated risk costs, and lower traffic delay costs. The cost is even more skewed for intact versus non-intact start case due to the relatively slower initial deterioration rate of pavements in comparison to bridges in this work. Thus, the pavements with intact starting state conditions do not need significant maintenance activity until the later stages of their life-cycle. Among pavements, primary highways absorb the largest part of the budget, as they outnumber the other components in the network (47 in total). Figs.~\ref{fig:pol_real_int} and \ref{fig:pol_real_current} also show the evolution of the system risk with time. In both cases, the risk is minimal initially and increases with time, with downward jumps mainly due to the maintenance actions taken for bridges, especially of type I. 

\section{Conclusions}
\noindent
This study develops a new stochastic optimum control deep reinforcement learning (DRL) paradigm for inspecting and maintaining deteriorating transportation network systems under partial information and constraints. It establishes a comprehensive modeling framework for pavements and bridges using Partially Observable  Markov Decision Process (POMDP) principles, in the yet most complex and detailed application of DRL to this type of system optimization problems, to the best of our knowledge. A large transportation network with 96 components situated in the state of Virginia, USA, is studied, and due to an exhaustively intractable number of system state and action combinations, the problem is solved with a newly developed here DRL algorithmic approach. In particular, a Deep Decentralized Multi-agent Actor-Critic with Centralized Training and Decentralized Execution (DDMAC-CTDE) is presented, which uses sparse parameterizations and local component state information for actor networks to obtain near-optimal solutions. Two indices, the Critical Condition Index (CCI) and the International Roughness Index (IRI), are used to determine the pavement structural condition and serviceability, and the associated transition probabilities for both indices are determined. For bridges, the deck performance is utilized since it deteriorates at a faster rate and has a higher maintenance cost. Many operational, budget, and performance-based constraints are also imposed, by means of hard and soft constraints in the DRL formulation. The framework is successfully demonstrated to offer significantly better solutions than heuristically optimized Condition Based Maintenance (CBM) and VDOT recommended policy baselines, which are 7.5\% and 31\% more expensive, respectively, starting from an intact condition state. The proposed framework effectively also handles all performance targets and constraints imposed by FHWA and VDOT. In addition to the intact starting condition state, the Hampton Roads network's 2021 conditions are also considered as starting states. In this case, DDMAC-CTDE policies again outperform the considered baselines with respect to the total cost, with the CBM and VDOT policies  being 7.3\% and 27\% more expensive, respectively. Overall, this work importantly contributes toward DRL-based inspection and maintenance polices for transportation networks and related infrastructure systems. Future works can also additionally consider dynamic traffic effects and incorporation of traffic control actions in the entire management framework.  

\section*{Acknowledgements}
\noindent
The authors acknowledge the support of the U.S. National Science Foundation under CAREER Grant No. 1751941, and the Center for Integrated Asset Management for Multimodal Transportation Infrastructure Systems, U.S. DOT Region 3 University Center. Dr. Andriotis would further like to acknowledge the support of the TU Delft AI Labs program.




\begin{thebibliography}{100}

\bibitem{ASCE2025}
{American Society of Civil Engineers}, ``{2025 Infrastructure Report Card},'' tech. rep., {American Society of Civil Engineers}, 2025.
\newblock Last accessed February 23, 2026: \url{https://infrastructurereportcard.org/wp-content/uploads/2025/03/Full-Report-2025-Natl-IRC-WEB.pdf}.

\bibitem{vdot2021a}
VDOT, ``Preservation, maintenance, repair, widening and rehabilitation, chapter 32,'' tech. rep., Virginia Department of Transportation, 2022.
\newblock Last accessed February 23, 2026: \url{https://www.virginiadot.org/business/resources/bridge/manuals/part2/chapter32.pdf}.

\bibitem{bellman1957a}
R.~Bellman, {\em Dynamic programming}.
\newblock Princeton University Press, 1957.

\bibitem{pineau2003a}
J.~Pineau, G.~Gordon, and S.~Thrun, ``Point-based value iteration: An anytime algorithm for {POMDPs},'' in {\em IJCAI}, vol.~3, pp.~1025--1032, 2003.

\bibitem{bocchini2011a}
P.~Bocchini and D.~Frangopol, ``A probabilistic computational framework for bridge network optimal maintenance scheduling,'' {\em Reliability Engineering \& System Safety}, vol.~96, no.~2, p.~332–49, 2011.

\bibitem{sanchez2016maintenance}
M.~S{\'a}nchez-Silva, D.~M. Frangopol, J.~Padgett, and M.~Soliman, ``Maintenance and operation of infrastructure systems,'' {\em Journal of Structural Engineering}, vol.~142, no.~9, p.~F4016004, 2016.

\bibitem{AJRUA60001104}
J.~Mendoza, E.~Bismut, D.~Straub, and J.~Köhler, ``Risk-based fatigue design considering inspections and maintenance,'' {\em ASCE-ASME Journal of Risk and Uncertainty in Engineering Systems, Part A: Civil Engineering}, vol.~7, no.~1, p.~04020055, 2021.

\bibitem{nguyen2015multi}
K.-A. Nguyen, P.~Do, and A.~Grall, ``Multi-level predictive maintenance for multi-component systems,'' {\em Reliability engineering \& system safety}, vol.~144, pp.~83--94, 2015.

\bibitem{yang2019life}
D.~Y. Yang and D.~M. Frangopol, ``Life-cycle management of deteriorating civil infrastructure considering resilience to lifetime hazards: A general approach based on renewal-reward processes,'' {\em Reliability Engineering \& System Safety}, vol.~183, pp.~197--212, 2019.

\bibitem{bismut2021optimal}
E.~Bismut and D.~Straub, ``Optimal adaptive inspection and maintenance planning for deteriorating structural systems,'' {\em Reliability Engineering \& System Safety}, vol.~215, p.~107891, 2021.

\bibitem{yang2021risk}
D.~Y. Yang and D.~M. Frangopol, ``Risk-based inspection planning of deteriorating structures,'' {\em Structure and Infrastructure Engineering}, vol.~18, no.~1, pp.~109--128, 2021.

\bibitem{han2022risk}
{Han, Xu and Frangopol, Dan M}, ``Risk-informed bridge optimal maintenance strategy considering target service life and user cost at project and network levels,'' {\em ASCE-ASME Journal of Risk and Uncertainty in Engineering Systems, Part A: Civil Engineering}, vol.~8, no.~4, p.~04022050, 2022.

\bibitem{han2023life}
X.~Han and D.~M. Frangopol, ``Life-cycle risk-based optimal maintenance strategy for bridge networks subjected to corrosion and seismic hazards,'' {\em Journal of Bridge Engineering}, vol.~28, no.~1, p.~04022128, 2023.

\bibitem{saydam2015risk}
D.~Saydam and D.~M. Frangopol, ``Risk-based maintenance optimization of deteriorating bridges,'' {\em Journal of Structural Engineering}, vol.~141, no.~4, p.~04014120, 2015.

\bibitem{CHU20121123}
J.~C. Chu and Y.-J. Chen, ``Optimal threshold-based network-level transportation infrastructure life-cycle management with heterogeneous maintenance actions,'' {\em Transportation Research Part B: Methodological}, vol.~46, no.~9, pp.~1123--1143, 2012.

\bibitem{frangopol2007}
D.~M. Frangopol and M.~Liu, ``Maintenance and management of civil infrastructure based on condition, safety, optimization, and life-cycle cost,'' {\em Structure and Infrastructure Engineering}, vol.~3, no.~1, pp.~29--41, 2007.

\bibitem{azad2016disruption}
N.~Azad, E.~Hassini, and M.~Verma, ``Disruption risk management in railroad networks: An optimization-based methodology and a case study,'' {\em Transportation Research Part B: Methodological}, vol.~85, pp.~70--88, 2016.

\bibitem{petchrompo2019review}
S.~Petchrompo and A.~K. Parlikad, ``A review of asset management literature on multi-asset systems,'' {\em Reliability Engineering \& System Safety}, vol.~181, pp.~181--201, 2019.

\bibitem{frangopol2007bridge}
D.~M. Frangopol and M.~Liu, ``Bridge network maintenance optimization using stochastic dynamic programming,'' {\em Journal of Structural Engineering}, vol.~133, no.~12, pp.~1772--1782, 2007.

\bibitem{papakonstantinou2014corrod}
K.~G. Papakonstantinou and M.~Shinozuka, ``Optimum inspection and maintenance policies for corroded structures using partially observable markov decision processes and stochastic, physically based models,'' {\em Probabilistic Engineering Mechanics}, vol.~37, p.~93–108, 2014.

\bibitem{papakonstantinou2018a}
K.~G. Papakonstantinou, C.~P. Andriotis, and M.~Shinozuka, ``{POMDP} and {MOMDP} solutions for structural life-cycle cost minimization under partial and mixed observability,'' {\em Structure and Infrastructure Engineering}, vol.~14, no.~7, p.~869–882, 2018.

\bibitem{kaelbling1998a}
L.~Kaelbling, M.~Littman, and A.~Cassandra, ``Planning and acting in partially observable stochastic domains,'' {\em Artificial Intelligence}, vol.~101, no.~1, p.~99–134, 1998.

\bibitem{papakonstantinou2014a}
K.~G. Papakonstantinou and M.~Shinozuka, ``Planning structural inspection and maintenance policies via dynamic programming and {Markov} processes. {Part I: Theory.},'' {\em Reliability Engineering \& System Safety}, vol.~130, p.~202–213, 2014.

\bibitem{papakonstantinou2014partII}
K.~G. Papakonstantinou and M.~Shinozuka, ``Planning structural inspection and maintenance policies via dynamic programming and markov processes. {Part II: POMDP} implementation,'' {\em Reliability Engineering \& System Safety}, vol.~130, p.~214–224, 2014.

\bibitem{papakonstantinou2016point}
K.~G. Papakonstantinou, C.~P. Andriotis, and M.~Shinozuka, ``Point-based {POMDP} solvers for life-cycle cost minimization of deteriorating structures,'' in {\em Proceedings of the 5th International Symposium on Life-Cycle Civil Engineering (IALCCE 2016), {Delft, The Netherlands}}, p.~427, 2016.

\bibitem{memarzadeh2016aIntegrated}
M.~Memarzadeh and M.~Pozzi, ``Integrated inspection scheduling and maintenance planning for infrastructure systems,'' {\em Computer‐Aided Civil and Infrastructure Engineering}, vol.~31, no.~6, p.~403–415, 2016.

\bibitem{schoebi2016a}
R.~Schöbi and E.~N. Chatzi, ``Maintenance planning using continuous-state partially observable {Markov} decision processes and non-linear action models,'' {\em Structure and Infrastructure Engineering}, vol.~12, no.~8, p.~977–994, 2016.

\bibitem{memarzadeh2015optimal}
M.~Memarzadeh, M.~Pozzi, and J.~Zico~Kolter, ``Optimal planning and learning in uncertain environments for the management of wind farms,'' {\em Journal of Computing in Civil Engineering}, vol.~29, no.~5, p.~04014076, 2015.

\bibitem{MORATO2022102140}
P.~G. Morato, K.~G. Papakonstantinou, C.~P. Andriotis, J.~S. Nielsen, and P.~Rigo, ``Optimal inspection and maintenance planning for deteriorating structural components through dynamic {Bayesian} networks and {Markov} decision processes,'' {\em Structural Safety}, vol.~94, p.~102140, 2022.

\bibitem{arcieri2023bridging}
G.~Arcieri, C.~Hoelzl, O.~Schwery, D.~Straub, K.~G. Papakonstantinou, and E.~Chatzi, ``{Bridging {POMDP}s and {Bayesian} decision making for robust maintenance planning under model uncertainty: An application to railway systems},'' {\em Reliability Engineering \& System Safety}, vol.~239, p.~109496, 2023.

\bibitem{madanat1993a}
S.~Madanat, ``Optimal infrastructure management decisions under uncertainty,'' {\em Transportation Research Part C: Emerging Technologies}, vol.~1, no.~1, p.~77–88, 1993.

\bibitem{madanat1994a}
S.~Madanat and M.~Ben-Akiva, ``Optimal inspection and repair policies for infrastructure,'' {\em Transportation Science}, vol.~28, no.~1, p.~55–62, 1994.

\bibitem{guillaumot2003adaptive}
V.~M. Guillaumot, P.~L. Durango-Cohen, and S.~M. Madanat, ``Adaptive optimization of infrastructure maintenance and inspection decisions under performance model uncertainty,'' {\em Journal of Infrastructure Systems}, vol.~9, no.~4, pp.~133--139, 2003.

\bibitem{ZHANG2023104054}
L.~Zhang, W.~Gu, Y.-J. Byon, and J.~Lee, ``Condition-based pavement management systems accounting for model uncertainty and facility heterogeneity with belief updates,'' {\em Transportation Research Part C: Emerging Technologies}, vol.~148, p.~104054, 2023.

\bibitem{SHON2021103328}
H.~Shon and J.~Lee, ``Integrating multi-scale inspection, maintenance, rehabilitation, and reconstruction decisions into system-level pavement management systems,'' {\em Transportation Research Part C: Emerging Technologies}, vol.~131, p.~103328, 2021.

\bibitem{spaan2005a}
M.~Spaan and N.~Vlassis, ``Perseus: Randomized point-based value iteration for {POMDPs},'' {\em Journal of Artificial Intelligence Research}, vol.~24, p.~195–220, 2005.

\bibitem{shani2013a}
G.~Shani, J.~Pineau, and R.~Kaplow, ``A survey of point-based {POMDP} solvers,'' {\em Autonomous Agents and Multi-Agent Systems}, vol.~27, no.~1, p.~1–51, 2013.

\bibitem{lovejoy1991survey}
W.~S. Lovejoy, ``A survey of algorithmic methods for partially observed {Markov} decision processes,'' {\em Annals of Operations Research}, vol.~28, no.~1, pp.~47--65, 1991.

\bibitem{mnih2015human}
V.~Mnih, K.~Kavukcuoglu, D.~Silver, A.~A. Rusu, J.~Veness, M.~G. Bellemare, A.~Graves, M.~Riedmiller, A.~K. Fidjeland, G.~Ostrovski, {\em et~al.}, ``Human-level control through deep reinforcement learning,'' {\em Nature}, vol.~518, no.~7540, pp.~529--533, 2015.

\bibitem{schulman2017proximal}
J.~Schulman, F.~Wolski, P.~Dhariwal, A.~Radford, and O.~Klimov, ``Proximal policy optimization algorithms,'' {\em arXiv preprint arXiv:1707.06347}, 2017.

\bibitem{andriotis2019managing}
C.~P. Andriotis and K.~G. Papakonstantinou, ``Managing engineering systems with large state and action spaces through deep reinforcement learning,'' {\em Reliability Engineering \& System Safety}, vol.~191, p.~106483, 2019.

\bibitem{lei2022deep}
X.~Lei, Y.~Xia, L.~Deng, and L.~Sun, ``A deep reinforcement learning framework for life-cycle maintenance planning of regional deteriorating bridges using inspection data,'' {\em Structural and Multidisciplinary Optimization}, vol.~65, no.~5, p.~149, 2022.

\bibitem{DU2022102221}
A.~Du and A.~Ghavidel, ``Parameterized deep reinforcement learning-enabled maintenance decision-support and life-cycle risk assessment for highway bridge portfolios,'' {\em Structural Safety}, vol.~97, p.~102221, 2022.

\bibitem{HAMIDA2023109214}
Z.~Hamida and J.-A. Goulet, ``Hierarchical reinforcement learning for transportation infrastructure maintenance planning,'' {\em Reliability Engineering \& System Safety}, vol.~235, p.~109214, 2023.

\bibitem{MOHAMMADI2022108615}
R.~Mohammadi and Q.~He, ``A deep reinforcement learning approach for rail renewal and maintenance planning,'' {\em Reliability Engineering \& System Safety}, vol.~225, p.~108615, 2022.

\bibitem{arcieri2023pomdp}
G.~Arcieri, C.~Hoelzl, O.~Schwery, D.~Straub, K.~G. Papakonstantinou, and E.~Chatzi, ``Pomdp inference and robust solution via deep reinforcement learning: An application to railway optimal maintenance,'' {\em Machine Learning}, vol.~113, no.~10, pp.~7967--7995, 2024.

\bibitem{yang2022deep}
D.~Y. Yang, ``Deep reinforcement learning--enabled bridge management considering asset and network risks,'' {\em Journal of Infrastructure Systems}, vol.~28, no.~3, p.~04022023, 2022.

\bibitem{lei2023sustainable}
X.~Lei, Y.~Dong, and D.~M. Frangopol, ``Sustainable life-cycle maintenance policymaking for network-level deteriorating bridges with a convolutional autoencoder--structured reinforcement learning agent,'' {\em Journal of Bridge Engineering}, vol.~28, no.~9, p.~04023063, 2023.

\bibitem{Lai2024SynergeticDRL}
L.~Lai, Y.~Dong, C.~P. Andriotis, A.~Wang, and X.~Lei, ``Synergetic-informed deep reinforcement learning for sustainable management of transportation networks with large action spaces,'' {\em Automation in Construction}, vol.~160, p.~105302, 2024.

\bibitem{sresakoolchai2023railway}
J.~Sresakoolchai and S.~Kaewunruen, ``Railway infrastructure maintenance efficiency improvement using deep reinforcement learning integrated with digital twin based on track geometry and component defects,'' {\em Scientific Reports}, vol.~13, no.~1, p.~2439, 2023.

\bibitem{LEE2023108908}
J.~Lee and M.~Mitici, ``Deep reinforcement learning for predictive aircraft maintenance using probabilistic remaining-useful-life prognostics,'' {\em Reliability Engineering \& System Safety}, vol.~230, p.~108908, 2023.

\bibitem{ZHANG2023104063}
E.~Zhang, R.~Zhang, and N.~Masoud, ``Predictive trajectory planning for autonomous vehicles at intersections using reinforcement learning,'' {\em Transportation Research Part C: Emerging Technologies}, vol.~149, p.~104063, 2023.

\bibitem{YAZDANI2023103991}
M.~Yazdani, M.~Sarvi, S.~{Asadi Bagloee}, N.~Nassir, J.~Price, and H.~Parineh, ``Intelligent vehicle pedestrian light ({IVPL}): A deep reinforcement learning approach for traffic signal control,'' {\em Transportation Research Part C: Emerging Technologies}, vol.~149, p.~103991, 2023.

\bibitem{wu2024dynamic}
W.~Wu, Y.~Zhu, and R.~Liu, ``Dynamic scheduling of flexible bus services with hybrid requests and fairness: Heuristics-guided multi-agent reinforcement learning with imitation learning,'' {\em Transportation Research Part B: Methodological}, vol.~190, p.~103069, 2024.

\bibitem{fardyuan}
A.~Fard and A.~X.~X. Yuan, ``Multi-year maintenance planning for large-scale infrastructure systems: A novel network deep {Q}-learning approach,'' 2025.
\newblock arXiv preprint arXiv:2507.18732.

\bibitem{andriotis2019lifecycle}
C.~P. Andriotis and K.~G. Papakonstantinou, ``Life-cycle policies for large engineering systems under complete and partial observability,'' in {\em 13th International Conference on Applications of Statistics and Probability in Civil Engineering (ICASP), Seoul, South Korea}, 2019.

\bibitem{wang2016a}
Z.~Wang, V.~Bapst, N.~Heess, R.~Munos, K.~Kavukcuoglu, and N.~Freitas, ``Sample efficient actor-critic with experience replay,'' 2016.
\newblock arXiv preprint arXiv:1611.01224,.

\bibitem{degris2012off}
T.~Degris, M.~White, and R.~S. Sutton, ``Off-policy actor-critic,'' {\em arXiv preprint arXiv:1205.4839}, 2012.

\bibitem{zhou2022reinforcement}
W.~Zhou, E.~Miller-Hooks, K.~G. Papakonstantinou, S.~Stoffels, and S.~McNeil, ``A reinforcement learning method for multi-asset roadway improvement scheduling considering traffic impacts,'' {\em Journal of Infrastructure Systems}, vol.~28, no.~4, p.~04022033, 2022.

\bibitem{EM194378890002028}
D.~Y. Yang, ``Adaptive risk-based life-cycle management for large-scale structures using deep reinforcement learning and surrogate modeling,'' {\em Journal of Engineering Mechanics}, vol.~148, no.~1, p.~04021126, 2022.

\bibitem{zhang2024multi}
J.~Zhang, X.~Li, Y.~Yuan, D.~Yang, P.~Xu, and F.~T. Au, ``A multi-agent ranking proximal policy optimization framework for bridge network life-cycle maintenance decision-making,'' {\em Structural and Multidisciplinary Optimization}, vol.~67, no.~11, p.~194, 2024.

\bibitem{andriotis2021deep}
C.~P. Andriotis and K.~G. Papakonstantinou, ``Deep reinforcement learning driven inspection and maintenance planning under incomplete information and constraints,'' {\em Reliability Engineering \& System Safety}, vol.~212, p.~107551, 2021.

\bibitem{MORATO2023109144}
P.~G. Morato, C.~P. Andriotis, K.~G. Papakonstantinou, and P.~Rigo, ``Inference and dynamic decision-making for deteriorating systems with probabilistic dependencies through {Bayesian} networks and deep reinforcement learning,'' {\em Reliability Engineering \& System Safety}, vol.~235, p.~109144, 2023.

\bibitem{andriotis2022optimizing}
C.~P. Andriotis and K.~G. Papakonstantinou, ``Optimizing policies for deteriorating systems considering ordered action structuring and value of information,'' in {\em 13th International Conference on Structural Safety \& Reliability (ICOSSAR)}, 2022.

\bibitem{SaifIcasp14}
M.~Saifullah, C.~P. Andriotis, and K.~G. Papakonstantinou, ``The role of value of information in multi-agent deep reinforcement learning for optimal decision-making under uncertainty,'' in {\em 14th International Conference on Applications of Statistics and Probability in Civil Engineering (ICASP14), Dublin}, 2023.

\bibitem{nguyen2022artificial}
V.-T. Nguyen, P.~Do, A.~Vosin, and B.~Iung, ``Artificial-intelligence-based maintenance decision-making and optimization for multi-state component systems,'' {\em Reliability Engineering \& System Safety}, vol.~228, p.~108757, 2022.

\bibitem{nguyen2022weighted}
V.-T. Nguyen, P.~Do, A.~Voisin, and B.~Iung, ``Weighted-qmix-based optimization for maintenance decision-making of multi-component systems,'' in {\em PHM Society European Conference}, vol.~7, pp.~360--367, 2022.

\bibitem{do2024multi}
P.~Do, V.-T. Nguyen, A.~Voisin, B.~Iung, and W.~A.~F. Neto, ``Multi-agent deep reinforcement learning-based maintenance optimization for multi-dependent component systems,'' {\em Expert Systems with Applications}, vol.~245, p.~123144, 2024.

\bibitem{rashid2020weighted}
T.~Rashid, G.~Farquhar, B.~Peng, and S.~Whiteson, ``Weighted qmix: Expanding monotonic value function factorisation for deep multi-agent reinforcement learning,'' {\em Advances in Neural Information Processing Systems}, vol.~33, pp.~10199--10210, 2020.

\bibitem{lee2023risk}
D.~Lee and J.~Song, ``Risk-informed operation and maintenance of complex lifeline systems using parallelized multi-agent deep q-network,'' {\em Reliability Engineering \& System Safety}, p.~109512, 2023.

\bibitem{su2021value}
J.~Su, S.~Adams, and P.~Beling, ``Value-decomposition multi-agent actor-critics,'' in {\em Proceedings of the AAAI conference on Artificial Intelligence}, vol.~35, pp.~11352--11360, 2021.

\bibitem{SU2022116323}
J.~Su, J.~Huang, S.~Adams, Q.~Chang, and P.~A. Beling, ``Deep multi-agent reinforcement learning for multi-level preventive maintenance in manufacturing systems,'' {\em Expert Systems with Applications}, vol.~192, p.~116323, 2022.

\bibitem{wang2016multi}
X.~Wang, H.~Wang, and C.~Qi, ``Multi-agent reinforcement learning based maintenance policy for a resource constrained flow line system,'' {\em Journal of Intelligent Manufacturing}, vol.~27, pp.~325--333, 2016.

\bibitem{rodriguez2022multi}
M.~L.~R. Rodr{\'\i}guez, S.~Kubler, A.~de~Giorgio, M.~Cordy, J.~Robert, and Y.~Le~Traon, ``Multi-agent deep reinforcement learning based predictive maintenance on parallel machines,'' {\em Robotics and Computer-Integrated Manufacturing}, vol.~78, p.~102406, 2022.

\bibitem{leroy2023imp}
P.~Leroy, P.~G. Morato, J.~Pisane, A.~Kolios, and D.~Ernst, ``Imp-marl: a suite of environments for large-scale infrastructure management planning via marl,'' {\em arXiv preprint arXiv:2306.11551}, 2023.

\bibitem{oliehoek2016concise}
F.~A. Oliehoek and C.~Amato, {\em A concise introduction to decentralized POMDPs}.
\newblock Springer, 2016.

\bibitem{bernstein2002complexity}
D.~S. Bernstein, R.~Givan, N.~Immerman, and S.~Zilberstein, ``The complexity of decentralized control of {Markov} decision processes,'' {\em Mathematics of Operations Research}, vol.~27, no.~4, pp.~819--840, 2002.

\bibitem{sondik1971a}
E.~Sondik, {\em The optimal control of partially observable Markov processes}.
\newblock Stanford, CA: Stanford University, Stanford Electronics Labs, 1971.

\bibitem{kurniawati2008sarsop}
H.~Kurniawati, D.~Hsu, and W.~S. Lee, ``Sarsop: Efficient point-based {POMDP} planning by approximating optimally reachable belief spaces,'' in {\em Robotics: Science and systems}, vol.~2008, Citeseer, 2008.

\bibitem{sutton2018reinforcement}
R.~S. Sutton and A.~G. Barto, {\em Reinforcement learning: An introduction}.
\newblock MIT press, 2018.

\bibitem{sutton2000policy}
R.~S. Sutton, D.~A. McAllester, S.~P. Singh, and Y.~Mansour, ``Policy gradient methods for reinforcement learning with function approximation,'' in {\em Advances in Neural Information Processing Systems}, pp.~1057--1063, 2000.

\bibitem{van2016deep}
H.~Van~Hasselt, A.~Guez, and D.~Silver, ``Deep reinforcement learning with double q-learning,'' in {\em Proceedings of the AAAI conference on Artificial Intelligence}, vol.~30, 2016.

\bibitem{williams1992simple}
R.~J. Williams, ``Simple statistical gradient-following algorithms for connectionist reinforcement learning,'' {\em Machine Learning}, vol.~8, no.~3-4, pp.~229--256, 1992.

\bibitem{konda2000actor}
V.~R. Konda and J.~N. Tsitsiklis, ``Actor-critic algorithms,'' in {\em Advances in Neural Information Processing Systems}, pp.~1008--1014, 2000.

\bibitem{mnih2016asynchronous}
V.~Mnih, A.~P. Badia, M.~Mirza, A.~Graves, T.~Lillicrap, T.~Harley, D.~Silver, and K.~Kavukcuoglu, ``Asynchronous methods for deep reinforcement learning,'' in {\em International Conference on Machine Learning}, pp.~1928--1937, PMLR, 2016.

\bibitem{lillicrap2016continuous}
T.~P. Lillicrap, J.~J. Hunt, A.~Pritzel, N.~Heess, T.~Erez, Y.~Tassa, D.~Silver, and D.~Wierstra, ``Continuous control with deep reinforcement learning,'' in {\em Proceedings of the International Conference on Learning Representations (ICLR)}, 2016.

\bibitem{sutton1988learning}
R.~S. Sutton, ``Learning to predict by the methods of temporal differences,'' {\em Machine Learning}, vol.~3, no.~1, pp.~9--44, 1988.

\bibitem{sunehag2017value}
P.~Sunehag, G.~Lever, A.~Gruslys, W.~M. Czarnecki, V.~Zambaldi, M.~Jaderberg, M.~Lanctot, N.~Sonnerat, J.~Z. Leibo, K.~Tuyls, and T.~Graepel, ``Value-decomposition networks for cooperative multi-agent learning,'' {\em arXiv preprint arXiv:1706.05296}, 2017.

\bibitem{oliehoek2008optimal}
F.~A. Oliehoek, M.~T. Spaan, and N.~Vlassis, ``Optimal and approximate q-value functions for decentralized pomdps,'' {\em Journal of Artificial Intelligence Research}, vol.~32, pp.~289--353, 2008.

\bibitem{foerster2018counterfactual}
J.~Foerster, G.~Farquhar, T.~Afouras, N.~Nardelli, and S.~Whiteson, ``Counterfactual multi-agent policy gradients,'' in {\em Proceedings of the AAAI Conference on Artificial Intelligence}, vol.~32, 2018.

\bibitem{rashid2018qmix}
T.~Rashid, M.~Samvelyan, C.~Schroeder, G.~Farquhar, J.~Foerster, and S.~Whiteson, ``Qmix: Monotonic value function factorisation for deep multi-agent reinforcement learning,'' in {\em International Conference on Machine Learning}, pp.~4295--4304, PMLR, 2018.

\bibitem{gronauer2022multi}
S.~Gronauer and K.~Diepold, ``Multi-agent deep reinforcement learning: a survey,'' {\em Artificial Intelligence Review}, pp.~1--49, 2022.

\bibitem{lowe2017multi}
R.~Lowe, Y.~Wu, A.~Tamar, J.~Harb, P.~Abbeel, and I.~Mordatch, ``Multi-agent actor-critic for mixed cooperative-competitive environments,'' {\em arXiv preprint arXiv:1706.02275}, 2017.

\bibitem{bertsekas1997nonlinear}
D.~P. Bertsekas, ``Nonlinear programming,'' {\em Journal of the Operational Research Society}, vol.~48, no.~3, pp.~334--334, 1997.

\bibitem{hasnat2023comparative}
M.~Hasnat, R.~R. Singh, M.~E. Kutay, J.~Bryce, S.~W. Haider, and B.~Cetin, ``Comparative study of different condition indices using {Michigan} department of transportation’s flexible distress data,'' {\em Transportation Research Record}, p.~03611981231156917, 2023.

\bibitem{katicha2016a}
S.~W. Katicha, S.~Ercisli, G.~W. Flintsch, J.~M. Bryce, and B.~K. Diefenderfer, ``Development of enhanced pavement deterioration curves,'' tech. rep., Virginia Transportation Research Council, 2016.

\bibitem{v2016a}
T.~Chowdhury, ``Supporting document for the development and enhancement of the pavement maintenance decision matrices used in the needs-based analysis,'' tech. rep., Virginia Transportation Research Council, Maintenance Division, Richmond, VA, 2016.

\bibitem{papakonstantinou2023hamiltonian}
K.~G. Papakonstantinou, H.~Nikbakht, and E.~Eshra, ``Hamiltonian {MCMC} methods for estimating rare events probabilities in high-dimensional problems,'' {\em Probabilistic Engineering Mechanics}, vol.~74, p.~103485, 2023.

\bibitem{m2014a}
{Dye Management Group Inc.}, ``Monitoring highway assets with remote technology,'' technical report, Michigan Department of Transportation, 2014.

\bibitem{a2020a}
``American road and transportation builders association.''
\newblock Last accessed February 23, 2026: \url{https://www.artba.org/}.

\bibitem{f2020a}
``Cost per mile models by {FDOT},'' 2022.
\newblock Last accessed February 23, 2026: \url{https://www.fdot.gov/}.

\bibitem{penndot2017a}
PennDOT, ``{Road Maintenance and Preservation (MaP)},'' 2017.
\newblock Last accessed February 23, 2026: \url{https://www.pa.gov/content/dam/copapwp-pagov/en/penndot/documents/about-us/documents/penndot%20road%20map%20initiative.pdf}.

\bibitem{a2018a}
ADOT, ``Production rates guidelines for {Arizona} highway construction,'' 2018.
\newblock Last accessed February 23, 2026: \url{https://azdot.gov/sites/default/files/2019/06/adot-typical-production-rates.pdf}.

\bibitem{penndot2019a}
PennDOT, ``Publication 242 pavement policy manual, ({May} 2015 edition) change no. 5,'' tech. rep., Pennsylvania Department of Transportation, 2019.
\newblock Last accessed February 23, 2026: \url{https://www.butlercountypa.gov/DocumentCenter/View/1555/PennDOT-Publication-242-Pavement-Policy-Manual-PDF}.

\bibitem{sayer1986a}
T.~D. Gillespie, W.~Paterson, and M.~W. Sayers, ``Guidelines for conducting and calibrating road roughness measurements,'' tech. rep., The World Bank, 1986.

\bibitem{faddoul2013a}
R.~Faddoul, W.~Raphael, A.-H. Soubra, and A.~Chateauneuf, ``Incorporating {Bayesian} networks in {Markov} decision processes,'' {\em Journal of Infrastructure Systems}, vol.~19, no.~4, p.~415–424, 2013.

\bibitem{f1999a}
{United States Department of Transportation}, ``1999 {Status} of the nation's highways, bridges and transit: Conditions \& performance, a report to {Congress}: Executive summary.'' 2000.
\newblock Last accessed February 23, 2026: \url{https://rosap.ntl.bts.gov/view/dot/15379}.

\bibitem{penndot2009a}
{Pennsylvania Department of Transportation and Bureau of Design}, ``{Bridge Management System 2 (BMS2) coding manual: Publication 100A, 2022 edition},'' 2022.

\bibitem{manafpour2018a}
A.~Manafpour, I.~Guler, A.~Radli{\'n}ska, F.~Rajabipour, and G.~Warn, ``Stochastic analysis and time-based modeling of concrete bridge deck deterioration,'' {\em Journal of Bridge Engineering}, vol.~23, no.~9, p.~04018066, 2018.

\bibitem{papakonstantinou2023hamiltonian2}
K.~G. Papakonstantinou, E.~Eshra, and H.~Nikbakht, ``Hamiltonian {MCMC} based framework for time-variant rare event uncertainty quantification,'' in {\em 14th International Conference on Applications of Statistics and Probability in Civil Engineering (ICASP), Dublin, Ireland}, 2023.

\bibitem{f2019a}
{Federal Highway Administration}, {\em Bridge Replacement Unit Costs}.
\newblock Washington D.C: FHWA, 2019.

\bibitem{wells1995a}
D.~T. Wells, ``Technical assistance report: Maintenance, repair, and rehabilitation unit costs for {PONTIS},'' tech. rep., Virginia Transportation Research Council, 1995.
\newblock Last accesed February 23, 2026: \url{https://rosap.ntl.bts.gov/view/dot/20088}.

\bibitem{goodspeed2017a}
C.~Goodspeed and J.~Brown, ``Gilford rapid bridge deck replacement, a project summary,'' tech. rep., University of New Hampshire, 2017.
\newblock Last accessed February 23, 2026: \url{https://mm.nh.gov/files/uploads/dot/remote-docs/nhdot-research-15680x-2017-summary-report.pdf}.

\bibitem{oakgrove2013a}
Oakgrove, {\em Accelerated Bridge Construction-Four Bridge Deck Replacements in Region 5}.
\newblock Oakgrove Construction, Inc, 2013.

\bibitem{trafficVDOT2019}
``{Daily Vehicle Miles Traveled {(DVMT)}} by maintenance jurisdiction,'' 2020.
\newblock Last accessed February 23, 2026: \url{http://166.67.200.218/info/resources/Traffic_2020/VMTReport_2100_2020.pdf}.

\bibitem{hamptonmaps}
{Hampton Roads Planning District Commision (HRPDC)}, ``Hampton roads maps.''
\newblock Last accessed February 23, 2026: \url{https://www.hrpdcva.gov/1009/Data-Maps}.

\bibitem{vadakpat2000road}
G.~Vadakpat, S.~Stoffels, and K.~Dixon, ``Road user cost models for network-level pavement management,'' {\em Transportation Research Record}, vol.~1699, no.~1, pp.~49--57, 2000.

\bibitem{fhwaUserCost}
{Federal Highway Administration}, ``{Work Zone Road User Costs - Concepts and Applications: Chapter 2}. {Work} zone road user costs.''
\newblock Last accessed February 23, 2026: \url{https://ops.fhwa.dot.gov/wz/resources/publications/fhwahop12005/sec2.htm}.

\bibitem{FHWAbridge}
{Federal Highway Administration and United States Department of Transportation}, ``23 {CFR} 490.411: {Establishment} of minimum level for condition for bridges.''
\newblock Last accessed February 23, 2026: \url{https://www.ecfr.gov/current/title-23/section-490.411}.

\bibitem{FHWApavement}
{Federal Highway Administration and United States Department of Transportation}, ``23 {CFR} 490.315: {Establishment} of minimum level for condition of pavements.''
\newblock Last accessed February 23, 2026: \url{https://www.ecfr.gov/current/title-23/part-490/section-490.315}.

\bibitem{VDOT2018pav}
{Virginia Department of Transportation}, ``Approval of the report on the commonwealth's large and unique bridge and tunnel structures: Vital infrastructure.'' \url{https://ctb.virginia.gov/media/ctb/agendas-and-meeting-minutes/2018/dec/res/14title-approval-of-the-report-on-the-commonwealths-large-and-unique-bridge-and-tunnel-structures-vital-infrastructure.pdf}, 2018.
\newblock Last accessed February 23, 2026.

\bibitem{vdot2019a}
{Virginia Department of Transportation}, ``Maintenance and operations comprehensive review,'' 2019.
\newblock Last accessed February 23, 2026: \url{https://www.vdot.virginia.gov/media/vdotvirginiagov/about/legislative-studies-and-reports/Maintenance_and_Operations_Comprehensive_Review_2019.pdf}.

\bibitem{hampton2020}
{Hampton Roads Transportation Planning Organization (HRTPO)}, ``The state of transportation in {Hampton} roads,'' 2024.
\newblock Last accessed February 23, 2026: \url{https://hrpdcva.gov/DocumentCenter/View/13727/T25-01-State-of-Transportation-2024-Final-PDF?bidId=}.

\bibitem{vdot_pav_map}
{Virginia Department of Transportation}, ``Pavement condition map.''
\newblock Last accessed February 23, 2026: \url{https://vdot.maps.arcgis.com/apps/webappviewer/index.html?id=d93cecc94efb4bda9213d0c3cef73ce4}.

\bibitem{vdot_str_map}
{Virginia Department of Transportation}, ``Structure condition map.''
\newblock Last accessed February 23, 2026: \url{https://dashboard.virginiadot.org/pages/maintenance/bridge.aspx}.

\end{thebibliography}

\newpage
\appendix
\renewcommand\thetable{\thesection.\arabic{table}}
  \renewcommand\thefigure{\thesection.\arabic{figure}}
  \renewcommand{\thealgocf}{\thesection.\arabic{algocf}}
\section*{APPENDIX}
\setcounter{figure}{0}  
\setcounter{table}{0} 
\setcounter{algocf}{0}

\section{Optimal Condition Based Maintenance (CBM) Policy}\label{CBM}

\begin{figure}[h!]
\scalebox{0.99}{%
\centering
    \begin{minipage}[b]{0.45\textwidth}
        \centering

     \begin{algorithm}[H]
\footnotesize
\SetAlgoLined
 initialize total cost metric $\bar c = 0$ \;
 \While{$i$ $<$ max\_episodes}{
 $t$ = 0;\\
 initialize component beliefs ${\mathbf{b}_t} = \left[ {{\bf{b}}_t^{(i)}} \right]_{i = 1}^N$, observations $\mathbf{o}_t=\left[ {{{o}}_t^{(i)}} \right]_{i = 1}^N$\;
  \While{$t$ $<$ T }{
  
  \uIf{$t$ is even}{
   select action $a_t=6$ for all components\;
   }
   \Else{
   \uIf{component = bridge}{
     select action $a_t=[0,6,8,6,8,8,9]$ if the observed states are $[9,8,7,6,5,\{4,...\},failure]$, respectively\; 
  
   }\uElseIf{component = interstate}{
     select action $a_t=[0,6,7,6,7,8]$ if the observed CCI states are $[6,5,4,3,2,1]$, respectively\; 
   }\uElseIf{component = primary}{
	 select action $a_{t,1}=[0,3,3,4,4,5]$ if the observed CCI states are $[6,5,4,3,2,1]$\;
	 select action $a_{t,2}=[0,3,4,4,5]$ if the observed IRI states are $[5,4,3,2,1]$\;
	
	 	\uIf {$a_{t,1}>=a_{t,2}$}{
	 	 	select action $a_t=a_{t,1}$ \;
	 	}\Else{
	 		select action $a_t=a_{t,2}$ \;}
     }\ElseIf{component = secondary}{
	 select action $a_t=[0,0,3,4,4,4]$ if the observed CCI states are $[6,5,4,3,2,1]$, respectively\;  
  }
  }
  
 Estimate total cost ${\bar c_{b,t}}$ based on the current belief $\mathbf{b}_t$ accounting for the budget and terminal cost ${\bar c}_{s_T}$ \; \eIf {$t\neq T$}{${\bar c_{b,t}} = {\bar c_b}$\;} 
{${\bar c_{b,t}} = {\bar c_b}+{\bar c}_{s_T}$\;}
Estimate next belief state  $\mathbf{b}_{t+1}$, and collect the next observation $\mathbf{o}_{t+1}$\; 
Assign $\mathbf{b}_{t} \leftarrow \mathbf{b}_{t+1}$, $\mathbf{o}_{t} \leftarrow \mathbf{o}_{t+1}$\; 
   }
$\bar c = \bar c + \sum_{t=0}^{T-1}{\bar c_{b,t}}$\;
 }
 $\bar c = \bar c/max\_episodes$\;  
 
\caption{Optimal Condition Based Maintenance (CBM) policy for intact starting condition state}
 \label{algo:2}
\end{algorithm}

    \end{minipage}%
    ~ 
    \begin{minipage}[b]{0.45\textwidth}
        \centering
\begin{algorithm}[H]
\footnotesize
\SetAlgoLined
 initialize total cost metric $\bar c = 0$ \;
 \While{$i$ $<$ max\_episodes}{
 $t$ = 0;\\
 initialize component beliefs ${\mathbf{b}_t} = \left[ {{\bf{b}}_t^{(i)}} \right]_{i = 1}^N$, observations $\mathbf{o}_t=\left[ {{{o}}_t^{(i)}} \right]_{i = 1}^N$\;
  \While{$t$ $<$ T }{
  
  \uIf{$t$ is even}{
   select action $a_t=6$ for all components\;
   }
   \Else{
   \uIf{component = bridge}{
     select action $a_t=[0,6,8,6,8,8,9]$ if the observed states are $[9,8,7,6,5,\{4,...\},failure]$, respectively\; 
  
   }\uElseIf{component = interstate}{
     select action $a_t=[0,6,7,8,8,8]$ if the observed CCI states are $[6,5,4,3,2,1]$, respectively\; 
   }\uElseIf{component = primary}{
	 select action $a_{t,1}=[0,3,3,4,4,5]$ if the observed CCI states are $[6,5,4,3,2,1]$\;
	 select action $a_{t,2}=[0,3,4,4,5]$ if the observed IRI states are $[5,4,3,2,1]$\;
	
	 	\uIf {$a_{t,1}>=a_{t,2}$}{
	 	 	select action $a_t=a_{t,1}$ \;
	 	}\Else{
	 		select action $a_t=a_{t,2}$ \;}
     }\ElseIf{component = secondary}{
	 select action $a_t=[0,0,3,4,4,5]$ if the observed CCI states are $[6,5,4,3,2,1]$, respectively\;  
  }
  }
  
 Estimate total cost ${\bar c_{b,t}}$ based on the current belief $\mathbf{b}_t$ accounting for the budget and terminal cost ${\bar c}_{s_T}$ \; \eIf {$t\neq T$}{${\bar c_{b,t}} = {\bar c_b}$\;} 
{${\bar c_{b,t}} = {\bar c_b}+{\bar c}_{s_T}$\;}
Estimate next belief state  $\mathbf{b}_{t+1}$, and collect the next observation $\mathbf{o}_{t+1}$\; 
Assign $\mathbf{b}_{t} \leftarrow \mathbf{b}_{t+1}$, $\mathbf{o}_{t} \leftarrow \mathbf{o}_{t+1}$\; 
   }
$\bar c = \bar c + \sum_{t=0}^{T-1}{\bar c_{b,t}}$\;
 }
 $\bar c = \bar c/max\_episodes$\; 
 
\caption{Optimal Condition Based Maintenance (CBM) policy for non-intact starting condition state}
 \label{algo:3}
\end{algorithm}

    \end{minipage}
}
\end{figure}

\begin{tablenotes}
\footnotesize
\item \parbox{435pt}{\emph{Note:} Actions 0,1,2 are \textit{Do Nothing, Minor Repair, Major Repair} with no-inspection; 3, 4, 5 are \textit{Do Nothing, Minor Repair, Major Repair} with low fidelity inspection; 6, 7, 8 are \textit{Do Nothing, Minor Repair, Major Repair} with high fidelity inspection; and 9 is \textit{Reconstruction}.}
\end{tablenotes}

\begin{table}[b]
\centering
\begin{threeparttable}
\caption{\em{Considered bridge CBM policy in the VDOT-based baseline case}}
\label{tab:VDOT_brdg}
\begin{tabular}{@{}p{2cm}p{5cm}p{6cm}@{}}
\toprule
\textbf{Starting condition} & \textbf{Policy based on bridge conditions} & \textbf{Prioritization} \\
\midrule
Intact & Select action \(a_t=[0,6,6,8,6,8,9]\) if the observed states are \([9,8,7,6,5,\{4,...\},\text{failure}]\), respectively & Not required \\
Non-intact & Select action \(a_t=[0,6,8,8,8,8,9]\) if the observed states are \([9,8,7,6,5,\{4,...\},\text{failure}]\), respectively & Due to insufficient budget in the non-intact starting condition case, bridges  are prioritized based on their associated individual risk cost. The first eight bridges with the  highest risk costs are then selected at each time step for implementation of inspection and maintenance actions, contingent upon budget availability. Relevant actions for all other components of the network are then selected in a randomized priority, until the entire budget is utilized. \\
\bottomrule
\end{tabular}
\end{threeparttable}
\end{table}

\section{Condition of the Hampton Roads network}\label{sec:current_cond}
\noindent
The existing condition of the Hampton Roads network can be obtained from the VDOT dashboard and ArcGIS map \cite{vdot_pav_map, vdot_str_map}. For bridge decks, these conditions are reported as a general condition rating from 9 to 1, as shown in Table \ref{tab:Deck_curr}. For pavements, the CCI ratings are provided for interstate pavements, however, only the condition distribution on an aggregate level is provided for primary and secondary pavements, as shown in Table \ref{tab:CCI_curr}. Random sampling is thus performed in this work, using the provided distribution, to obtain the initial condition states for the primary and secondary highways.

\begin{table}[]
\centering
\scalebox{0.93}{

\begin{threeparttable} 
\caption{\em{Bridge geometric information and condition state as of 2021}}
\label{tab:Deck_curr}
\begin{tabular}{@{}cccccccc@{}}
\toprule
S.no. &
  Name &
  Starting node &
  End node &
  Length (miles) &
  \begin{tabular}[c]{@{}c@{}}Number of\\  lanes\end{tabular} &
  \begin{tabular}[c]{@{}c@{}}Deck Condition \\ Rating \\ (GCR 1-9)\end{tabular} &
  \begin{tabular}[c]{@{}c@{}}Sampled deck age \\ (years) \end{tabular} \\ \midrule
1  & James river (type I)                                                                    & 18 & 21 & 4.42 & 4 & 6 & 5  \\ \midrule
2  & \begin{tabular}[c]{@{}c@{}}MMM bridge \\ tunnel (type I)\end{tabular}                   & 19 & 68 & 4.63 & 4 & 6 & 7  \\ \midrule
3  & \begin{tabular}[c]{@{}c@{}}Hampton roads \\ bridge tunnel (type I)\end{tabular}         & 20 & 63 & 3.50 & 4 & 5 & 5  \\ \midrule
4  & \begin{tabular}[c]{@{}c@{}}Coleman memorial \\ bridge (type II)\end{tabular}            & 5  & 6  & 0.71 & 4 & 6 & 9  \\ \midrule
5  & Bridge\_RD (type II)                                                                    & 21 & 69 & 0.71 & 2 & 6 & 9  \\ \midrule
6  & \begin{tabular}[c]{@{}c@{}}High rise \\ bridge (type II) \end{tabular}                  & 49 & 50 & 0.94 & 4 & 5 & 12 \\ \midrule
7  & \begin{tabular}[c]{@{}c@{}}Berkley \\ bridge (type II)\end{tabular}                      & 49 & 66 & 0.40 & 8 & 6 & 10 \\ \midrule
8  & \begin{tabular}[c]{@{}c@{}}Gilmerton \\ bridge (type II)\end{tabular}                    & 50 & 62 & 0.42 & 4 & 7 & 1  \\ \midrule
9  & \begin{tabular}[c]{@{}c@{}}W Norfolk\\ bridge (type II)\end{tabular}                    & 66 & 69 & 0.63 & 4 & 7 & 3  \\ \midrule
10 & \begin{tabular}[c]{@{}c@{}}Bridge on Nansemond \\ River Suffolk (type III)\end{tabular} & 31 & 45 & 0.20 & 4 & 6 & 13 \\ \midrule
11 & \begin{tabular}[c]{@{}c@{}}Great bridge \\ bypass (type III)\end{tabular}               & 53 & 54 & 0.56 & 4 & 5 & 7  \\ \bottomrule
\end{tabular}
\end{threeparttable}
}

\end{table}

\begin{table}[]
\centering
\begin{threeparttable} 
\caption{\em{CCI state distribution for primary and secondary pavements}}
\label{tab:CCI_curr}
\begin{tabular}{@{}ccccccc@{}}
\toprule
{Pavements} &
 {V. poor} &
  {Poor} &
 {Fair} &
 {Good} &
  {V. good} &
 {Excellent} \\
\hline
\begin{tabular}[c]{@{}c@{}}CCI distribution of\\  primary pavements\end{tabular} &
  0.091 &
  0.057 &
  0.135 &
  0.241 &
  0.215 &
  0.261 \\
  \hline
\begin{tabular}[c]{@{}c@{}}CCI distribution of \\ secondary pavements\end{tabular} &
  0.148 &
  0.102 &
  0.109 &
  0.272 &
  0.193 &
  0.176 \\
  \bottomrule
\end{tabular}
\end{threeparttable}
\end{table}

\begin{table}[b]
\centering
\begin{threeparttable} 
\caption{\em{Statistics of IRI values for different pavement types, and mean and standard \\ deviation of the fitted lognormal distribution.}}
\label{tab:IRI_curr}
\begin{tabular}{@{}ccccc@{}}
\toprule
{Pavement} & {Mean IRI   (m/km)} & {\begin{tabular}[c]{@{}c@{}}Deficient pavement (\%)\end{tabular}} & {$\mu$} & {$\sigma$} \\
\hline
Interstate & 1.264 & 4.30\%  & 0.169  & 0.361 \\
Primary    & 1.485 & 12.20\% & 0.3115 & 0.41  \\
Secondary  & 2.196 & 35.00\% & 0.4933 & 0.766\\ \bottomrule
\end{tabular}
\end{threeparttable}
\end{table} 

For IRI states, VDOT only provides the mean IRI value and percentage of deficient pavements for all types of highway systems (type I-III), as presented in Table \ref{tab:IRI_curr}. VDOT uses different definitions of deficiency for different pavement types. Here, we considered, for simplicity, deficient pavements if IRI $>$ 2.20 m/km or 140 inch/mile for all pavement types (I-III), which is consistent with the definition given by VDOT for interstate highways (i.e., if the pavement is in IRI state 1 or 2 will be classified as deficient). For sampling, a lognormal distribution is chosen here, as IRI has continuous positive values. The fitted lognormal distribution and the obtained mean ($\mu$), and standard deviation ($\sigma$) in the log space are reported in Table \ref{tab:IRI_curr}.     

\begin{figure}[t]
\centering
    \begin{subfigure}[b]{0.45\textwidth}
        \centering
   \includegraphics[width=\textwidth]{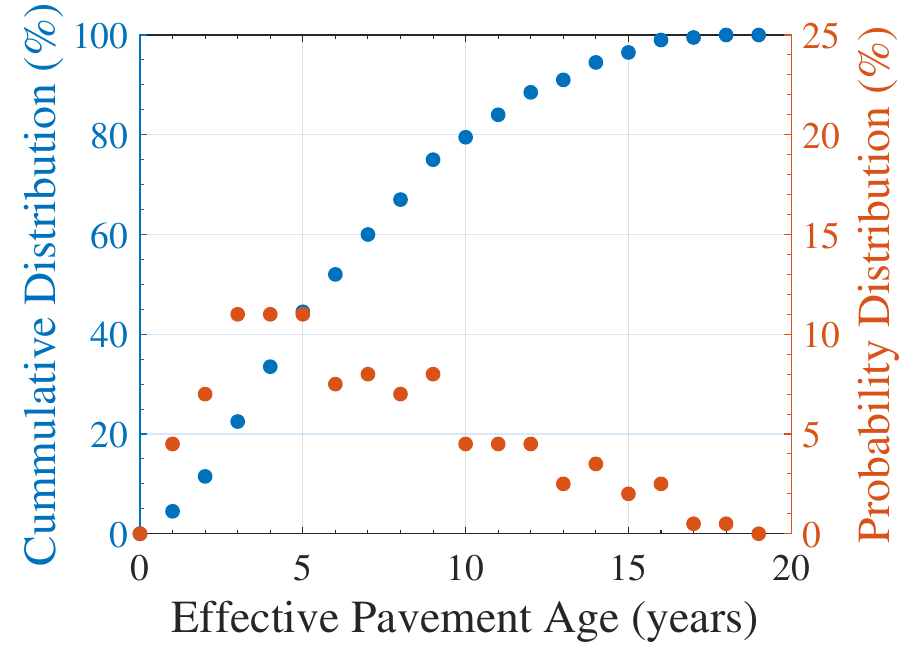}
        \caption{}
        \label{fig:age_dist1}
    \end{subfigure}%
    ~ 
    \begin{subfigure}[b]{0.45\textwidth}
        \centering
   \includegraphics[width=\textwidth]{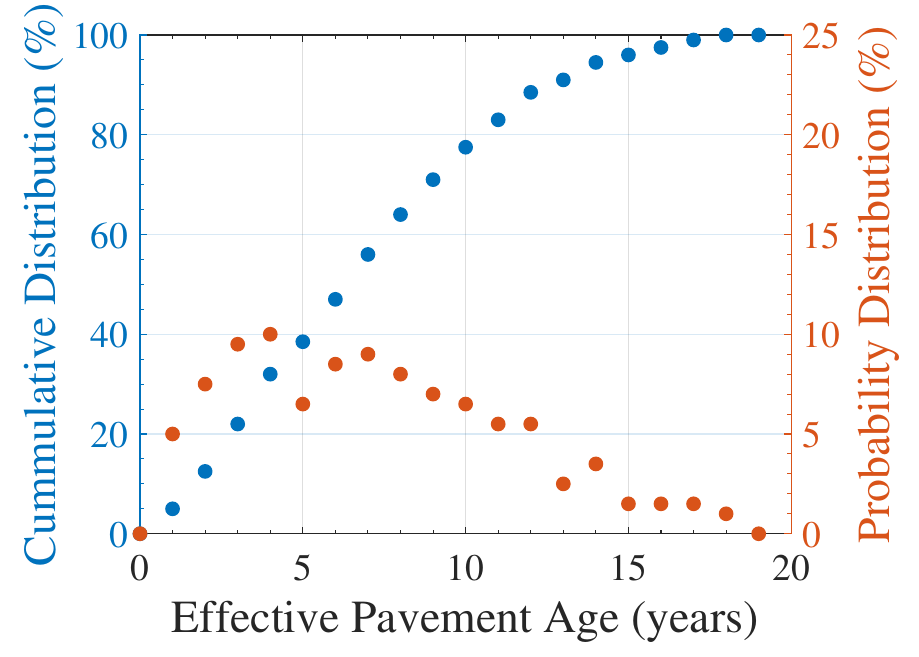}
        \caption{}
        \label{fig:age_dist2}
    \end{subfigure}
   
\caption[]{Age distribution of (a) interstate and (b) primary highways according to VDOT \citep{v2016a}.}
\label{fig:age_dist}
\end{figure}

Along with the condition states, the component effective age is needed to acquire the deterioration rates for non-stationary indices. For the CCI, the age distribution is taken from VDOT \citep{v2016a}, which is shown in Figs. \ref{fig:age_dist1} and \ref{fig:age_dist2} for interstate and primary highways, respectively. For secondary pavements, the same distribution as for primary pavements is utilized, but with a three-year shift. This assumption is based on the fact that VDOT inspects 20\% of secondary highways annually, compared to the annual assessment of primary pavements, thus accordingly assuming that maintenance activities for the secondary system will be at least 2.5 years behind. Because age data for decks are unavailable, the same age distribution as for the interstate pavements is used for type-I bridges, while the age distribution of primary pavements is utilized for the remaining bridges. Finally, sampling is performed over the corresponding distributions to determine the component effective ages.

\section{Gamma process coefficients}\label{sec:gamma_coef}
\noindent
The gamma process coefficients $f(t)$ and $g(t)$ from section \ref{subsubsec:gamma_process} are presented in Table \ref{tab:gammacoef} for $t = 1, 2, 3 \cdots 20$ years.

\begin{table}[h!]
\centering
\begin{threeparttable} 

\caption{\em{Variables $f(t)$ and $g(t)$  for gamma process for calculating CCI transition probabilities }}
\label{tab:gammacoef}
\begin{tabular}{@{}ccccccccccc@{}}
\toprule
\multirow{2}{*}{\begin{tabular}[c]{@{}c@{}}Time\\ (years)\end{tabular}} & \multicolumn{2}{c}{Traffic level (A)} & \multicolumn{2}{c}{Traffic level (B)} & \multicolumn{2}{c}{Traffic level (C)} & \multicolumn{2}{c}{Traffic level (D)} & \multicolumn{2}{c}{Traffic level (E)} \\ \cline{2-11} 
                                                                        & $f(t)$            & $g(t)$            & $f(t)$            & $g(t)$            & $f(t)$            & $g(t)$            & $f(t)$            & $g(t)$            & $f(t)$            & $g(t)$            \\ \cline{1-1}
                                                                        \hline 
1                                                                       & 0                 & 0.178             & 0                 & 0.280             & 0                 & 0.349             & 0                 & 0.420             & 0                 & 0.473             \\
2                                                                       & 0.922             & 0.101             & 1.217             & 0.158             & 1.394             & 0.197             & 1.560             & 0.237             & 1.679             & 0.267             \\
3                                                                       & 0.649             & 0.061             & 0.857             & 0.096             & 0.981             & 0.119             & 1.098             & 0.143             & 1.182             & 0.161             \\
4                                                                       & 0.475             & 0.038             & 0.628             & 0.059             & 0.718             & 0.074             & 0.804             & 0.089             & 0.866             & 0.100             \\
5                                                                       & 0.354             & 1.135             & 0.467             & 1.350             & 0.535             & 1.469             & 0.599             & 1.577             & 0.644             & 1.651             \\
6                                                                       & 12.759            & 0.952             & 12.759            & 1.133             & 12.759            & 1.233             & 12.759            & 1.323             & 12.759            & 1.386             \\
7                                                                       & 12.759            & 0.801             & 12.759            & 0.953             & 12.759            & 1.037             & 12.759            & 1.113             & 12.759            & 1.166             \\
8                                                                       & 12.759            & 0.675             & 12.759            & 0.804             & 12.759            & 0.874             & 12.759            & 0.938             & 12.759            & 0.982             \\
9                                                                       & 12.759            & 0.570             & 12.759            & 0.678             & 12.759            & 0.738             & 12.759            & 0.792             & 12.759            & 0.829             \\
10                                                                      & 12.759            & 0.481             & 12.759            & 0.573             & 12.759            & 0.623             & 12.759            & 0.669             & 12.759            & 0.700             \\
11                                                                      & 12.759            & 0.407             & 12.759            & 0.484             & 12.759            & 0.527             & 12.759            & 0.566             & 12.759            & 0.592             \\
12                                                                      & 12.759            & 0.344             & 12.759            & 0.410             & 12.759            & 0.446             & 12.759            & 0.479             & 12.759            & 0.501             \\
13                                                                      & 12.759            & 0.292             & 12.759            & 0.347             & 12.759            & 0.378             & 12.759            & 0.405             & 12.759            & 0.424             \\
14                                                                      & 12.759            & 0.247             & 12.759            & 0.294             & 12.759            & 0.320             & 12.759            & 0.343             & 12.759            & 0.359             \\
15                                                                      & 12.759            & 0.209             & 12.759            & 0.249             & 12.759            & 0.271             & 12.759            & 0.291             & 12.759            & 0.305             \\
16                                                                      & 12.759            & 0.177             & 12.759            & 0.211             & 12.759            & 0.230             & 12.759            & 0.247             & 12.759            & 0.258             \\
17                                                                      & 12.759            & 0.151             & 12.759            & 0.179             & 12.759            & 0.195             & 12.759            & 0.209             & 12.759            & 0.219             \\
18                                                                      & 12.759            & 0.128             & 12.759            & 0.152             & 12.759            & 0.165             & 12.759            & 0.177             & 12.759            & 0.186             \\
19                                                                      & 12.759            & 0.108             & 12.759            & 0.129             & 12.759            & 0.140             & 12.759            & 0.151             & 12.759            & 0.158             \\
20                                                                      & 12.759            & 0.092             & 12.759            & 0.109             & 12.759            & 0.119             & 12.759            & 0.128             & 12.759            & 0.134             \\ \bottomrule
\end{tabular}
\end{threeparttable} 
\end{table}

\section{Terminal costs}\label{sec:term_cost}
\noindent
The terminal cost may be defined as the cost borne due to the condition state of the components at the end of the policy horizon, mathematically this is expressed through the value function at the last step $V^{\pi}(s_T)$, where $s_T$ is the state at time step $T$, for finite-horizon problems of length $T$. If we consider no terminal costs/constraints (i.e., $V^{\pi}(s_T) = 0$), the agent usually withholds from performing any action at the end of the horizon, as there are no after-effects/benefits of these actions, except for the bridges in this work. However, in any engineering system problem, such as civil infrastructure maintenance problems, it is important to consider the aftermath of every decision-cycle. Thus, this terminal cost is added to incentivize the agents to still take inspection and maintenance actions at the end of the horizon. These actions also prevent violation of constraints, while seeking minimization of life-cycle costs and preservation of infrastructure beyond the current decision horizon.  

For an arbitrary system terminal state, the terminal cost can be theoretically connected to an infinite horizon problem, for all components. However, due to computational infeasibility, the terminal cost is calculated here using a representative component from each type of pavements and bridges, and, for further simplification, the CBM policy in Algorithm \ref{algo:2}, \& \ref{algo:3} are chosen, with an unlimited budget, to approximate the remaining value of the system components at the end of the decision horizon. The terminal cost/value function $V^{\pi_{CBM}}(s_T)$ is thus computed as:

\begin{equation}
c_{s_T} = \alpha V^{\pi_{CBM}}(s_T) = \alpha\E_{s\in S, a \sim \pi_{CBM}}\left[\sum\limits_{t=T}^\infty \gamma^t c_t|s_T \right]
\end{equation}
where $\pi_{CBM}$ is a CBM action policy, $c_t$ is the total cost that constitutes inspection and maintenance cost, user delay, and risk costs, and $\alpha$ is a hyperparameter determining the overall effect of the remaining value of the network. As this remaining cost has, naturally, a higher magnitude than other costs, the $\alpha$ value is carefully chosen to be 0.55 here, consistent with other utilized costs, and so that the computed policy is not influenced excessively, particularly at the end of the horizon. This cost is then directly added to the total cost at the last step of the life cycle, $t=T$.

\section{Tables related to modeling of bridges and pavements}
\noindent
\setcounter{table}{0} 
This section presents the rest of the matrices used for the pavement and deck models. Table \ref{tab:actions} gives the detailed maintenance actions for the pavements based on VDOT \cite{v2016a}. Tables \ref{tab:3.2} and \ref{tab:3.3} provide the observation probabilities for six CCI states, for low- and high-fidelity inspections, respectively, based on the model described in Section \ref{sec:obs_prob_CCI}. Similarly, Table \ref{tab:3.15} describes the observation probabilities for IRI with 3 different level of fidelities. Three maintenance actions ({\it{Minor Repair, Major Repair}}, and {\it{Reconstruction}}) transition probabilities are shown in Tables \ref{tab:3.5}-\ref{tab:3.7} for CCI states, and Tables \ref{tab:3.12}-\ref{tab:3.14} for IRI states. Table \ref{tab:3.9} shows the durations used for completion of the maintenance actions per lane-mile of pavement. Similarly, Tables \ref{tab:3.18}-\ref{tab:3.20} provide the action transition probabilities for the 7 deck states for maintenance actions {\it{Minor Repair, Major Repair}}, and {\it{Reconstruction}}, respectively. Table \ref{tab:3.22} reports the maintenance action durations in days for decks. Finally, Tables \ref{tab:3.24}-\ref{tab:3.25} provide the observation probabilities for 7 deck states, for low- and high-fidelity inspection techniques, respectively. 

\begin{table}[]
\centering
\begin{threeparttable} 
\caption[]{\em{Maintenance activities for interstate and primary pavements under different categories based on VDOT \cite{v2016a}}}
\label{tab:actions}
\begin{tabular}{|c|c|}
\toprule
\textbf{Activity Category}                             & \textbf{Activities}                                                                                                                                         \\ \hline
Do Nothing (DN)                                        & N/A                                                                                                                                                         \\ \hline
\multirow{3}{*}{Preventive   Maintenance (PM)}            & \begin{tabular}[c]{@{}c@{}}1. Minor patching (\textless{}5\% of pavement Area; \\ surface patching; depth 2")\end{tabular}                                  \\ 
                                                       & 2. Crack sealing                                                                                                                                            \\  
                                                       & \begin{tabular}[c]{@{}c@{}}3. Thin treatments (chip seal, slurry seal, latex, \\ Thin Hotmix Asphalt Concrete (THMACO), ‘Novachip’ etc.)\end{tabular}       \\ \hline
\multirow{4}{*}{Corrective   Maintenance (CM)}         & \begin{tabular}[c]{@{}c@{}}1. Moderate patching (\textless{}10\% of pavement area; \\ partial depth patching; depth 6")\end{tabular}                        \\ 
                                                       & \begin{tabular}[c]{@{}c@{}}2. Partial depth patching (\textless{}10\% of pavement area; \\ depth 4"-6"); and surface treatment\end{tabular}                  \\ 
                                                       & \begin{tabular}[c]{@{}c@{}}3. Partial depth patching (\textless{}10\% of pavement area; \\ depth 4"-6"); and thin ($\leq$ 2") AC overlay\end{tabular}        \\  
                                                       & 4. $\leq$ 2" Milling; and $\leq$ 2" AC overlay                                                                                                               \\ \hline
\multirow{4}{*}{Restorative   Maintenance (RM)}        & \begin{tabular}[c]{@{}c@{}}1. Heavy patching (20\% of pavement area;\\ full depth patching; depth 12")\end{tabular}                           \\ 
                                                       & 2. $\leq$ 4" Milling and replace with $\leq$ 4" AC overlay                                                                                                  \\ 
                                                       & \begin{tabular}[c]{@{}c@{}}3. Full depth patching (\textless{}20\% of pavement area; \\ full depth patching; depth 9"-12"); and 4"  AC overlay\end{tabular} \\ 
                                                       & 4. Cold in-place recycling                                                                                                                                  \\ \hline
\multirow{3}{*}{Rehabilitation   /Reconstruction (RC)} & 1. Mill, break and seat and 9"-12" AC overlay                                                                                                               \\ 
                                                       & 2. Reconstruction                                                                                                                                         \\ 
                                                       & 3. Full depth reclamation                                                                                                                                 \\ \bottomrule
\end{tabular}
\end{threeparttable} 
\end{table}

\begin{table}[]
\centering
\begin{threeparttable} 

\caption[]{\em{Observation probability $p(o_t|s_t)$ given actual CCI state $s_t$ with $\sigma _{error}^2 = 72$}}
\label{tab:3.2}
\begin{tabular}[tbh]{@{}ccccccc@{}}
\toprule
\begin{tabular}[c]{@{}c@{}}Actual \\ state\end{tabular} & $p(o_t = 6|s_t)$ & $p(o_t = 5|s_t)$ & $p(o_t = 4|s_t)$ & $p(o_t = 3|s_t)$ & $p(o_t = 2|s_t)$ & $p(o_t = 1|s_t)$ \\ \midrule
$s_t = 6$                                               & 0.687           & 0.259           & 0.054           & 0.000           & 0.000           & 0.000           \\
$s_t = 5$                                               & 0.276           & 0.422           & 0.297           & 0.005           & 0.000           & 0.000           \\
$s_t = 4$                                               & 0.023           & 0.139           & 0.648           & 0.167           & 0.022           & 0.001           \\
$s_t = 3$                                               & 0.000           & 0.003           & 0.266           & 0.455           & 0.248           & 0.028           \\
$s_t = 2$                                               & 0.000           & 0.000           & 0.031           & 0.224           & 0.486           & 0.259           \\
$s_t = 1$                                               & 0.000           & 0.000           & 0.000           & 0.005           & 0.059           & 0.936           \\ \bottomrule
\end{tabular} 
\end{threeparttable} 
\end{table}

\begin{table}[]
\centering
\begin{threeparttable} 
\caption[]{\em{Observation probability $p(o_t|s_t)$ given actual CCI state $s_t$ with $\sigma _{error}^2 = 18$}}
\label{tab:3.3}
\begin{tabular}{@{}ccccccc@{}}
\toprule
\begin{tabular}[c]{@{}c@{}}Actual \\ state\end{tabular} & $p(o_t = 6|s_t)$ & $p(o_t = 5|s_t)$ & $p(o_t = 4|s_t)$ & $p(o_t = 3|s_t)$ & $p(o_t = 2|s_t)$ & $p(o_t = 1|s_t)$ \\ \midrule
$s_t = 6$                                               & 0.801           & 0.197           & 0.002           & 0.000           & 0.000           & 0.000           \\
$s_t = 5$                                               & 0.153           & 0.664           & 0.183           & 0.000           & 0.000           & 0.000           \\
$s_t = 4$                                               & 0.001           & 0.078           & 0.822           & 0.099           & 0.000           & 0.000           \\
$s_t = 3$                                               & 0.000           & 0.000           & 0.149           & 0.693           & 0.158           & 0.000           \\
$s_t = 2$                                               & 0.000           & 0.000           & 0.001           & 0.137           & 0.718           & 0.144           \\
$s_t = 1$                                               & 0.000           & 0.000           & 0.000           & 0.000           & 0.042           & 0.958           \\ \bottomrule
\end{tabular} 
\end{threeparttable} 
\end{table}

\begin{table}[]
\centering
\begin{threeparttable} 

\caption[Minor Repair transition probabilities for 6 CCI states]{\em{Minor Repair transition probabilities for 6 CCI states.}}
\label{tab:3.5}
\begin{tabular}{@{}ccccccc@{}}
\toprule
\begin{tabular}[c]{@{}c@{}}Condition \\ State\end{tabular} & $s_{t+1} = 6$ & $s_{t+1} = 5$ & $s_{t+1} = 4$ & $s_{t+1} = 3$ & $s_{t+1} = 2$ & $s_{t+1} = 1$  \\ \midrule
$s_t = 6$                                               & 0.97      & 0.03      & 0.00      &           &           &           \\
$s_t = 5$                                               & 0.87      & 0.10      & 0.03      &           &           &           \\
$s_t = 4$                                               & 0.40      & 0.47      & 0.10      & 0.03      &           &           \\
$s_t = 3$                                               &           & 0.40      & 0.47      & 0.10      & 0.03      &           \\
$s_t = 2$                                               &           &           & 0.40      & 0.47      & 0.10      & 0.03      \\
$s_t = 1$                                               &           &           &           & 0.40      & 0.47      & 0.13      \\ \midrule
Deterioration rate                                      & \multicolumn{6}{c}{Does not change}                                   \\ \bottomrule
\end{tabular} 
\end{threeparttable}
\end{table}

\begin{table}[h!]
\centering
\begin{threeparttable} 

\caption[]{\em{Major Repair transition probabilities for 6 CCI states.}}
\label{tab:3.6}
\begin{tabular}{@{}ccccccc@{}}
\toprule
\begin{tabular}[c]{@{}c@{}}Condition \\ State\end{tabular} & $s_{t+1} = 6$ & $s_{t+1} = 5$ & $s_{t+1} = 4$ & $s_{t+1} = 3$ & $s_{t+1} = 2$ & $s_{t+1} = 1$ \\ \midrule
$s_t = 6$                                               & 1.00                          & 0.00                          & 0.00                          &                               &                               &                               \\
$s_t = 5$                                               & 0.96                          & 0.04                          & 0.00                          &                               &                               &                               \\
$s_t = 4$                                               & 0.80                          & 0.20                          & 0.00                          &                               &                               &                               \\
$s_t = 3$                                               & 0.65                          & 0.25                          & 0.10                          &                               &                               &                               \\
$s_t = 2$                                               & 0.50                          & 0.30                          & 0.20                          &                               &                               &                               \\
$s_t = 1$                                               & 0.40                          & 0.30                          & 0.30                          &                               &                               &                               \\ \midrule
Deterioration rate                                      & \multicolumn{6}{c}{Reset by 5 years}                                   \\ \bottomrule                                                          
\end{tabular} 
\end{threeparttable} 
\end{table}

\begin{table}[]
\centering
\begin{threeparttable} 
\caption[]{\em{Reconstruction transition probabilities for 6 CCI states.}}
\label{tab:3.7}
\begin{tabular}{@{}ccccccc@{}}
\toprule
\begin{tabular}[c]{@{}c@{}}Condition \\ State\end{tabular} & $s_{t+1} = 6$ & $s_{t+1} = 5$ & $s_{t+1} = 4$ & $s_{t+1} = 3$ & $s_{t+1} = 2$ & $s_{t+1} = 1$  \\ \midrule
$s_t = 6$                                               & 1.00                          & 0.00                          & 0.00                          &                               &                               &                               \\
$s_t = 5$                                               & 1.00                          & 0.00                          & 0.00                          &                               &                               &                               \\
$s_t = 4$                                               & 1.00                          & 0.00                          & 0.00                          &                               &                               &                               \\
$s_t = 3$                                               & 1.00                          & 0.00                          & 0.00                          &                               &                               &                               \\
$s_t = 2$                                               & 1.00                          & 0.00                          & 0.00                          &                               &                               &                               \\
$s_t = 1$                                               & 1.00                          & 0.00                          & 0.00                          &                               &                               &                               \\
\midrule
Deterioration rate                                      & \multicolumn{6}{c}{Reset to newly built}                                   \\ \bottomrule                                                                                                                       
\end{tabular} 
\end{threeparttable}
\end{table}

\begin{table}[]
\centering
\begin{threeparttable} 

\caption[Minor Repair transition probabilities for 5 IRI states]{\em{Minor Repair transition probabilities for 5 IRI states.}}
\label{tab:3.12}
\begin{tabular}{@{}cccccc@{}}
\toprule
Condition State & $s_{t+1} = 5$ & $s_{t+1} = 4$ & $s_{t+1} = 3$ & $s_{t+1} = 2$ & $s_{t+1} = 1$ \\ \midrule
$s_t = 5$       & 0.97          & 0.03          & 0.00          &               &               \\
$s_t = 4$       & 0.85          & 0.12          & 0.03          &               &               \\
$s_t = 3$       & 0.45          & 0.40          & 0.12          & 0.03          &               \\
$s_t = 2$       &               & 0.45          & 0.40          & 0.12          & 0.03          \\
$s_t = 1$       &               &               & 0.45          & 0.40          & 0.15          \\ \bottomrule
\end{tabular} 
\end{threeparttable} 
\end{table}

\begin{table}[]
\centering
\begin{threeparttable} 

\caption[Major Repair transition probabilities for 5 IRI states]{\em{Major Repair transition probabilities for 5 IRI states.}}
\label{tab:3.13}
\begin{tabular}{@{}cccccc@{}}
\toprule
Condition State & $s_{t+1} = 5$ & $s_{t+1} = 4$ & $s_{t+1} = 3$ & $s_{t+1} = 2$ & $s_{t+1} = 1$ \\ \midrule
$s_t = 5$       & 1.00          & 0.00          & 0.00          &               &               \\
$s_t = 4$       & 0.95          & 0.05          & 0.00          &               &               \\
$s_t = 3$       & 0.80          & 0.20          & 0.00          &               &               \\
$s_t = 2$       & 0.70          & 0.25          & 0.05          &               &               \\
$s_t = 1$       & 0.45          & 0.35          & 0.20          &               &               \\ \bottomrule
\end{tabular} 
\end{threeparttable} 
\end{table}

\begin{table}[]
\centering
\begin{threeparttable} 
\caption[Reconstruction transition probabilities for 5 IRI states]{\em{Reconstruction transition probabilities for 5 IRI states.}}
\label{tab:3.14}
\begin{tabular}{@{}cccccc@{}}
\toprule
Condition State & $s_{t+1} = 5$ & $s_{t+1} = 4$ & $s_{t+1} = 3$ & $s_{t+1} = 2$ & $s_{t+1} = 1$ \\ \midrule
$s_t = 5$       & 1.00          & 0.00          & 0.00          &               &               \\
$s_t = 4$       & 1.00          & 0.00          & 0.00          &               &               \\
$s_t = 3$       & 1.00          & 0.00          & 0.00          &               &               \\
$s_t = 2$       & 1.00          & 0.00          & 0.00          &               &               \\
$s_t = 1$       & 1.00          & 0.00          & 0.00          &               &               \\ \bottomrule
\end{tabular} 
\end{threeparttable} 
\end{table}

\begin{table}[]
\centering
\begin{threeparttable} 

\caption[Inspection action costs (in USD/lane-mile or USD/m2) for CCI using 3 different techniques]{{\em{Inspection action costs (in}} USD/lane-mile or USD/m\textsuperscript{2}) {\it{for CCI using 3 fidelity levels.}}}
\label{tab:3.4}
\begin{tabular}{@{}ccccc@{}}
\toprule
\begin{tabular}[c]{@{}c@{}}Inspection\\ Technique\end{tabular} & Description   & \begin{tabular}[c]{@{}c@{}}Observation error \\($\sigma _{error}^2$)\end{tabular} & \begin{tabular}[c]{@{}c@{}}Cost (USD/lane-mile)\end{tabular} & \begin{tabular}[c]{@{}c@{}}Cost (USD/m\textsuperscript{2})\end{tabular} \\ \midrule
$i_2$                                                          & High fidelity & 18                                                                               & 0.57                                                             & 0.16                                                        \\
$i_1$                                                          & Low fidelity  & 72                                                                               & 0.27                                                             & 0.08                                                        \\
$i_0$                                                          & No Inspection & $\infty$                                                                      & 0.00                                                             & 0.00 \\    \bottomrule                                                  
\end{tabular} 
\end{threeparttable} 
\end{table}

\begin{table}[]
\centering
\begin{threeparttable} %
\caption[Observation probability $p(o_t|s_t)$ for IRI given state $s_t$]{\em{Observation probability $p(o_t|s_t)$ for IRI given actual state $s_t$}}
\label{tab:3.15}
\begin{tabular}{@{}cccccccc@{}}
\toprule
\begin{tabular}[c]{@{}c@{}}Inspection   \\ Technique\end{tabular} & \begin{tabular}[c]{@{}c@{}}SD \\ (m/km)\end{tabular} & \begin{tabular}[c]{@{}c@{}}$p(o_t = j-$\\ $2|s_t = j$\end{tabular} & \begin{tabular}[c]{@{}c@{}}$p(o_t = j-$\\ $1|s_t = j$\end{tabular} & \begin{tabular}[c]{@{}c@{}}$p(o_t = j$\\ $|s_t = j$\end{tabular} & \begin{tabular}[c]{@{}c@{}}$p(o_t = j+$\\ $1|s_t = j$\end{tabular} & \begin{tabular}[c]{@{}c@{}}$p(o_t = j+$\\ $2|s_t = j$\end{tabular} & \begin{tabular}[c]{@{}c@{}}Cost \\(USD/m\textsuperscript{2})\end{tabular} \\ \midrule
$i_2$                                                             & 0.08                                                 & 0                                                                & 0.05                                                             & 0.90                                                           & 0.05                                                             & 0                                                                & 0.10                                                            \\
$i_1$                                                             & 0.32                                                 & 0                                                                & 0.20                                                             & 0.60                                                           & 0.20                                                             & 0                                                                & 0.03                                                            \\
$i_0$                                                             & $\infty$                                             & 0.20                                                             & 0.20                                                             & 0.20                                                           & 0.20                                                             & 0.20                                                             & 0                                                               \\ \bottomrule
\end{tabular} 
\end{threeparttable} 
\end{table}

\begin{table}[]
\centering
\begin{threeparttable} 

\caption[Maintenance actions duration for pavements (in days per lane-mile)]{\em{Maintenance actions duration for pavements (in days per lane-mile).}}
\label{tab:3.9}
\begin{tabular}{@{}ccc@{}}
\toprule
Actions        & Days per lane-mile & \begin{tabular}[c]{@{}c@{}}Additional days \\ per mile for shoulder, etc.\end{tabular} \\ \midrule
Do Nothing     & 0                  & 0                                                                                      \\
Minor Repair   & 3.5                & 1                                                                                      \\
Major Repair   & 6.5                & 2                                                                                      \\
Reconstruction & 32                 & 10                                                                                     \\ \bottomrule
\end{tabular} 
\end{threeparttable} %
\end{table}

\begin{table}[tbh]
\centering
\begin{threeparttable}
\caption[Minor Repair transition probabilities for deck states]{\em{Minor Repair transition probabilities for deck states.}}
\label{tab:3.18}
\begin{tabular}{@{}cccccccc@{}}
\toprule
\begin{tabular}[c]{@{}c@{}}Condition \\ State\end{tabular} & $s_{t+1} = 9$ & $s_{t+1} = 8$ & $s_{t+1} = 7$ & $s_{t+1} = 6$ & $s_{t+1} = 5$ & \begin{tabular}[c]{@{}c@{}}$s_{t+1}$ \\ $= 4,...$\end{tabular} & \begin{tabular}[c]{@{}c@{}}$s_{t+1}$\\ $ = failed$\end{tabular} \\ \midrule
$s_t = 9$                                                  & 0.97          & 0.03          & 0.00          &               &               &                                                                &                                                                 \\
$s_t = 8$                                                  & 0.85          & 0.12          & 0.03          &               &               &                                                                &                                                                 \\
$s_t = 7$                                                  & 0.40          & 0.45          & 0.12          & 0.03          &               &                                                                &                                                                 \\
$s_t = 6$                                                  &               & 0.40          & 0.45          & 0.12          & 0.03          &                                                                &                                                                 \\
$s_t = 5$                                                  &               &               & 0.40          & 0.45          & 0.12          & 0.03                                                           &                                                                 \\
$s_t = 4,...$                                              &               &               &               & 0.40          & 0.45          & 0.15                                                           &                                                                 \\
$s_t = failed$                                             &               &               &               &               &               &                                                                & 1.0                                                             \\ \midrule
Deterioration rate                                         & \multicolumn{7}{c}{Does not change}                                                                                                                                                                              \\ \bottomrule
\end{tabular} 
\end{threeparttable} 
\end{table}

\begin{table}[]
\centering
\begin{threeparttable} 

\caption[Major Repair transition probabilities for deck states]{\em{Major Repair transition probabilities for deck states.}}
\label{tab:3.19}
\begin{tabular}{@{}cccccccc@{}}
\toprule
\begin{tabular}[c]{@{}c@{}}Condition \\ State\end{tabular} & $s_{t+1} = 9$ & $s_{t+1} = 8$ & $s_{t+1} = 7$ & $s_{t+1} = 6$ & $s_{t+1} = 5$ & \begin{tabular}[c]{@{}c@{}}$s_{t+1}$ \\ $= 4,...$\end{tabular} & \begin{tabular}[c]{@{}c@{}}$s_{t+1}$\\ $ = failed$\end{tabular} \\ \midrule
$s_t = 9$                                                  & 1.00          & 0.00          & 0.00          &               &               &                                                                &                                                                 \\
$s_t = 8$                                                  & 0.95          & 0.05          & 0.00          &               &               &                                                                &                                                                 \\
$s_t = 7$                                                  & 0.80          & 0.20          & 0.00          &               &               &                                                                &                                                                 \\
$s_t = 6$                                                  & 0.60          & 0.30          & 0.10          &               &               &                                                                &                                                                 \\
$s_t = 5$                                                  & 0.40          & 0.40          & 0.20          &               &               &                                                                &                                                                 \\
$s_t = 4,...$                                              & 0.30          & 0.40          & 0.30          &               &               &                                                                &                                                                 \\
$s_t = failed$                                             &               &               &               &               &               &                                                                & 1.0                                                             \\ \midrule
Deterioration rate                                         & \multicolumn{7}{c}{Reset by 5 years}                                                                                    \\ \bottomrule
\end{tabular} 
\end{threeparttable} 
\end{table}

\begin{table}[tbh]
\centering
\begin{threeparttable} 

\caption[Reconstruction transition probabilities for deck states]{\em{Reconstruction transition probabilities for deck states.}}
\label{tab:3.20}
\begin{tabular}{@{}cccccccc@{}}
\toprule
\begin{tabular}[c]{@{}c@{}}Condition \\ State\end{tabular} & $s_{t+1} = 9$ & $s_{t+1} = 8$ & $s_{t+1} = 7$ & $s_{t+1} = 6$ & $s_{t+1} = 5$ & \begin{tabular}[c]{@{}c@{}}$s_{t+1}$ \\ $= 4,...$\end{tabular} & \begin{tabular}[c]{@{}c@{}}$s_{t+1}$\\ $ = failed$\end{tabular} \\ \midrule
$s_t = 9$                                                  & 1.00          &           &          &               &               &                                                                &                                                                 \\
$s_t = 8$                                                  & 1.00          &         &           &               &               &                                                                &                                                                 \\
$s_t = 7$                                                  & 1.00          &          &          &               &               &                                                                &                                                                 \\
$s_t = 6$                                                  & 1.00          &          &         &               &               &                                                                &                                                                 \\
$s_t = 5$                                                  & 1.00          &         &         &               &               &                                                                &                                                                 \\
$s_t = 4,...$                                              & 1.00          &          &          &               &               &                                                                &                                                                 \\
$s_t = failed$                                             & 1.00          &          &          &               &               &                                                                &                                                                 \\ \midrule
Deterioration rate                                         & \multicolumn{7}{c}{Reset to newly built deck}                                                                                                                                                                    \\ \bottomrule
\end{tabular} 
\end{threeparttable} 
\end{table}

\begin{table}[]
\centering
\begin{threeparttable} 

\caption[Observation probability $p(o_t|s_t)$ given state $s_t$, for low fidelity inspection techniques]{ \em{Observation probability $p(o_t|s_t)$ given state $s_t$, for low fidelity inspection techniques for decks. }}
\label{tab:3.24}
\begin{tabular}{cccccccc}
\hline
{ \begin{tabular}[c]{@{}c@{}}Condition\\ State\end{tabular}} & \begin{tabular}[c]{@{}c@{}}$p(o_t = 9$\\ $|s_t)$\end{tabular} & \begin{tabular}[c]{@{}c@{}}$p(o_t = 8$\\ $|s_t)$\end{tabular} & \begin{tabular}[c]{@{}c@{}}$p(o_t = 7$\\ $|s_t)$\end{tabular} & \begin{tabular}[c]{@{}c@{}}$p(o_t = 6$\\ $|s_t)$\end{tabular} & \begin{tabular}[c]{@{}c@{}}$p(o_t = 5$ \\ $|s_t)$\end{tabular} & \begin{tabular}[c]{@{}c@{}}$p(o_t = $\\ $4,...|s_t)$\end{tabular} & \begin{tabular}[c]{@{}c@{}}$p(o_t = $ \\ $failed|s_t)$\end{tabular} \\ \hline
$s_t = 9$                                                                        & 0.80                                                           & 0.15                                  & 0.05                                                           &                                                               &                                                               &                                                                   &                                                                     \\
$s_t = 8$                                                                        & 0.15                                                           & 0.65                                                          & 0.15                                                           & 0.05                                                          &                                                               &                                                                   &                                                                     \\
$s_t = 7$                                                                        & 0.05                                                           & 0.15                                                          & 0.60                                                           & 0.15                                                          & 0.05                                                          & 0.0                                                               &                                                                     \\
$s_t = 6$                                                                        &                                                                & 0.05                                                          & 0.15                                                           & 0.60                                                          & 0.15                                                          & 0.05                                                              &                                                                     \\
$s_t = 5$                                                                        &                                                                &                                                               & 0.05                                                           & 0.15                                                          & 0.65                                                          & 0.15                                                              &                                                                     \\
$s_t = 4,...$                                                                    &                                                                &                                                               &                                                                & 0.05                                                          & 0.15                                                          & 0.80                                                              &                                                                     \\
$s_t = failed$                                                                   &                                                                &                                                               &                                                                &                                                               &                                                               &                                                                   & 1.0                                                                 \\ \hline
\end{tabular} 
\end{threeparttable} 
\end{table}

\begin{table}[]
\centering
\begin{threeparttable} 
\caption[Observation probability $p(o_t|s_t)$ given state $s_t$, for high fidelity inspection techniques]{ \em{Observation probability $p(o_t|s_t)$ given state $s_t$, for high fidelity inspection techniques for decks. }}
\label{tab:3.25}
\begin{tabular}{cccccccc}
\hline
{ \begin{tabular}[c]{@{}c@{}}Condition\\ State\end{tabular}} & \begin{tabular}[c]{@{}c@{}}$p(o_t = 9 $\\ $|s_t)$\end{tabular} & \begin{tabular}[c]{@{}c@{}}$p(o_t = 8 $\\ $|s_t)$\end{tabular} & \begin{tabular}[c]{@{}c@{}}$p(o_t = 7$\\$|s_t)$\end{tabular} & \begin{tabular}[c]{@{}c@{}}$p(o_t = 6$\\ $|s_t)$\end{tabular} & \begin{tabular}[c]{@{}c@{}}$p(o_t = 5$ \\ $|s_t)$\end{tabular} & \begin{tabular}[c]{@{}c@{}}$p(o_t = $\\ $4,...|s_t)$\end{tabular} & \begin{tabular}[c]{@{}c@{}}$p(o_t = $ \\ $failed|s_t)$\end{tabular} \\ \hline
$s_t = 9$                                                                        & 0.90                                                           & 0.10                                  &                                                                &                                                               &                                                               &                                                                   &                                                                     \\
$s_t = 8$                                                                        & 0.10                                                           & 0.80                                                          & 0.10                                                           &                                                               &                                                               &                                                                   &                                                                     \\
$s_t = 7$                                                                        &                                                                & 0.10                                                          & 0.80                                                           & 0.10                                                          &                                                               &                                                                   &                                                                     \\
$s_t = 6$                                                                        &                                                                & 0.05                                                          & 0.10                                                           & 0.80                                                          & 0.10                                                          &                                                                   &                                                                     \\
$s_t = 5$                                                                        &                                                                &                                                               &                                                                & 0.10                                                          & 0.80                                                          & 0.10                                                              &                                                                     \\
$s_t = 4,...$                                                                    &                                                                &                                                               &                                                                &                                                               & 0.10                                                          & 0.90                                                              &                                                                     \\
$s_t = failed$                                                                   &                                                                &                                                               &                                                                &                                                               &                                                               &                                                                   & 1.0                                                                 \\ \hline
\end{tabular} 
\end{threeparttable} 
\end{table}

\begin{table}[tbh]
\centering
\begin{threeparttable} 

\caption[]{ \em{ Deck maintenance action durations in days.}}
\label{tab:3.22}
\begin{tabular}{@{}cccc@{}}
\toprule
{Actions} & Type I Bridges & Type II Bridges           & Type III Bridges \\ \midrule
Do Nothing                     & 0              & 0 & 0                \\
Minor Repair                   & 25             & 12                        & 6                \\
Major Repair                   & 70             & 30                        & 15               \\
Reconstruction                 & 300            & 150                       & 70               \\ \bottomrule
\end{tabular} 
\end{threeparttable} 
\end{table}

\end{document}